\documentclass[12pt]{article}
\pdfoutput=1
\usepackage[margin=3cm]{geometry}
\usepackage[utf8]{inputenc}
\usepackage{amsmath}
\usepackage{amssymb}
\usepackage{authblk} % for authors
\usepackage{xcolor}
\usepackage{physics}
\usepackage{graphicx} % for graphics in pdf_tex figures
\usepackage{float}
\usepackage{tikz}
\usetikzlibrary{calc}
\usepackage{comment}

\usepackage{hyperref}
\usepackage{varioref}       
\usepackage{cleveref}     

\crefname{table}{table}{tables}
\Crefname{table}{Table}{Tables}
\crefname{figure}{figure}{figures}
\Crefname{figure}{Figure}{Figures}

\definecolor{tealblue}{rgb}{0.21, 0.56, 0.63}
\hypersetup{colorlinks=true,allcolors = tealblue,linktocpage=true}

\usepackage{pgf} % for pgf pics, might have to remove on publication if pgfs have to be prerendered

\usepackage[
        backend=bibtex,
        sorting=none,
        style=phys,
        eprint=true,
        doi=false,
        biblabel=brackets
        ]{biblatex}

\bibliography{bibfilename}

\usepackage[backgroundcolor=white!95!red, linecolor=cyan, bordercolor=cyan, textsize=footnotesize]{todonotes}

\newcommand{\commie}[1]{}

\numberwithin{equation}{section}
\numberwithin{table}{section}

\newenvironment{eqaed}
    {\begin{equation}
    \begin{aligned}
    }
    { 
    \end{aligned}
    \end{equation}
    \ignorespacesafterend
    }

\newcommand{\modZ}[2]{Z \big[ {\tiny  {\textstyle{#1 \atop #2}} } \big]  }

\newcommand{\Jtheta}[4]{ \theta \big[ {\tiny  {\textstyle{#1 \atop #2}} } \big] 	\small( #3 | #4 \small) }

\newcommand{\Jbartheta}[4]{ \overline{\theta} \big[ {\tiny  {\textstyle{#1 \atop #2}} } \big]  \small( #3 | #4 \small) }

\begin{document}

\title{Anomaly constraints for heterotic strings \\ and supergravity in six dimensions}
\date{}

\author{Ivano Basile\thanks{ivano.basile@lmu.de}}
\affil{\emph{Arnold-Sommerfeld Center for Theoretical Physics}\\ \emph{Ludwig Maximilians Universit\"at M\"unchen}\\ \emph{Theresienstraße 37, 80333 M\"unchen, Germany}}
\author{Giorgio Leone\thanks{giorgio.leone@unito.it}}
\affil{ {\em Dipartimento di Fisica, Universit\`a di Torino and INFN Sezione di Torino}
        \\
        {\em Via Pietro Giuria 1, 10125 Torino, Italy}}

\maketitle

\begin{abstract}
    
    \noindent The landscape of six-dimensional supergravities is dramatically constrained by the cancellation of gauge and gravitational anomalies, but the full extent of its implications has not been uncovered. We explore the cancellation of global anomalies of the Dai-Freed type in this setting with abelian and simply laced gauge groups, finding novel constraints. In particular, we exclude arbitrarily large abelian charges in an infinite family of theories for certain types of quadratic refinements, including a specific one defined in the literature. We also show that the Gepner orientifold with no tensor multiplets is anomaly-free for a different choice, as well as a number of heterotic models with and without spacetime supersymmetry in six dimensions. The latter analysis extends previous results in ten dimensions to some lower-dimensional settings in the heterotic landscape.
    
\end{abstract}

\thispagestyle{empty}

\newpage

\tableofcontents

\thispagestyle{empty}

\newpage

\pagenumbering{arabic}

\section{Introduction}\label{sec:introduction}

Despite decades of impressive progress, the lack of an all-encompassing formulation of string theory remains an obstacle toward a thorough understanding of the theory and its physical implications. While the ultraviolet (UV) physics is universal, at least perturbatively \cite{Gross:1987kza, Gross:1987ar, Mende:1989wt}, the infrared (IR) properties of the vacuum can be manifold. This property, expected to arise in any theory of gravity by compactification, gives rise to the string landscape. The many partial but complementary formulations at our disposal \cite{Witten:1995ex} cover some corners of the landscape, and thus the important task of exploring it further dovetails with the search for the underlying physical principles of quantum gravity. In light of this state of affairs, top-down constructions are faced with a ``lamppost'' effect, and it is sometimes unclear which properties pertain to the whole landscape and which cannot be extrapolated beyond the limited cases where one can achieve quantitative control. In order to address this ``missing corner problem'' \cite{Brennan:2017rbf}, one can follow a complementary bottom-up approach, seeking general properties entailed by the consistency of quantum gravity rather than extrapolating from top-down examples. This is the essence of the swampland program \cite{Vafa:2005ui} (see also \cite{Palti:2019pca, vanBeest:2021lhn, Grana:2021zvf, Agmon:2022thq} for reviews). Ultimately, one would like to find a perfect match between these consistency conditions and the string landscape of effective field theories (EFTs) of gravity. This daunting task is enormously simplified in settings with several supercharges, where the pool of potentially allowable supergravities can be classified. The constraining power of (extended) supersymmetry can be combined with unitarity, which requires that anomalies of various types cancel. The results are quite striking: the swampland task is essentially complete in supersymmetric theories in dimensions $d > 6$ \cite{Kim:2019vuc, Montero:2020icj, Hamada:2021bbz, Bedroya:2021fbu}, and there are many constraints in lower dimensions $d \leq 6$ as well \cite{Kumar:2010ru, Park:2011wv, Taylor:2018khc, Kim:2019vuc, Lee:2019skh, Kim:2019ths, Katz:2020ewz, Angelantonj:2020pyr, Tarazi:2021duw, Martucci:2022krl, Baykara:2023plc, Hayashi:2023hqa}.

\noindent In the spirit of continuing the exploration of the landscape and consistency conditions within the simplest and most constrained settings, one is led to minimal supergravity in six dimensions, denoted $\mathcal{N}=(1,0)$ or $\mathcal{N}=1$ when no ambiguities arise. This is because purely gravitational anomalies can exist in six and ten dimensions\footnote{In this paper we only consider $d \geq 4$, where the consistency conditions of gravity are much more restrictive.}, and while the latter case comprises just two theories the former is much richer. There are several indications that the landscape of consistent $6d$ $\mathcal{N}=1$ supergravities is finite \cite{Kumar:2010ru, Park:2011wv, Taylor:2018khc, Kim:2019vuc, Lee:2019skh, Tarazi:2021duw}, but a full classification is lacking. Still, this framework provides an ideal context to study more refined constraints, such as the ones coming from the cancellation of global anomalies. Dai-Freed anomalies \cite{Garcia-Etxebarria:2018ajm} comprise both local \cite{Alvarez-Gaume:1983ict, Alvarez-Gaume:1983ihn, Alvarez-Gaume:1984zlq, Alvarez-Gaume:1984zst, Alvarez-Gaume:2022aak} and global \cite{Witten:1982fp, Witten:1985xe, Witten:1985mj, Witten:2019bou} anomalies in the traditional sense, but they also include more general anomalies arising from spacetime topology change. Demanding that these anomalies cancel can entail additional, stronger constraints \cite{Garcia-Etxebarria:2018ajm, Monnier:2018nfs, Lee:2022spd}, and their cancellation in (non-)supersymmetric string theories in ten dimensions provides further evidence for their non-perturbative consistency \cite{Basile:2023knk}.

\noindent In this paper we study Dai-Freed anomalies in $6d$ $\mathcal{N} = 1$ supergravity, focusing on families of theories with $n_T = 1$ tensor multiplets and simply laced gauge groups. We will also discuss $U(1)$ gauge groups and a model with $n_T = 0$, as well as some non-supersymmetric heterotic models, showing that the constraints arising from Dai-Freed anomaly cancellation extend beyond the lamppost of supersymmetry along the lines of \cite{Basile:2023knk}. The upshot of our analysis shows that the cancellation of Dai-Freed anomalies excludes some theories that were not previously ruled out, up to an assumption on the structure of the anomaly theory which we will specify. In the case of $U(1)$, this argument can exclude arbitrarily large abelian charges in the hypermultiplets discussed in \cite{Taylor:2018khc}. The mentioned assumption concerns the structure of the anomaly theory of the chiral $2$-form fields, which in six dimensions play an important role. According to our analysis, the ultimate fate of these theories thus depends on these data, which is inaccessible from their local Lagrangian formulation. At present, there is no understanding on how to access such information from stringy inputs. Given this state of affairs, we can limit ourselves to assume the construction that has already been discussed in literature \cite{Hsieh:2020jpj}, allowing us to exclude some models with abelian and simply-laced gauge groups. Another point of view, closer in spirit to the swampland program, is to consider this result as novel and powerful restrictions on possible EFTs, if one includes global data in their specification. A possible way to better understand these matters is to analyse the case in which chiral fields contribute to the anomaly theory of six-dimensional heterotic orbifolds and orientifolds models, since in these cases it expected to vanish. Therefore, global data must be also fixed by the choice of vacuum \cite{Dierigl:2022zll}. As it turns out, the choices that cancel the anomaly sometimes differ from the one discussed in \cite{Hsieh:2020jpj}, and thus further investigation is required to settle the issue. Nevertheless, when the $2$-form fields of interest are non-chiral, such ambiguities disappear and the anomaly theory turns out to vanish for all the models at stake, including a number of non-supersymmetric heterotic orbifolds\footnote{More precisely, there could be additional potentially anomalous backgrounds that we have not found. All the anomalies that we have computed using Lens spaces vanish.}. This analysis thus extends the results of \cite{Tachikawa:2021mvw, Tachikawa:2021mby} to some non-supersymmetric settings and those of \cite{Basile:2023knk} to lower dimensions.

\noindent The contents of the paper are summarised as follows. In \cref{sec:anomaly_intro} we present the formalism with which we describe global anomalies of chiral fields. In \cref{sec:dai-freed_anomalies} we introduce Dai-Freed anomalies in the specific context of $6d$ supergravity, where the Green-Schwarz mechanism plays a central role. Then we move on to the main part of the paper, computing anomalies on Lens spaces for supergravity theories for the simply laced gauge groups appearing in heterotic string constructions: in \cref{sec:anomaly_SUn} we discuss special unitary groups, in \cref{sec:anomaly_Spin2n} we discuss Spin groups and in \cref{sec:anomaly_E7E8} we discuss the exceptional groups $E_7$ and $E_8$. When describing anomalies for special unitary groups, we provide a separate, more detailed exposition for $SU(2)$ in \cref{sec:anomaly_SU2}, where a comparison with the string landscape and the general properties of elliptic genera is performed. Abelian charges behave differently, and we discuss them in \cref{sec:anomaly_U1} where we also show in \cref{sec:abelian_no_tensors} that an infinite family with arbitrary large charges are excluded, at least for certain types of quadratic refinement for the self-dual tensor. Finally, as a first step to extend our approach beyond the main setting of supergravity with $n_T = 1$, in \cref{sec:non-susy_models} we show that a number of non-supersymmetric heterotic models are devoid of Dai-Freed anomalies on Lens spaces. Similarly, in \cref{sec:gepner} we show that the anomaly cancels for the peculiar ``non-geometric'' Gepner orientifold with $n_T = 0$ for certain choices of quadratic refinement which however differ from the one defined in \cite{Hsieh:2020jpj}. These consistency checks are not \emph{a priori} trivial, since the analysis of \cite{Tachikawa:2021mby} covers heterotic settings (assuming the Stolz-Teichner conjecture \cite{Stolz:2011zj}) while the analysis of \cite{Basile:2023knk} covers smooth geometric compactifications of ten-dimensional models.

\section{Anomalies and unitarity in (super)gravity}\label{sec:anomaly_intro}

Unitarity is perhaps the most fundamental requirement that a quantum field theory should possess in order to be consistent. When chiral fields are present, quantum effects can spoil this property by breaking gauge invariance of an internal or local Lorentz symmetry when gravity is involved \cite{Alvarez-Gaume:1983ict, Alvarez-Gaume:1984zlq, Alvarez-Gaume:1983ihn, Alvarez-Gaume:1984zst, Adler:1969gk,Bell:1969ts, Fujikawa:2004cx} (see \emph{e.g.} \cite{Bilal:2008qx, Alvarez-Gaume:2022aak,Harvey:2005it,10.1093/acprof:oso/9780198507628.003.0004} for reviews). Thus, the lack of unitarity manifests itself into an anomalous transformation of the one-loop effective action and can be ultimately traced back to the ambiguities occurring in the definitions of the partition function of the chiral fermionic and bosonic fields. Recently, it has been shown \cite{Witten:2019bou, Yonekura:2016wuc, Hsieh:2020jpj} that these ambiguities can be clarified by describing the chiral fields on a given spacetime $X$ as a boundary mode of a gapped Dirac spinor or a non-chiral $(p+1)$-form living in a $(d+1)$-dimensional space $Y$ such that $\partial Y=X$ with suitable elliptic boundary conditions\footnote{These results have been obtained working in Euclidean signature. The case of Minkowski signature has been covered in \cite{Fukaya:2019qlf}.} \cite{Witten:2019bou, Yonekura:2016wuc, Hsieh:2020jpj, Witten:2018lgb}, required in order for the Dirac operator to be self-adjoint on $X$. The dependence of the partition function on the choices of $Y$ and on the boundary conditions $L$ is given by the {\em  Dai-Freed theorems} on the index of the Dirac operator \cite{Dai:1994kq,Freed:1986hv}. As a result, the partition function for a chiral fermion and a given spacetime extension $Y$ is well-defined and reads \cite{Witten:2019bou,Yonekura:2016wuc} 
	\begin{equation}\label{ferpart}
		\mathcal{Z}_{\frac{1}{2},+} \big (L, Y \big )= \left | \text{Pf} \left ( D_{\text{D}}^{+}(X) \right ) \right | e^{-2 i \pi \eta_\text{D} (Y)}
	\end{equation}
	where $L$ localises a chiral fermion of a given chirality on the boundary $X$ and the Pfaffian of its Dirac operator corresponds to the regularised product of the positive eigenvalues\footnote{When the Dirac operator in the bulk $Y$ has zero modes, the boundary conditions should be properly modified (see for instance \cite{Yonekura:2016wuc,Dai:1994kq} for details). However for the analysis of anomalies such subtleties play no role and we shall skip them.} whose phase is given by the so-called {\em eta invariant} $\eta (Y)$ \cite{Dai:1994kq,Freed:1986hv}. 
	
\noindent In similar fashion one can write the partition function of a chiral gravitino. However, one has to take into account that the degrees of freedom of a gravitino comprise a Rarita-Schwinger field and a Weyl fermion of opposite chirality. A Rarita-Schwinger field in $d+1$ dimensions localises to a chiral Rarita-Schwinger field and a Weyl fermion of opposite chirality on the $d$-dimensional boundary \cite{Debray:2023yrs, Debray:2021vob}, while the Dirac field localises to a Weyl fermion as above. As a result, the structure of the partition function is dictated by
	\begin{equation}\label{gravitinopart}
		\mathcal{Z}_{\text{grav},+} \big (L, Y \big )= \frac{ \mathcal{Z}_{\frac{3}{2},+} \big (L, Y \big ) }{\mathcal{Z}_{\frac{1}{2},+} \big (L, Y \big )^2} = \frac{\left | \text{Pf} \left ( D_{\text{RS}}^{+}(X) \right ) \right |}{\left | \text{Pf} \left ( D_{\text{D}}^{+}(X) \right ) \right |^2 } \, e^{-2 i \pi \eta_{\text{RS}} (Y)+ 4 i \pi \eta_{\text{D}}(Y) } \,  ,
	\end{equation}
	where the operator $ D_{\text{RS}}^{+}(X)$ is the Rarita-Schwinger operator defined on the product between the spin and the tangent bundle, and its eta invariant is related to the density index via the Atiyah-Patodi-Singer (APS) index theorem \cite{Atiyah:1975jf,Debray:2023yrs, Debray:2021vob} 
	\begin{equation}
		\text{Index}^{\text{RS}}= \eta_{\text{RS}}(\partial Z) + \int_Z \left ( I^{\text{RS}} - I^D \right ).
	\end{equation} 
	However this is not the end of the story, since string theory and supergravity provide additional ingredients in their spectra, namely $p$-form fields. Although the action and partition function of non-chiral forms is well-understood and under control, one faces unavoidable difficulties in the definition of a similar action for chiral forms, thus preventing one from writing a consistent partition function straightforwardly. Furthermore, the standard techniques that describe abelian gauge fields (whether chiral or not) in terms of differential forms miss all the restrictions arising from a consistent coupling to matter in topologically non-trivial backgrounds. Indeed when the topology of the manifold is non-trivial, a consistent coupling to the wavefunction naturally leads to the presence of magnetic monopoles implying the quantisation of the electric charge, the celebrated {\em Dirac quantisation condition} \cite{PhysRevD.12.3845, PhysRev.74.817}. More generally, it restricts charges to an integer lattice \cite{PhysRev.144.1087, PhysRev.176.1489}. Therefore a $p$-form field is not only described by the De Rham cohomology classes of the electric and magnetic currents with the suitable support conditions \cite{Freed:2000ta,Szabo:2012hc}, but it should properly incorporate the holonomy of the $U(1)$ gauge bundle encoding, eventually, the restriction imposed on the corresponding charges by Dirac quantisation. From the point of view of the functional integral this is apparent, since one sums over gauge bundles and their isomorphism classes correspond to elements in integer cohomology. More precisely, these consideration mean that a $p$-form gauge field is mathematically described by a (generalised) differential cohomology known as Cheeger-Simons \cite{10.1007/BFb0075216} or, equivalently, Deligne cohomology \cite{Deligne}. These cohomology groups are defined as the set of homomorphisms \cite{Hsieh:2020jpj, Freed:2006yc, Freed:2006ya, Freed:2000ta, Hopkins:2002rd}
	\begin{equation} \label{CheegerSimons}
		\chi \in \check{H}^p(X) \subset \text{Hom}(Z_{p-1}(X,\mathbb{Z}), U(1) )
	\end{equation}
    such that there exists a globally defined closed $p$-form with integer periods $F_\chi \in \Omega_{\mathbb{Z}}^p(X)$ which on exact cycles satisfies
    \begin{equation}
    	\chi(\partial \gamma)= \text{exp} \bigg (2 \pi i \int_{\gamma} F_\chi \bigg ) \, , \qquad \text{with} \qquad  \gamma \ \ \in \  C_{p}(X,\mathbb{Z}) \, .
    \end{equation}
    Although the result so far encodes all the additional properties that were missing, the relationship with the usual treatment is far from obvious. Indeed, to make contact with the standard approach to gauge fields, we have to clarify how the gauge connection and gauge transformations emerge in this setup. The gauge connection $A \in C^{p-1}(X, \mathbb{R})$ is related to the holonomy for a general cycle $\Gamma \in Z_{p-1}(X, \mathbb{Z})$ via
    \begin{equation} \label{gaugeconn}
    	\chi(\Gamma)= \text{exp} \bigg ( 2 \pi i\int_\Gamma A \bigg ) \, ,
    \end{equation}     
     which however it is not required to be a globally defined $(p-1)$-form. The definition in  \cref{gaugeconn} allows a different but equivalent characterisation of exact cycles allowing to identify the coboundary $\delta A$ with the field strength $F_\chi$ up to an integer cochain $N_\chi \in C^p(X, \mathbb{Z})$ that defines the {\em characteristic class of $\chi$} 
     \begin{equation} \label{charclass}
     	N_\chi= F_\chi - \delta A \, .
     \end{equation}
      However the latter definition and \cref{gaugeconn} have a gauge redundancy, as they are unaffected by the transformations
      	\begin{equation} \label{gaugered}
      	A \to A + \delta a + n \, , \qquad 	N \to N - \delta n
      \end{equation}
      with $ a \in C^{p-2}(X,\mathbb{R}) $ and $ n \in C^{p-1}(X, \mathbb{Z})$, providing equivalence classes forming a cohomology group isomorphic to the one in \cref{CheegerSimons}. Therefore, in this picture the non-trivial topological data is encoded in the characteristic integer cohomology element $[N] \in H^p(X,\mathbb{Z})$, which carries torsional information invisible to De Rham cohomology curvature classes $[F] \in H_{\text{dR}}^p(X)$. 
	
\noindent This fact has deep consequences for the structure and dynamics of the theory. If $p$-form fields are present in the spectrum, the action and partition function should make sense at the level of differential cohomology, requiring the associated action to be invariant under the transformations in \cref{gaugered}. For non-chiral $p$-form fields $\check{\mathbf{A}} \in \check{H}^{p}(X)$ this can be done following \cite{Hsieh:2020jpj, Gaiotto:2014kfa}, namely extending the spacetime $X$ to the ``bulk'' manifold $Y$ such that $\partial Y=X$. Without background fields, gauge invariance comes automatically. However, when the gauge field is coupled both electrically to $\check{\mathbf{B}} \in \check{H}^{p+1}(X)$ and magnetically to $\check{\mathbf{C}} \in \check{H}^{d-p+1}(X)$, one has to introduce an SPT phase \cite{ Baez:1995xq, Freed:2007vy, Lurie:2009keu,  Freed:2016rqq, Schommer-Pries:2017sdd, Yonekura:2018ufj} (and local counterterms) on the $Y$. This leads to the action
	\begin{eqaed} \label{nonchiralb}
		S & =\frac{\pi}{g^2} \int_X \left ( F_{\check{\mathbf{A}}} + A_{\check{\mathbf{B}}} \right ) \wedge * \left ( F_{\check{\mathbf{A}}} + A_{\check{\mathbf{B}}} \right ) \\
        & - 2 \pi i \big ( \check{\mathbf{A}}, \check{\mathbf{C}} \big )_X - 2 \pi i\int_X q(F_{\check{\mathbf{C}}}, A_{\check{\mathbf{B}}}) - 2 \pi i (-1)^{d-p} \big ( \check{\mathbf{C}}, \check{\mathbf{B}} \big )_Y \, ,
	\end{eqaed}     
	where the homotopy cochain $q(\omega_1,\omega_2)$ acts on differential forms $\omega_1$ and $\omega_2$ according to	
	\begin{equation}
	\omega_1 \wedge \omega_2 - \omega_1 \cup \omega_2 = q(\delta \omega_1, \omega_2)+ (-1)^{p_1} q(\omega_1, \delta \omega_2) + \delta q( \omega_1,  \omega_2) \, .
	\end{equation}
	\cref{nonchiralb} also contains the cohomology pairing \cite{Hsieh:2020jpj}
	\begin{equation} \label{eq:cohomology_pairing}
		\big ( \check{\mathbf{A}}_1, \check{\mathbf{A}}_2 \big )_Y= \int_Y A_{\check{\mathbf{A}}_1 *\check{\mathbf{A}}_2} \, ,
	\end{equation} 
where the product between differential characters is defined by
\begin{equation}
	\begin{aligned}
	   &N_{\check{\mathbf{A}}_1 *\check{\mathbf{A}}_2}= N_{\check{\mathbf{A}}_1} \cup N_{\check{\mathbf{A}}_2} \, ,
	   \\
	   & F_{\check{\mathbf{A}}_1 *\check{\mathbf{A}}_2}= F_{\check{\mathbf{A}}_1} \wedge F_{\check{\mathbf{A}}_2} \, ,
	   \\
	   &A_{\check{\mathbf{A}}_1 *\check{\mathbf{A}}_2}= A_{\check{\mathbf{A}}_1} \cup F_{\check{\mathbf{A}}_2} + (-1)^{p_1+1} F_{\check{\mathbf{A}}_1} \cup A_{\check{\mathbf{A}}_2} + q(F_{\check{\mathbf{A}}_1}, F_{\check{\mathbf{A}}_2} ) \, .
	\end{aligned}
\end{equation}
Notice that the definition in \cref{nonchiralb} holds only when the electrically coupled field $\check{\mathbf{B}}$ is topologically trivial on $X$. Although such requirement seems to be rather restrictive, it actually corresponds to the unique choice available, since whenever $\check{\mathbf{B}}$ is non-trivial the partition function vanishes (see \cite{Hsieh:2020jpj} for details). Furthermore, the action in \cref{nonchiralb} is composed of two pieces, one depending on $\check{\mathbf{A}}$ defined on $X$ and another, the SPT phase, encoding the cohomology pairing of the background fields on $Y$. All in all, the partition function is
 \begin{equation}\label{eq:nonchiral_partition_function} 
 	\mathcal{Z}_{\check{\mathbf{A}}} (Y,L)= e^{2 \pi i (-1)^{d-p} \big ( \check{\mathbf{C}}, \check{\mathbf{B}} \big )_Y } \int \big [ D \check{\mathbf{A}} \big ] e^{-S[\check{\mathbf{A}},X]} \, ,
 \end{equation} 
where the boundary conditions $L$ imply that the field $\check{\mathbf{B}}$ extended to the bulk $Y$ shall be topologically trivial on the boundary.   

\noindent The analysis performed so far holds for non-chiral $p$-form fields but it is not suited to describe the case of chiral (self-dual) forms. These kinds of fields are ubiquitous in string theory and supergravity and thus require a proper discussion. In the na\"ive language of differential forms, a proper description of the action and consequently of the partition function seems hard to achieve. Following \cite{Hsieh:2020jpj}, it is possible to express the action and the partition function of a chiral bosonic field by localising on the boundary a chiral mode $\mathbf{\check{B}}$ of a gapped non-chiral form $\mathbf{\check{A}}$ on a $(d+1)$-dimensional bulk $Y$ with $\partial Y = X$. The presence of the boundary requires specifying suitable elliptic \cite{Witten:2018lgb} boundary conditions $L$, \emph{e.g.} that $(p+1)$-forms vanish on the boundary\footnote{Such a condition is motivated by the existence of the free propagator in order to build a well-defined perturbation theory \cite{Witten:2018lgb}.}. As a result, we can write the action as 
\begin{equation} \label{eq:chiralbaction}
    S=  \frac{\pi}{g^2} \int_Y F_{\mathbf{\check{A}}} \wedge * F_{\mathbf{\check{A}}} - 2 \pi i \left ( \mathcal{Q}_Y( \mathbf{\check{A}} ) - \mathcal{Q}_Y( 0) \right ) - 2 \pi i \big ( \mathbf{\check{A}},  \mathbf{\check{C}} \big )_Y \, .
\end{equation}
In the latter expression, we have introduced a so-called (inhomogeneous) {\em quadratic refinement} \cite{Hopkins:2002rd} of the cohomology pairing, defined via the characteristic equation (see \emph{e.g.} \cite{Dierigl:2022zll} for an application in a physical context)
\begin{equation} \label{eq:quadrefcharacteristiceqinh}
    \mathcal{Q}_Y( \mathbf{\check{A}}_1+ \mathbf{\check{A}}_2 ) -  \mathcal{Q}_Y( \mathbf{\check{A}}_1 ) -  \mathcal{Q}_Y(  \mathbf{\check{A}}_2 ) + \mathcal{Q}_Y ( 0 )= \big ( \mathbf{\check{A}}_1,  \mathbf{\check{A}}_2 \big )_Y \, .
\end{equation} 
In \cref{eq:chiralbaction} the quadratic refinement appears subtracted by the piece $\mathcal{Q}( 0)$, in which the background field is turned off. One is thus led to define the quantity $\widetilde{\mathcal{Q}}( \mathbf{\check{A}} )=\mathcal{Q}( \mathbf{\check{A}} ) - \mathcal{Q}( 0) $ satisfying the homogeneous version of the characteristic equation,
\begin{equation} \label{eq:quadrefcharacteristiceq}
    \widetilde{\mathcal{Q}}_Y( \mathbf{\check{A}}_1+ \mathbf{\check{A}}_2 ) -  \widetilde{\mathcal{Q}}_Y( \mathbf{\check{A}}_1 ) -  \widetilde{\mathcal{Q}}_Y(  \mathbf{\check{A}}_2 ) = \big ( \mathbf{\check{A}}_1,  \mathbf{\check{A}}_2 \big )_Y \, .
\end{equation} 
In general, the choice of quadratic refinements is dictated by a general solution of the characteristic equation, subject to the constraint that whenever $Y= \partial Z$
\begin{equation}
    Q_{\partial Z}(\mathbf{\check{A}})= \int_Z \left \{ \frac{1}{2} \left ( w + F_{\mathbf{\check{A}}} \right )^2 - \frac{1}{8} L \right \} \, ,
\end{equation}
where we have introduced an integral lift $w$ of the {\em Wu class} defined as the characteristic element of the cup product pairing \cite{Monnier:2018nfs}, corresponding in six dimensions to $w=- \frac{p_1}{4}$. In general, these requirements do not admit a unique solution and up to now it is unclear how to choose the right quadratic refinement systematically. Moreover, in all the cases analysed in this paper, chiral fields take values also in a lattice of signature $(1,n_T)$, depending on the number $n_T$ of tensor multiplets. Hence, the previous considerations have to be slightly modified. Specifically, chiral forms arise as boundary modes of $\check{H}^4(Y) \otimes \Lambda$ \cite{Michelangelo}, where $\Lambda$ is the $SO(1,n_T)$ lattice determined by the cancellation of local anomalies. In this scenario, the lattice-valued quadratic refinement on $Y= \partial Z$ reads
\begin{equation}\label{eq:lattice_quadratic_refinement}
    Q_{\partial Z}^{\Lambda}(\mathbf{\check{c}})= \int_Z \left \{ \frac{1}{2} \left ( a_0 w +  b_0 F_{\mathbf{\check{c}}} \right )^2 -  \frac{1}{2} \sum_{i=1}^{n_T} \left ( a_i w +  b_i F_{\mathbf{\check{c}}} \right )^2 + \frac{n_T-1}{8} \, L \right \} \, ,
\end{equation}
where $a=(a_0,\{ a_i \}_{i=1,\ldots, n_T})$ and $b=(b_0,\{b_i \}_{i=1,\ldots, n_T})$ span the anomaly lattice $\Lambda$ and $\mathbf{\check{c}}$ is the differential character encoding the gauge bundle: its field strength is the invariant polynomial in the curvature $F$, while its integer flux is the Chern characteristic class $c_2$. It is worth noting that \cref{eq:lattice_quadratic_refinement} can be equivalently interpreted as the contribution of one antiself-dual and $n_T$ self-dual fields, as befits the traditional additive view on anomalies for chiral $p$-form fields.

\noindent The action in \cref{eq:chiralbaction} is suited for our purpose since, whenever the gapped field is topologically trivial and thus can be safely described by differential forms, the equations of motion lead to the chirality condition, as well as the usual dynamical equations for $p$-forms, for the boundary mode $\mathbf{\check{B}}$, as shown explicitly in \cite{Hsieh:2020jpj}. 

\noindent The partition function can be written as 
\begin{equation}\label{eq:chiral_partition_function}
    \mathcal{Z}_{\mathbf{\check{A}}} (L,Y) = N_0 N_1 N_2 \, e^{-2 \pi i  \left ( \widetilde{\mathcal{Q}}(\mathbf{\check{C}}) + \frac{1}{8} 2 \eta(\Tilde{D}_Y^{\text{sig}}) - \text{Arf}_w(Y) \right ) } \, ,
\end{equation}
where the term $ N_1 \, e^{- \frac{1}{8} 2 \eta(\Tilde{D}_Y^{\text{sig}})} $ associated to the differential operator
\begin{eqaed} \label{eq:signature_operator}
    \Tilde{D}_Y^{\text{sig}}=-*_Y d_Y i^{m(m+2)} \, : \, d_Y^{\dagger} \Omega^m(Y) \longrightarrow d_Y^{\dagger} \Omega^m(Y)
\end{eqaed}
comes from the evaluation of the one-loop determinant with zero modes removed. 
The expression also involves the so-called  {\em Arf invariant}, defined via the Gauss sum \cite{DELOUP2005105, Taylor2022GaussSI} associated to the quadratic refinement $\widetilde{\mathcal{Q}}$
\begin{equation}
   N_2 \, e^{2\pi i\text{Arf}_w(Y)}= \sum_{j \in J_w} e^{2 \pi i \widetilde{\mathcal{Q}}(\mathbf{\check{a}}_j)} \, ,
\end{equation}
where $J_w$ denotes the set of the flat gauge fields whose characteristic classes $\left [ N_{\mathbf{\check{a}}} \right ]_{\mathbb{Z}}$ identify the differential Wu class in De Rham cohomology $\left [ N_{\mathbf{\check{a}}} \right ]_{dR}= - \left [ w \right ]_{dR} $. Moreover, among flat gauge fields, the topologically trivial ones do not depend on the topological sector identified by $J_w$ and thus their contribution can be factorised in \eqref{eq:chiral_partition_function} as $N_0$, leaving only a sum over flat $p$-forms which are topologically non-trivial and take values in a group isomorphic to the torsional subgroup of $H^{p+1}(Y, \mathbb{Z})$ \cite{Freed:2006ya,  Freed:2006yc, Hsieh:2020jpj}. 

\noindent In the above presentation we have not discussed the physical role of the bulk $Y$ extending spacetime. In this formulation, the (phase) ambiguities leading to anomalies are reflected by the arbitrariness of $Y$. In order for a theory to be consistent, physical quantities including the partition function cannot depend on the choice of $Y$. This requirement leads to Dai-Freed anomalies. In the following, we introduce them in the specific context of $6d$ supergravity. These fields play a crucial role analogous to the case of type IIB supergravity in ten dimensions \cite{Debray:2021vob}.

\subsection{Dai-Freed anomalies in \texorpdfstring{$6d$}{6d} supergravity} \label{sec:dai-freed_anomalies}

As we have anticipated in the introduction, in the following we are going to study minimal supergravity theories in $d=6$ dimensions with gauge groups\footnote{When discussing supergravity, we will mostly implicitly work at the level of the gauge Lie algebra. When computing anomalies, we will choose specific gauge groups.} $SU(n) \, , \, SO(2N) \, , \, E_7$ and $E_8$, and later on $U(1)$ as well. With this setup, the landscape of putative allowable EFTs has been thoroughly analysed and much progress has been made toward bounding and classifying it. In particular, ordinary local anomalies already impose severe constraints on the available theories in which matter hypermultiplets transform in a miscellany of fundamental, adjoint, symmetric and antisymmetric representations, with degeneracies collectively denoted by $d_R$ for each representation $R$.

\noindent Local anomalies are encoded in a characteristic class of degree 8, the anomaly polynomial $I_8$ \cite{Alvarez-Gaume:1983ict, Alvarez-Gaume:1984zlq, Alvarez-Gaume:1983ihn}. The structure of $I_8$ involves an $SO(1,n_T)$-invariant bilinear form of components $\Omega_{a b}$, as well as classes $X_4^a$ of degree 4. Writing them in terms of their Chern-Weil representatives, determined by the curvatures $R$ and $F$, the anomaly polynomial for anomaly-free theories takes the factorised form
\begin{equation}
	I_8= \frac{1}{2} \, \Omega_{a b} \, X_4^a \, X_4^b \, .
\end{equation}
The irreducible contribution has to cancel by consistency, while the remaining factorised contribution can be compensated by the celebrated Green-Schwarz mechanism \cite{Green:1984sg,Green:1984bx,Sagnotti:1992qw}, since at least the gravity multiplet contains a (chiral) 2-form field. In six dimensions the requirement of factorisation is particularly stringent, since purely gravitational anomalies can arise. As we shall see, this well-known framework is reproduced by the more general construction of the ``anomaly theory'' $\mathcal{A}(Y)$ evaluating it on 7-dimensional backgrounds $Y = \partial Z$ that are boundaries of $8$-dimensional manifolds $Z$, and then applying the APS theorem.

\noindent Generally speaking, the anomaly theory $\mathcal{A}(Y)$ associated to a given theory $\mathcal{T}$ on a spacetime $X = \partial Y$ is a topological field theory whose invertible phase $e^{2\pi i \, \mathcal{A}(Y)}$ has a gauge variation opposite to that of the partition function of $\mathcal{T}$. According to our preceding discussion, chiral fermions contribute to the anomaly theory with eta invariants, schematically $\mathcal{A}_\text{fermions}(Y) = \sum \eta(Y)$. Since the phase of the partition function is additively encoded in $\mathcal{A}(Y)$ with $\partial Y = X$, the phase difference for different bulk extensions $Y_1 \, , \, Y_2$ is $\mathcal{A}(Y_1 \sqcup \overline{Y_2})$. Thus, the anomaly is encoded by $\mathcal{A}(Y)$ where now $Y$ is a \emph{closed manifold} of dimension $d+1$. The APS index theorem then ensures that, for a given theory $\mathcal{T}$,
\begin{eqaed}\label{eq:local_anomaly_reproduced}
    \mathcal{A}_{\mathcal{T}}(\partial Z) = \text{Index}_\mathcal{T} - \int_Z I_\mathcal{T}
\end{eqaed}
reproduces the anomaly polynomial in $d+2$ dimensions As explained in \cite{Garcia-Etxebarria:2018ajm}, global anomalies \cite{Witten:1982fp, Witten:1985xe, Witten:1985mj, Witten:2019bou} of a theory on a spacetime $X$ are encoded in $\mathcal{A}(T_X)$ with $T_X$ a mapping torus over $X$. This describes paths in the space of background gauge fields which may be non-trivial in homotopy, and thus cannot be expressed in terms of infinitesimal transformations at the level of the Lie algebra. Requiring that these anomalies also cancel imposes additional constraints in our setup \cite{Bershadsky:1997sb}. Dai-Freed anomalies involve a further step: allowing spacetime topology change along the path in field space, the special role of mapping tori is lost and the anomaly need vanish on \emph{any} closed manifold $Y$ \cite{Garcia-Etxebarria:2018ajm}\footnote{An earlier reference to this idea was actually given by Witten in the lecture ``Anomalies Revisited'' held at Strings 2015. We are not aware of a paper containing an explicit mention of this.}. This formulation serves as a stepping stone to derive additional constraints on EFTs. To this end, the cancellation of local anomalies is crucial not only for consistency but also for pragmatic reasons: because of the APS index theorem, the anomaly theory only depends on the \emph{bordism equivalence class} of the background manifold $Y$. One can thus restrict to representatives of nontrivial bordism classes, namely manifolds $Y$ that are not the boundary of a higher-dimensional manifold $Z$. Furthermore, since the set of bordism classes is a finitely generated abelian group (loosely denoted $\Omega_{d+1}$) under the disjoint union, it is sufficient to identify a set of generators on which to evaluate the anomaly theory.

\noindent In theories where the Green-Schwarz mechanism takes place, the relevant notion of bordism is more subtle: to define a cobordism $\partial Z = Y_1 \sqcup \overline{Y_2}$ between two manifolds $Y_1 \, , \, Y_2$, the relevant structures need to extend over $Z$ compatibly with their restrictions to the appropriate ones on $Y_{1,2}$. In the case of gauge theories with fermions this structure includes a principal bundle and a spin structure\footnote{In certain settings the spin structure can be modified to a pin$^\pm$, spin$^c$ or spin$^{\mathbb{Z}_4}$ structure. This is crucial to find the correct consistency conditions, for instance when reducing M-theory over non-orientable manifolds where I-fold defects emerge \cite{Montero:2020icj} and certain topological symmetries are broken \cite{McNamara:2019rup, McNamaraThesis}.}. In the settings at stake there is an additional ingredient, which is dictated by the Bianchi identities
\begin{equation} \label{twistedstringcond}
	\begin{aligned}
		d H_3^a = X_4^a =  \frac{a^a}{4} \, p_1(R) - \sum_i \frac{b_i^a}{\lambda_i} \, c_2(F_i) 
	\end{aligned}
\end{equation} 
for the (gauge-invariant) field strengths $H^a = dB^a + \omega_\text{CS}^a$ associated to the 2-forms $B^a$ in the gravity and tensor multiplets. Here, as summarised in \cref{appendix:anomaly_formulae}, in our conventions $ \frac{1}{4} p_1(R)= -\frac{1}{2}\text{tr} R^2$, $c_2(F)= \frac{1}{2}  \text{tr} F^2 $ are the Chern-Weil expressions for the first Pontryagin class of the tangent bundle and the second Chern class of the (complexified, if need be) gauge bundle for simply laced groups\footnote{For the abelian case studied in \cref{sec:anomaly_U1}, the second Chern class should be replaced by $-c_1^2/2$.}, while the $\lambda_i$ are coefficients chosen such that the minimal instanton number for the corresponding gauge bundle is $1$.

\noindent The right-hand side of \cref{twistedstringcond}, which we dub the Bianchi class, is thus trivial in cohomology. More precisely, it turns out that it is trivial in integral cohomology, which can include torsional classes undetected by differential forms. This was originally argued in \cite{Witten:1985mj}, and (at least for perturbative heterotic strings) it follows from Dai-Freed anomaly cancellation on the worldsheet \cite{Basile:2023knk}. This very stringent trivialisation requirement is tantamount to the existence of a \emph{twisted string structure} (or string-$G$, for a given gauge group $G$) on spacetime and on the background $Y$ for the anomaly theory. Turning off any gauge contribution one reduces to the ordinary string structure, for which the bordism group $\Omega_7^\text{string} = 0$. This shows that there are no purely gravitational Dai-Freed anomalies in six dimensions when the Green-Schwarz mechanism cancels the local anomaly. In the presence of gauge fields, twisted string structures can generate novel anomalies \cite{Dierigl:2022zll} that need to vanish, lest the theory be in the swampland. If such an anomaly is found it may still be possible to cancel it with a ``topological'' version of the Green-Schwarz mechanism \cite{Debray:2021vob, Dierigl:2022zll}, but we will not consider this subtle possibility in this paper. When tensor multiplets are present, there are multiple Bianchi identities. As we shall discuss later on, it seems that the correct requirement to impose on allowable anomaly backgrounds is that at least one identity be satisfied (at the level of integral cohomology). This turns out to eliminate all anomalies in string theory examples, and it can produce non-trivial bordism groups. Imposing multiple identities simultaneously generically kills all independent characteristic classes. Nevertheless, when presenting examples of anomalous theories, we will include backgrounds where all Bianchi classes are trivialised.

\noindent As we have discussed, for two given $d$-dimensional manifolds to be bordant, the existence of the twisted string structure requires that their disjoint union (with reversed orientation) be the boundary of a $(d+1)$-dimensional manifold over which the twisted string structure extends. Therefore, a systematic strategy to proceed would be to evaluate these theories on a set of generators of the twisted string bordism group $\Omega^\text{string-$G$}_7$ to check whether their anomalies vanish, and thus if their cancellation provide further constraints on the landscape of (super)gravity theories. In order to fully exploit these novel anomalies, it is efficient to first impose all the constraints known up to now, to understand if these anomalies actually entail new consistency conditions or merely disguise what has already been discovered. 

\noindent These constraints reflect different properties that a consistent quantum theory coupled to gravity should have, although they ultimately stem from unitarity. In particular, there are local anomalies that do not pertain to the bulk EFT, but rather live on the worldvolumes of defects \cite{Kim:2019vuc}. The completeness principle \cite{Polchinski:2003bq, Banks:2010zn} is intimately connected with the absence of (non-invertible) symmetries \cite{Heidenreich:2020pkc, Heidenreich:2021xpr} in gravity \cite{McNamara:2019rup, McNamara:2021cuo, McNamaraThesis}, and it requires the presence of unitary defects in the theory. In the settings at stake, these defects are (effective) strings coupled to the $2$-forms entering the Green-Schwarz mechanism. Their presence generates new anomalies localised on their worldsheets, which have to cancel by inflow from the bulk theory \cite{Kim:2019vuc}. Furthermore, unitarity of the worldsheet dynamics imposes nontrivial constraints on the possible bulk gauge groups, which manifest themselves as current algebras on the worldsheet. Such requirements have been used to exclude some supergravities for which there is no UV completion, such as the $SU(n) \times SU(n)$ theory with $9$ tensor multiplets and two hypermultiplets in the bifundamental representation with $n>9$ \cite{Kim:2019vuc}. Another example is the $SU(24)\times SO(8)$ theory with $1$ tensor multiplet and $3$ hypermultiplets in the antisymmetric representation of $SU(24)$. Furthermore, the factorisation of the anomaly polynomial $I_8$, required for the Green-Schwarz mechanism to take place, should not only cancel the anomaly of the microscopic theory living on the defect, but it should also be such that the ``anomaly lattice''\footnote{In \cite{Kumar:2010ru} it has been shown that cancellation of local anomalies and global anomalies on mapping tori with fixed topology, discussed in \cite{Witten:1985mj}, implies that the coefficients $a$, $b_i$ span a lattice, dubbed the ``anomaly lattice''.} be embedded into the self-dual lattice spanned by the charges coupling the defects to the $2$-forms \cite{Seiberg:2011dr}. As emphasised in \cite{Seiberg:2011dr}, this translates to a consistent Dirac quantization for higher $p$-form gauge theories, thus taking into account global properties of the manifold. A more precise way of describing the situation is saying that gauge fields, described na\"ively as differential forms, should rather be lifted to elements of a (possibly generalised) cohomology theory, such as integer cohomology, differential cohomology or K-theory. This means that a well-defined (twisted) string structure should be defined by $3$-forms understood as elements of generalised cohomology, in order to encode global properties. Such conditions form the backbone on which our analysis is based and thus should always be imposed before proceeding. In practice this means that, in order for the lift to exist, the Bianchi identities at the level of differential forms should make sense at least as elements of integer cohomology. It is not \emph{a priori} obvious that this condition can be satisfied, since the first Pontryagin class appears with a $\frac14$ in front and integral classes can have torsion. However, as explained in \cite{Hsieh:2020jpj}, the image in integer cohomology of $\frac14 p_1$ in $7d$ is the (integral lift of the) Wu class $w_{\mathbb{Z}} \in H^4(Y,\mathbb{Z})$, a well-defined integral cohomology class.

\noindent In this paper, we are mostly going to focus on supergravities with one tensor multiplet, since they are closely connected to perturbative heterotic constructions which are well-understood. With a single tensor multiplet, there exist only two possible self-dual lattices:
\begin{enumerate}
	\item \label{sec:offdiag}The ``off-diagonal'' lattice, in which the bilinear form in a suitable basis is given by
	\begin{equation}
		\Omega=	\begin{pmatrix}
			0 & 1 \\
			1 & 0
		\end{pmatrix} \ ,
	\end{equation}
	thus dictating a factorisation of the anomaly polynomial of the type
	\begin{equation} \label{offfact}
		\begin{aligned}
			I_8 &= \frac{1}{4} \left ( 2 \text{tr} R^2 +  \tfrac{b^{(1)}}{\lambda} \text{tr} F^2  \right )  \left ( 2 \text{tr} R^2 +   \tfrac{b^{(2)}}{\lambda} \text{tr} F^2 \right  )
			\\
			&=     \left ( \tfrac{1}{2} p_1 - \tfrac{b^{(1)}}{\lambda} c_2 \right ) \left ( \tfrac{1}{2} p_1 - \tfrac{b^{(2)}}{\lambda} c_2 \right  ) \, .
		\end{aligned}
	\end{equation}
	\item \label{item:diag} The ``diagonal'' lattice, in which the bilinear form in a suitable basis is given by
	\begin{equation}
		\Omega=	\begin{pmatrix}
			1 & 0 \\
			0 & -1
		\end{pmatrix} \ ,
	\end{equation}
	which gives a factorisation of the anomaly polynomial of the type
	\begin{equation} \label{diagfact}
		\begin{aligned}
			I_8 &= \frac{1}{8} \left \{   \left ( 3 \text{tr} R^2 +  \tfrac{b^{(1)}}{\lambda} \text{tr} F^2  \right )^2 -  \left ( \text{tr} R^2 +   \tfrac{b^{(2)}}{\lambda} \text{tr} F^2 \right  )^2 \right \}
			\\
			&= \frac{1}{2} \left \{   \left ( \tfrac{3}{4} p_1 - \tfrac{b^{(1)}}{\lambda} c_2 \right )^2 -  \left ( \tfrac{1}{4} p_1 - \tfrac{b^{(2)}}{\lambda} c_2 \right  )^2 \right \}.
		\end{aligned}
	\end{equation}
\end{enumerate}  
Although one can derive the above expressions following the same steps for both cases, there are very different considerations to be made. Indeed, the difference can be traced back to the fact that the 2-forms in the gravity and tensor multiplets are chiral in six dimensions, and thus contribe directly to the total anomaly. However, as explained in the preceding section, there is no consistent action at the level of differential forms for these chiral fields, unless one can interpret them as boundary modes of non-chiral fields living in one dimension higher using a quadratic refinement $\mathcal{Q}$. When gauge fields are non-chiral, such terms simply reproduce the well-known action of form fields in seven dimensions, but it is crucial to allow for a generalisation to the case of self-dual or antiself-dual gauge fields \cite{Hsieh:2020jpj}. Adopting this view, the anomaly contribution from (anti)self-dual forms is given by the quadratic refinement of the Chern-Simons character $c_2(F)$ included into the Green-Schwarz term $\int_Y \, \Omega_{a b} \, H_a \wedge X_4^b$, and reads
\begin{equation}\label{eq:chiral_anomaly_theory}
	\mathcal{A} \big ( Y \big ) = \sum_{\psi \text{ chiral}} \eta_\psi \big ( Y \big )  + {\mathcal{Q}_+}(b^{(1)} \, \mathbf{\check{c}}) - {\mathcal{Q}_-}(b^{(2)} \, \mathbf{\check{c}}) \, .
\end{equation}
The latter two terms, ${\mathcal{Q}}_+$ and ${\mathcal{Q}}_-$, arise from the tensor and gravity multiplets with opposite chiralities. In particular, on Lens spaces these terms evaluate to appropriate fractional values\footnote{This is ultimately related to the fact that integral cohomology classes are purely torsional for Lens spaces. These can be thought of as flat gauge fields with fractional fluxes.}, and thus they may cancel rational anomalies in a discrete version of the Green-Schwarz mechanism \cite{Dierigl:2022zll, Hsieh:2020jpj}. The differential Chern-Simons character $\mathbf{\check{c}}$ encodes the gauge bundle: its associated field strength is the Chern-Weil representation of the second Chern class, while its integer flux is given by the characteristic class.

\noindent Determining the quadratic refinement can be cumbersome. A starting point is that whenever the manifold $Y$ can be extended to a higher-dimensional space $Z$ the quadratic refinement should reproduce both the Green-Schwarz and the gravitational irreducible contributions. Therefore, for null-bordant $Y$ the quadratic refinement is uniquely fixed, as shown in \cite{Hsieh:2020jpj} for Spin-bordism. In the settings at stake we do not know if the twisted string bordism group vanishes, nor if the relevant backgrounds actually lie in the trivial class, so that {\em a priori} $\mathcal{Q}_Y$ could be different. Nevertheless, we can exploit the fact that in the purely gravitational case $\Omega_7^{\text{string}}=0$ to constrain $\mathcal{Q}_Y(0)$. Applying the APS index theorem to \cref{eq:lattice_quadratic_refinement}, this gravitational contribution can be cast in the form\footnote{Strictly speaking, since the Bianchi class is either $\tfrac{p_1}{4}$ or $\tfrac{3}{4} p_1$, it is not obvious that the expression of $\mathcal{Q}(0)$ is unique as in the case of the standard string structure trivialising $\tfrac{p_1}{2}$.}
\begin{equation}\label{eq:quadratic_refinement_0}
    \mathcal{Q}_{Y} ( 0)= \int_Z \left ( \frac{1}{2} \left ( \frac{a}{4} p_1  \right )^2-\, \frac{1}{8} L \right ) = - \, \frac{7(35 a^2-3) }{8} \, \eta_D(Y) +\frac{(a^2-1) }{8} \, \eta_{\text{grav}}(Y) \, ,
\end{equation}
where $a$ is the relevant coefficient in the Bianchi classes in \cref{twistedstringcond}. Then one can use the formulae in \cref{appendix:systematics} to evaluate this expression on Lens spaces. Let us anticipate that, for models with $n_T > 0$ tensor multiplets, a general bottom-up analysis of quadratic refinements is prohibitive. We managed to complete it in the simple setting of $SU(2)$ models, but in general one would need some restriction on the possible quadratic refinements. We will expand upon this point in \cref{sec:anomaly_SU2} and \cref{sec:diag_SUn}. Nevertheless, we provide a systematic analysis of these models for simply laced groups, and study the anomaly for the particular quadratic refinement defined in \cite{Hsieh:2020jpj}, leaving a more complete classification of anomalies for future work. For models with $n_T=0$ the analysis is simpler, and we carry it out in \cref{sec:abelian_no_tensors} and \cref{sec:gepner}.

\noindent It is not always possible to consistently implement a quadratic refinement: such terms can be added whenever the chiral gauge fields can be lifted to elements in integer cohomology, which requires that the coefficients in $I_8$ in diagonal form be integers. For the off-diagonal family in \cref{sec:offdiag}, diagonalisation often produces irrational coefficients, and thus no quadratic refinement is possible in those cases. On the other hand, in this case the two chiral forms can be thought of as the two chiral components of a non-chiral form $B$, whose Bianchi identity singles out a Green-Schwarz coupling $B \wedge X_4$. Therefore, along the lines of \cite{Basile:2023knk}, in this case the anomaly theory is simply given by 
\begin{equation}\label{eq:nonchiral_anomaly_theory}
	\mathcal{A} \big ( Y \big ) = \sum_{\psi \text{ chiral}} \eta_\psi \big ( Y \big ) - \, \int_Y H \wedge X_4
\end{equation}
in accord with \cref{eq:nonchiral_partition_function}. Such expression can be also obtained from a different but equivalent consideration: the dependence on the extended manifold $Y$ of the partition function in \cref{eq:nonchiral_partition_function} is contained in the cohomology pairing of the background fields $\big ( \mathbf{\check{B}}, \mathbf{\check{C}} \big)_Y$ defined in \cref{eq:cohomology_pairing}. In the present case the background fields corresponding to $\mathbf{\check{B}}$ and $ \mathbf{\check{C}}$ are given by combinations of the Chern-Simons and the Wu characters identified by the Bianchi identities. However, requiring the twisted string structure to hold means that the corresponding combination of characters is trivial in integer cohomology and thus the pairing reduces to the one included in \cref{eq:nonchiral_anomaly_theory}. As mentioned before, the consistency of the EFT implies the cancellation of all the anomalies on every available background dictated by the twisted string structure \eqref{twistedstringcond}. Unfortunately, twisted string structures are particularly unwieldy, and there is no known result on a full classification. Nevertheless, we can try to approach the problem from another perspective, checking some backgrounds that are under control where the anomaly can be computed. For us these are Lens spaces, schematically building blocks for spaces with torsional (co)cycles. For our purposes, the construction of these spaces, as well as evaluation of eta invariants, is detailed in \cite{Debray:2023yrs}. In the following we will consider supergravity theories with gauge groups $SU(n)$ (with particular emphasis on the $SU(2)$ case), $\text{Spin}(N), \, E_7 \, , \, E_8$ and later on $U(1)$, evaluating their anomaly theories on Lens spaces $L^7_p$. For these theories, the fermion anomaly is given by
\begin{eqaed}\label{eq:fermion_anomaly_general}
    \mathcal{A}_\text{fermions} = \sum_{\text{hypers}} \eta_\text{hyper} + \eta_T - \, \eta_{\text{\bf adj}} - \, \eta_\text{gravitino} \, ,
\end{eqaed}
where $\eta_T$ is the eta invariant of a Dirac fermion in the trivial representation $T$, which arises from the tensor multiplet. No confusion should arise with the notation, since we fix $n_T=1$ tensor multiplets for the ensuing analysis except for \cref{sec:abelian_no_tensors} and \cref{sec:gepner}. The eta invariant for the gravitino is $\eta_\text{gravitino} = \eta^\text{RS} - 2 \, \eta^\text{D}$ in terms of the eta invariants of a Rarita-Schwinger and Dirac field on the bulk $Y$. On Lens spaces,
\begin{eqaed}
    \eta_\text{gravitino}(L^{2k-1}_p) = k \left( \eta_1(L^{2k-1}_p) + \eta_{-1}(L^{2k-1}_p) \right) - \, 3 \, \eta_0(L^{2k-1}_p) \, ,
\end{eqaed}
where $\eta_q$ denotes the eta invariant of a fermion of charge $q$ under the $\mathbb{Z}_p$ action on $L^{2k-1}_p$, and is given by \cite{Debray:2021vob, Debray:2023yrs}
\begin{eqaed}
    \eta_q(L^{2k-1}_p) = - \, \frac{1}{(2i)^k \, p} \sum_{l = 1}^{p-1} \frac{e^{2\pi i q \, l/p}}{\sin^k(\pi l/p)} \, .
\end{eqaed}
In \cref{appendix:systematics} we provide explicit expressions for the eta invariant of the Dirac and the Rarita-Schwinger operators, relevant for our discussion.  Turning off the gauge bundle probes the gravitational Dai-Freed anomaly, which simplifies \cref{eq:fermion_anomaly_general} to
\begin{eqaed}\label{eq:fermion_grav_anomaly_general}
    \mathcal{A}_\text{fermions}^\text{grav} & = (n_H - n_V + 1) \, \eta_T - \, \eta_\text{gravitino} \\
    & = 245 \, \eta_T - \, \eta_\text{gravitino} \\
    & \overset{\text{Lens}}{=} 248 \, \eta_0 - 4 \left( \eta_1 + \eta_{-1} \right) \\
    & = - \, \frac{p^4 + 11p^2 - 12}{3p} \, .
\end{eqaed}
For anomaly lattices embedded in the diagonal lattice, the full gravitational anomaly on Lens spaces then evaluates to
\begin{eqaed}\label{eq:grav_anomaly_vanishes}
    \mathcal{A}(L_p^7) = - \, \frac{p^4 + 11p^2 - 12}{3p} + \mathcal{Q}_+(0) - \mathcal{Q}_-(0) = 0
\end{eqaed}
on account of eqs. \eqref{eq:fermion_grav_anomaly_general} and \eqref{eq:quadratic_refinement_0}. In particular, the anomaly vanishes for the unique Lens space which can trivialise a Bianchi class, namely $L_3^7$.

\noindent For anomaly lattices embedded in the off-diagonal lattice there is no quadratic refinement in general, but the only allowed backgrounds are ordinary string manifolds with $\tfrac{p_1}{2} = 0$. For Lens space this requires $p=2$, for which $L_2^7 = \mathbb{R}P^7$ is the real projective 7-space. The resulting gravitational anomaly
\begin{eqaed}\label{eq:grav_anomaly_vanishes_off-diag}
    \mathcal{A}(\mathbb{R}P^7) = -8 \equiv_1 0
\end{eqaed}
also vanishes, consistently with the fact that it is a bordism invariant of $\Omega_7^\text{string} = 0$. Therefore, in the following analysis of $n_T=1$ models, we will not consider trivial gauge bundles.

\section{Dai-Freed anomalies for unitary groups}\label{sec:anomaly_SUn}

In this section we begin our analysis of supergravity EFTs with special unitary gauge groups $SU(n)$. The resulting consistency conditions also apply (with some caveats discussed below) when the full gauge gauge group $G = SU(n) \times H$ contains an $SU(n)$ factor, but they are \emph{a priori} weaker than what one would find including the contributions to the anomaly pertaining to the remaining gauge group $H$. This simple consideration applies to the rest of the analysis in this paper. To warm up, let us first consider $SU(2)$, since it behaves slightly differently and it allows a simple comparison with a class of F-theory and heterotic string constructions.

\subsection{Warm-up with \texorpdfstring{$SU(2)$}{SU(2)}}\label{sec:anomaly_SU2}

Let us begin considering supergravity theories in $6d$ with $SU(2)$ gauge group and one tensor multiplet. For simplicity, we only consider hypermultiplets in the trivial ($T$) and fundamental ($F$) representations\footnote{In the following the multiplicities of the various representations $R$ will be denoted $d_R$.}. In this setup, the landscape of models is already highly constrained by local anomalies. Indeed, the cancellation of the irreducible purely gravitational piece of the anomaly polynomial implies that the number of trivial and fundamental hypermultiplets satisfy   
\begin{equation} \label{irrgravsu2}
	d_T +  2 d_F =247 \, . 
\end{equation}     
A comment is however in order. Here we are looking at constraints for {\em would-be } UV complete theories with gauge group $SU(2)$. If $SU(2)$ is embedded into a larger gauge group, the condition in \eqref{irrgravsu2} implicitly takes into account the contribution from the vector multiplets whose gauge fields are turned off, and thus contribute as ``hypermultiplets of wrong chirality''. Allowing $d_T < 0$ yields a much larger family to consider, and the control over these theories is lost. For the purpose of uncovering new constraints to exclude EFTs, as well as for testing particular (top-down) examples, this consideration is not relevant. Therefore, taking into account the cancellation of the irreducible gravitational anomaly the polynomial takes the form
\begin{equation} \label{anpolsu2}
    I_8 = \left ( \text{tr} R^2 \right )^2- \frac{1}{24} \text{tr} R^2 \, \text{tr} F^2 \left ( d_F-4\right ) + \frac{1}{48} \left ( \text{tr} F^2 \right )^2  \left ( d_F - 16 \right ) , 
\end{equation}
where we used the decomposition $ \text{tr} \ F^4= \frac{1}{2} \left ( \text{tr} \ F^2 \right )^2$ valid for $SU(2)$. From \cref{anpolsu2} we can proceed computing the possible embeddings in the two available self-dual lattices: 

\subsubsection{Diagonal lattice}\label{sec:diag_lattice}
 
	For the diagonal lattice the number of fundamentals allowing an integral factorisation  of the anomaly polynomial has to be of the form
	\begin{equation} \label{su2famdiag}
		d_F=10 + 12 s \, .
	\end{equation}
	Indeed, for such values it is easy to see that the polynomial factorises according to 
	\begin{equation} \label{poldiagsu2}
		I_8 = \frac{1}{8} \left \{   \left ( 3 \text{tr} R^2 - s \text{tr} F^2  \right )^2 -  \left ( \text{tr} R^2 +  (1 - s)  \text{tr} F^2 \right  )^2 \right \}.	
	\end{equation} 
 
\noindent A complete study of Dai-Freed anomalies of such theories would require to know the proper $SU(2)$-twisted string bordism group, where the twisted string structure is spelled out by the Bianchi identities
	\begin{equation} \label{bianchidiagsu2} 
		\begin{aligned}
			& d  H_1= \tfrac{3}{4} p_1 + s \ c_2 \, ,
			\\
			& d  H_2= \tfrac{1}{4} p_1 + (s-1) c_2 \, .
		\end{aligned}
	\end{equation} 
As described above, the task is considerably simplified by restricting to Lens spaces $L_p^7$. We do not know if these exhaust all the nontrivial bordism representatives, but they provide particularly simple families of candidate backgrounds to work with. The complete expression for the anomaly theory in \cref{eq:chiral_anomaly_theory} involves the knowledge of the quadratic refinement on the allowed Lens spaces. The latter, thus, has to satisfy the characteristic equation \eqref{eq:quadrefcharacteristiceqinh}, given the cohomology pairing 
	\begin{equation}
		\big ( \mathbf{\check{A}}, \mathbf{\check{B}} \big )_{L_p^7}= - \, \frac{a b}{p} \, ,
	\end{equation}  
where $\left [ N_{\mathbf{\check{A}}} \right ]= a y$ and $\left [ N_{\mathbf{\check{B}}} \right ]= b y$, with $y$ a generator of the integral cohomology group.

\noindent Computing the anomaly we can conveniently work with $\widetilde{\mathcal{Q}}({\mathbf{\check{A}}} )=\mathcal{Q}({\mathbf{\check{A}}} ) -\mathcal{Q}(0)$, which \emph{a priori} is a general solution of \cref{eq:quadrefcharacteristiceq}. It can be parametrised by an integer $m=0, \ldots, 2p-1$ as \cite{Hsieh:2020jpj}
\begin{equation}\label{eq:tildeQ_lens}
    \widetilde{\mathcal{Q}}({\mathbf{\check{A}}} )= - \, \frac{a(a+m)}{2p} \, . 
\end{equation}
In top-down examples, the choice of the quadratic refinement may not be arbitrary. For instance, in \cite{Dierigl:2022zll} it has been argued that F-theory selects a single possibility for $\mathbb{Z}_3$ gauge groups, although a general rule to determine $\widetilde{\mathcal{Q}}$ is not known.
	
\noindent We build $SU(2)$ bundles on $L_p^7$ backgrounds by including the defining line bundle of $\mathbb{Z}_p$ into $SU(2)$ according to the rules detailed in \cref{appendix:systematics}. For $SU(2)$, there are only two possibilities, encoded by a parameter $k = 0 \, , \, 1$. For $k=0$ the gauge bundle is trivial and one probes the purely gravitational anomaly, which vanishes on account of \cref{eq:grav_anomaly_vanishes}. For $k=1$, charged fermions decompose into representations with different $\mathbb{Z}_p$ charges, for which the eta invariants $\eta_q$ are easily calculated. The anomaly for $k=1$ is thus given by
	\begin{equation} \label{ansu2Lp7diag1}
		\begin{aligned}
			\mathcal{A} \big (L_p^7 \big) &= d_F \left ( \eta_1 + \eta_{-1} \right )- \left ( \eta_2 + \eta_{-2}+\eta_0 \right ) + (d_T+1) \eta_0- \eta_{\text{gravitino}} + {\mathcal{Q}}_+ - {\mathcal{Q}}_-
			\\
			&= d_F \widetilde{\eta}_1- \widetilde{\eta}_2 + 244 \eta_0 - 4  \widetilde{\eta}_1 + {\mathcal{Q}}_+ - {\mathcal{Q}}_-
			\\
			&= \frac{1}{12 p} \left ( d_F (p^2-1) -4 (p-2)(p-1) -  (p^2-1)(4 p^2 + 48) \right ) + {\mathcal{Q}}_+ - {\mathcal{Q}}_-
             \\
             &\equiv_1 \frac{d_F (p^2-1) - 4p^4 + 40}{12p} + {\mathcal{Q}}_+ - {\mathcal{Q}}_- \, ,
		\end{aligned}
	\end{equation}
	where $\equiv_1$ denotes equivalence modulo 1 and we have used the useful shorthand notation 
	\begin{equation}
		\widetilde{\eta}_q= \eta_q + \eta_{-q} - 2 \eta_0 \, .
	\end{equation}
	In particular, we need the values
	\begin{equation} \label{etatilde}
		\begin{aligned}
			&\widetilde{\eta}_1= \frac{p^2-1}{12 p} \ , \qquad  \qquad \widetilde{\eta}_2= \frac{(p-1)(p-2)}{3 p} \ , \qquad \eta_0=-\frac{(p^2-1)(p^2+11)}{720p} \, .
		\end{aligned}
	\end{equation}  
We now need to find Lens spaces which satisfy the twisted string structure. From the Bianchi identities in \cref{bianchidiagsu2}, using the facts (reviewed in \cref{appendix:systematics}) that $p_1 = y$ generates the fourth integral cohomology $H^4(L^7_p \, , \, \mathbb{Z}) = \mathbb{Z}_p$ and $c_2 = -k \, y$, the first Bianchi class is $(3 - k \, s)y$, while the second is $(1 - k (s-1))y$. For $k=1$ the Bianchi classes are $(3-s)y$ and $(2-s)y$. The gravitational contribution to the the quadratic refinement is the opposite of the gravitational anomaly in \cref{eq:fermion_grav_anomaly_general}, while the remainder is given by \cref{eq:tildeQ_lens} according to
\begin{eqaed}
    \widetilde{\mathcal{Q}}_\pm(b_\pm \, \check{\mathbf{c}}) \,  = - \, \frac{b_\pm (b_\pm + m_\pm)}{2p} \, .
\end{eqaed}
One can show that for any Lens space trivialising either Bianchi class the anomaly vanishes with an appropriate choice of $m_\pm$. This is crucial, since the F-theory landscape includes models in this family\footnote{We thank P. Oehlmann for correspondence on this point.}.

\subsubsection{Off-diagonal lattice}\label{sec:off-diag_lattice}	
	
Embedding in the off-diagonal lattice requires that the number of fundamental hypermultiplets be of the form 
	\begin{equation} \label{su2famoff}
		d_F=4 + 12 s.
	\end{equation} 
	For these values the anomaly polynomial indeed factorises according to
	\begin{equation}\label{poloffsu2}
		I_8 = \frac{1}{4} \left ( 2 \text{tr} R^2 -  \text{tr} F^2  \right )  \left ( 2 \text{tr} R^2 +  (1 - s)  \text{tr} F^2 \right  ) .	
	\end{equation} 
	The condition in \cref{su2famoff} ensures the Seiberg-Taylor conditions are satisfied.
	
\noindent With respect to the diagonal case the $SU(2)$-twisted string structure is different, since the Bianchi identities now read 
	\begin{equation} \label{bianchioffsu2}
		\begin{aligned}
			& d  H_1= \tfrac{1}{2} p_1 + c_2
			\, , \\
			& d  H_2= \tfrac{1}{2} p_1 +(s-1) c_2 \, .
		\end{aligned}
	\end{equation}  
The anomaly theory now takes the form of \cref{eq:nonchiral_anomaly_theory}, and for Lens spaces we can set the Green-Schwarz term to zero since there are no non-trivial cocycles of degree 3. Alternatively, $H=0$ is consistent with a trivialised twisted string structure.   

\noindent The family with $d_F=16+24 s$, namely theories with odd $s$ in \cref{su2famoff}, also affords an embedding in the diagonal lattice, since the anomaly polynomial can be recast as
	\begin{equation}\label{poldiagoffsu2}
		I_8 = \frac{1}{8} \left \{\left ( 3 \text{tr} R^2 -   (1 + s)  \text{tr} F^2  \right )^2-\left (  \text{tr} R^2 +  ( s-1)  \text{tr} F^2 \right  )^2 \right \} .	
	\end{equation} 
The corresponding Bianchi identities are
	\begin{equation} \label{bianchidiagoffsu2}
		\begin{aligned}
			& d  H_1= \tfrac{3}{4} p_1 +  (1 + s) c_2 \, ,\\
			& d  H_2= \tfrac{1}{4} p_1 - (s-1) c_2 \, .
		\end{aligned}
	\end{equation}  
	The computation of the anomaly is the same as in \cref{ansu2Lp7diag1}, so that
	\begin{equation} \label{ansu2Lp7off}
		\begin{aligned}
        \mathcal{A} \big (L_p^7 \big) \equiv_1 \frac{d_F (p^2-1) - 4p^4 + 40}{12p} \, .
		\end{aligned}
	\end{equation}

\noindent The Bianchi classes are now encoded in \cref{bianchioffsu2}, and evaluate to $(2-k)y$ and $(2-k(s-1))y$. For $k=1$, the only Bianchi class that can be trivialised on a Lens space is $(3-s)y$. Thus, setting $s = 3 + m p$ for some integer $m$ the Lens background has the appropriate twisted string structure, and one finds
\begin{eqaed}
    \mathcal{A}(L^7_p) \equiv_1 - \, \frac{(p-1) \, p \, (p + 1)}{3} \, ,
\end{eqaed}
which also vanishes because the numerator is always divisible by three.

\noindent All in all, we have found that for $SU(2)$ there are no Dai-Freed anomalies when the twisted string structure is satisfied. It would thus seem useless to seek examples in the string landscape, since there can be no anomalies regardless of the UV completion. However, when restricting to the perturbative heterotic landscape a pattern seems to emerge: at least certain types of free fermion constructions appear to single out settings in which $d_F$ is divisible by eight. In F-theory other values arise\footnote{We thank M. Dierigl, M. Kang and P. Oehlmann for discussions on this point.}, but we have no intuitive explanation for the pattern that we observe in the specific perturbative vacua that we discuss in the following.

\subsubsection{Comparison with the string landscape}\label{sec:worldsheet_scan_F-theory}

To compare the above discussion to the landscape of perturbative heterotic strings, we seek constructions where the gauge group admits the general decomposition 
\begin{eqaed}\label{eq:gauge_group}
    SU(2) \times G 
\end{eqaed}
and the matter hypermultiplets only comprise trivial and fundamental representations to match the analysis in the preceding section. To begin with, the constructions in \cite{Walton, Honecker:2006qz} feature numerous suitable models. It turns out that for the models that we analysed $d_F \in \{ 46 \, , \, 88 \, , \, 136 \, , \, 154 \, , \, 184 \, , \, 280 \}$, which are indeed divisible by eight for off-diagonal embeddings, \emph{i.e.} whenever the model falls under the case of \cref{su2famoff}. This is suggestive, since the anomaly theory for $k=1$ on $L_2^7 = \mathbb{R}P^7$ evaluates to
\begin{eqaed}
    \mathcal{A}(\mathbb{R}P^7) \equiv_1 \frac{d_F}{8} \, .
\end{eqaed}
As we have discussed, this background does not always have the appropriate twisted string structure, although it is of course a string manifold. Thus, somehow, it appears that this family of vacua selects models where the twisted string structure exists on this background and the anomaly vanishes (as shown above), rather than models where the twisted string structure forbids this background altogether. It is unclear to us whether the heterotic worldsheet may secretly allow this background in particular, since the argument in \cite{Basile:2023knk} shows that the twisted string structure ought to hold at the integral level in spacetime. We do not understand whether in this particular case a $7d$ bulk of this type is actually allowed or this pattern is just a coincidence of this specific corner of the landscape.

\noindent We now carry out a systematic analysis for worldsheet constructions of the free-fermion type. Namely, we consider worldsheet models where the internal sector of the superconformal field theory is described by level-one chiral affine algebras of the form
\begin{eqaed}\label{eq:chiral_algebra_ansatz}
    \widehat{\mathfrak{su}(2)}_1 \oplus \widehat{\mathfrak{g}} = \widehat{\mathfrak{su}(2)}_1 \oplus \bigoplus_{n} d_n \, \widehat{\mathfrak{so}(2n)}_1
\end{eqaed}
with some multiplicities $d_n \geq 0$ such that the total central charge be critical. Our strategy is the following: we will write down the most general $SU(2)$-refined elliptic genus that can arise from such worldsheet constructions, and then compare it to the one arising from \emph{modular invariant} torus partition functions of six-dimensional supersymmetric models. Automating this procedure yields a scan of potentially allowed worldsheet models, where the number of fundamental hypermultiplets can be extracted. The refined partition function and elliptic genus contain the refined affine $SU(2)$ characters
\begin{eqaed} \label{eq: su2 characters}
    &\chi_0^{(1)}(z;\tau)= \frac{\Theta_0^{(1)}(z;\tau)}{\eta(\tau)} \sim_{\substack{z = 0}} \frac{1}{q^{\frac{1}{24}}} + 3 q^{\frac{23}{24}}
    \\
    &\chi_1^{(1)}(z;\tau)= \frac{\Theta_1^{(1)}(z;\tau)}{\eta(\tau)} \sim_{\substack{z = 0}} 2 q^{\frac{5}{24}}
\end{eqaed}
where $\eta$ is the Dedekind eta function (there will be no ambiguity with eta invariants in this section) and
\begin{eqaed} \label{eq: general theta}
    \Theta_m^{(k)}(z;\tau)= \sum_{n \in \mathbb{Z}} q^{k(n+\frac{m}{2k})} e^{2 \pi i k z(n+\frac{m}{2k}) } \, , \qquad q=e^{2 \pi i \tau} \, .
\end{eqaed}
The chemical potential $z$, or more precisely its associated fugacity $\xi$, counts $SU(2)$ charges in the Cartan subalgebra generated by $J_3$. We shall schematically denote all the (affine) characters pertaining to the gauge factor $G$ with the symbols $\mathcal{O}, \ \mathcal{A}_0$ and $\mathcal{A}_1$. In principle these factors could be completely general, but in order to perform our analysis we will consider the specific cases arising from orthogonal chiral affine algebras. These characters multiply different spacetime contributions in the partition function. In particular, $\mathcal{O}$ is the identity of the chiral algebra, and thus should admit a $q$-expansion whose first two terms are 
\begin{equation}
    \mathcal{O} \sim \frac{1}{q^{\frac{19}{24}}} + \text{dim}(G) \, q^{\frac{5}{24}} \, ,
\end{equation}
whereas $\mathcal{A}_0$ and $\mathcal{A}_1$ multiply the refined $SU(2)$ characters $\chi^{(1)}_0$ and $\chi^{(1)}_1$ respectively. Thus, their $q$-expansion encodes the number $d_F$ of $SU(2)$ fundamental hypermultiplets, according to
\begin{eqaed}
    &\mathcal{A}_0 \sim \left ( H- 2 \, d_F \right ) q^{\frac{5}{24}} \, ,
    \\
    &\mathcal{A}_1 \sim d_F \ \frac{1}{q^{\frac{1}{24}}} \, .
\end{eqaed}
The spacetime $\mathcal{N}=(1,0)$ superconformal algebra fixes the contribution $Z_0$ to the partition function of a heterotic string arising from massless degrees of freedom, which for a level-$k$ $SU(2)$ affine algebra takes the form\footnote{The factors of 2 in the hypermultiplets count the degeneracies of half-hypermultiplets. In the spacetime characters we retain only the contributions from massless states that make (transverse) isometries manifest. The remaining terms, necessary for modular invariance, comprise the massive contributions to the full partition function.} 
\begin{eqaed}\label{eq:partition_function_ansatz}
    Z_0 = \frac{1}{\overline{\eta}^4 \eta^4} \left[ \overline{\left(q^{-\frac{1}{4}} \, V_4 - 2 \, S_4\right)} \, \mathcal{O} \, \chi_0^{(k)} + \overline{\left(2 \, q^{\frac{1}{4}} \, O_4 - C_4\right)} \left(2\mathcal{A}_1 \, \chi_1^{(k)} + 2\mathcal{A}_0 \, \chi_0^{(k)}\right) \right] \, ,
\end{eqaed}
where we explicitly implemented the assumption that the only charged hypermultiplets under $SU(2)$ be fundamentals. The corresponding refined elliptic genus\footnote{We suppress $q$-dependence. When $z=0$ we also do not write it as an argument.}
\begin{eqaed}\label{eq:refined_ell_genus}
    Z_\text{ell} = \frac{2}{\eta^4} \left(\mathcal{O} \, \chi_0^{(k)}(z) - \mathcal{A}_1 \, \chi_1^{(k)}(z) - \mathcal{A}_0 \, \chi_0^{(k)}(z)\right) \equiv 2 \, \frac{\Phi_{10,k}(z)}{\Delta}
\end{eqaed}
is a weak Jacobi form of weight 10 and index $m=k$ \cite{Lee:2020gvu} determined by the level of the $\widehat{\mathfrak{su}(2)}_k$ affine algebra realised on the internal worldsheet conformal field theory. In the above expression and in the following $\Delta = \eta^{24}$ denotes the modular discriminant.

\noindent The ring of weak Jacobi forms and a convenient set of generators is described in detail in \cite{Lee:2020gvu}. Over the ring of modular forms $\text{MF} = \mathbb{Q}[E_4, E_6, \Delta]/(E_4^3 - E_6^2 - 1728 \Delta)$ generated by the holomorphic Eisenstein series $E_4$ and $E_6$, where we allow rational coefficients, weak Jacobi forms are generated by
\begin{eqaed}
    & \varphi_{0,1}(\tau, z) \equiv 4 \left( \frac{\theta_2^2(\tau,z)}{\theta_2^2(\tau,0)} + \frac{\theta_3^2(\tau,z)}{\theta_3^2(\tau,0)} + \frac{\theta_4^2(\tau,z)}{\theta_4^2(\tau,0)} \right) \, , \\
    & \varphi_{-2,1}(\tau, z) \equiv - \, \frac{\theta_1^2(\tau,z)}{\eta^6(\tau)} \, ,
\end{eqaed}
where the suffix contains the weight and index. For our purposes, the appropriate ansatz for $\Phi_{10,k}$ reads
\begin{eqaed}\label{eq:jacobi_ansatz}
    \Phi_{10,k} = \frac{1}{12^k} \, E_4 \, E_6 \, \varphi_{0,1}^k + \frac{a_4}{12^{k-1}} \, E_4^3 \, \varphi_{-2,1} \, \varphi_{0,1}^{k-1} + \frac{a_6}{12^{k-1}} \, E_6^2 \, \varphi_{-2,1} \, \varphi_{0,1}^{k-1} + \dots
\end{eqaed}
where the extra terms contribute to $\mathcal{O}(z^4)$ upon expanding in the $SU(2)$ chemical potential $z$. Expanding
\begin{eqaed}\label{eq:z-exp_su2}
    \chi_\ell^{(k)}(z) = \chi_\ell^{(k)} + \widetilde{\chi}_{k,\ell} \, z^2 + \dots
\end{eqaed}
and comparing the ansatz to \cref{eq:refined_ell_genus} one finds, to order $\mathcal{O}(z^2)$,
\begin{eqaed}\label{eq:ansatz_matching_z2}
    & \frac{\Delta}{\eta^4} \left[ \mathcal{O} \, \chi_0^{(k)} - \mathcal{A}_1 \, \chi_1^{(k)} - \mathcal{A}_0 \, \chi_0^{(k)} \right] = E_4 \, E_6 \, , \\
    & \frac{\Delta}{\eta^4} \left[ \mathcal{O} \, E_2 , \widetilde{\chi}_{k,0} - \mathcal{A}_1 \, E_2 \, \widetilde{\chi}_{k,1} - \mathcal{A}_0 \, E_2 \, \widetilde{\chi}_{k,0} \right] = - \, \frac{k}{12} \, E_2 E_4 E_6 - \, a_4 \, E_4^3 - \, a_6 \, E_6^2 \, .
\end{eqaed}
Therefore, one reconstructs the characters $\mathcal{A}_0$ and $\mathcal{A}_1$, finding
\begin{eqaed}\label{eq:hyper_characters}
    & \frac{\mathcal{A}_0}{\eta^4} = \frac{\mathcal{O}}{\eta^4} + \frac{\left(\frac{k}{12} \, E_2 E_4 E_6 + a_4 \, E_4^3 + a_6 \, E_6^2 \right) \chi_1^{(k)} + E_4 E_6 \, \widetilde{\chi}_{k,1}}{(\chi_1^{(k)} \widetilde{\chi}_{k,0} - \, \chi_0^{(k)} \widetilde{\chi}_{k,1})\Delta} \, , \\
    & \frac{\mathcal{A}_1}{\eta^4} = - \, \frac{\left(\frac{k}{12} \, E_2 E_4 E_6 + a_4 \, E_4^3 + a_6 \, E_6^2 \right) \chi_0^{(k)} + E_4 E_6 \, \widetilde{\chi}_{k,0}}{(\chi_1^{(k)} \widetilde{\chi}_{k,0} - \, \chi_0^{(k)} \widetilde{\chi}_{k,1})\Delta} \, . \\
\end{eqaed}
Specialising to the case of interest $k=1$, we further obtain\footnote{For $k=1$ the ansatz in \cref{eq:jacobi_ansatz} is complete, without extra terms.}
\begin{eqaed}\label{eq:q-expansions}
    & \frac{\mathcal{A}_0 \, \chi_0^{(1)}}{\eta^4} = \frac{\mathcal{O} \, \chi_0^{(1)}}{\eta^4} + \left(- \, \frac{5}{6} + 2 \, a_4 + 2 \, a_6 \right) \frac{1}{q} + \left(\frac{568}{3} + 1504 \, a_4 - 1952 \, a_6 \right) + \dots \\
    & \frac{\mathcal{A}_1 \, \chi_1^{(1)}}{\eta^4} = \left(- \, \frac{1}{6} - 2 \, a_4 - 2 \, a_6 \right) \frac{1}{q} + \left(\frac{152}{3} - 1504 \, a_4 + 1952 \, a_6 \right) + \dots
\end{eqaed}
On the other hand, by counting chiral degrees of freedom in supergravity, it should be the case that
\begin{eqaed}\label{eq:q-expansions_ansatz}
    & \frac{\mathcal{O} \, \chi_0^{(1)}}{\eta^4} = \frac{1}{q} + 7 + \dim(G) + \dots \, , \\
    & \frac{\mathcal{A}_0 \, \chi_0^{(1)}}{\eta^4} = H-2d_F + \dots \, , \\
    & \frac{\mathcal{A}_1 \, \chi_1^{(1)}}{\eta^4} = 2d_F + \dots \, ,
\end{eqaed}
where $7 = 4+3$ counts gravitini and the adjoint representation of $SU(2)$. Solving these constraints and using that $H = 247 + \dim(G)$ by anomaly cancellation, one fixes the unknown coefficient $a_6$ to
\begin{eqaed}\label{eq:a_6_fixed}
    & a_6 = \frac{47}{61} \, a_4 + \frac{3 \, d_F - 76}{2928} \, .
\end{eqaed}
The remaining $q^{-1}$ terms are proportional to $56 + N_\text{f} + 1728 \, a_4$, which has to vanish on account of \cref{eq:q-expansions_ansatz}. Therefore, the last unknown coefficient $a_4$ is fixed to
\begin{eqaed}\label{eq:a_4_fixed}
    & a_4 = - \, \frac{56 + d_F}{1728} \, .
\end{eqaed}
Substituting these values, one finds the full refined elliptic genus
\begin{eqaed}\label{eq:full_ell_genus}
    Z_\text{ell} & = 2 \, \frac{E_4 \, E_6}{\Delta} \, \frac{\varphi_{0,1}}{12} - \, 2 \, \frac{d_F + 56}{1728} \, \frac{E_4^3}{\Delta} \, \varphi_{-2,1} + 2 \, \frac{d_F - 88}{1728} \, \frac{E_6^2}{\Delta} \, \varphi_{-2,1} \\
    & = \frac{2}{q} + 2 \left( - \, 242 + 2 d_F - d_F \, \xi^{\pm 1} + \xi^{\pm 2} \right) + \dots \\
    & = \frac{2}{q} - \, 480 + \left(2 d_F - 8\right) z^2 + \left(\frac{8}{3} - \, \frac{d_F}{6} \right) z^4 + \frac{d_F - \, 64}{180} \, z^6 + \dots
\end{eqaed}
where $\xi^{\pm n} \equiv \xi^n + \xi^{-n}$ counts the Cartan $SU(2)$ charge.

\noindent With these preparations, we have automated a scan over chiral algebras of the form of \cref{eq:chiral_algebra_ansatz}, writing all possible partition functions with critical central charges and imposing modular invariance. The resulting refined elliptic genera turn out to always yield $d_F \equiv_8 0$, which is consistent with anomaly cancellation. This is a nontrivial consistency check, since a priori \cref{eq:full_ell_genus} could contain any value of $d_F$ without imposing modular invariance at the level of the full partition function. It would be interesting to extend this scan to more general worldsheet constructions, in order to see whether modular invariance and criticality are powerful enough to exclude Dai-Freed anomalies (which are \emph{a priori} a non-perturbative quantum gravity effect) by themselves.

\subsection{The general case}\label{sec:anomaly_SUn_general}

Having established the general methodology, we can proceed to study Dai-Freed anomalies on Lens spaces for a general $SU(n)$ gauge group. For $n>2$ there can be irreducible local gauge anomalies in addition to purely gravitational ones. Allowing hypermultiplets in trivial, fundamental, symmetric and antisymmetric representations one finds the condition 
\begin{equation} \label{irrgrav}
	d_T + d_F \ n + d_A \ \frac{n^2-n}{2} + d_S \ \frac{n^2+n}{2} = 243+n^2,
\end{equation} 	
from the cancellation of $\text{tr}R^4$, and 
\begin{equation} \label{irrgauge}
	d_F   +   d_A \ (n-8)   +  d_S \ (n+8)  =  2 n,
\end{equation} 
from the cancellation of $\text{tr}F^4$. Solving for $d_A$ and $d_S$, the resulting anomaly polynomial is given by
\begin{equation} \label{anpolsun}
	I_8=\left (\text{tr} R^2 \right )^2-\frac{1}{4} \text{tr} R^2 \ \text{tr} F^2 \left ( d_A  -  d_S \right ) +	\frac{1}{8}  \left (  \text{tr} F^2 \right )^2 \left ( d_A  +  d_S -2 \right ).
\end{equation}
Once again, we impose the conditions in \cite{Seiberg:2011dr} for the diagonal and off-diagonal lattice embeddings of the anomaly lattice.

\subsubsection{Diagonal lattice}\label{sec:diag_SUn}

In order to have a consistent embedding into the diagonal lattice, one can express the multipliticies in terms of two integers $r \, , \, s$ according to
\begin{equation} \label{sundiag}
		\begin{aligned}
			&d_S=1 \ - \ \frac{s}{2} \ - \ \frac{s^2}{2} \ - \ 3 \ r \ s \ - \ 4  r^2 \ ,
			\\
			&d_A=1 \ + \ \frac{s}{2} \ - \ \frac{s^2}{2} \ - \ 3 \ r \ s \ - \ 4  r^2 \ ,
			\\
			& d_F=  s^2 n \ + \ 2 n r \  ( 4 r \ + \ 3 s) \ + \ 8 s   \ ,
			\\
			& d_T = 243 \ - \ \frac{n}{2} \left [ s \ ( 15 \ + \ n s) \ + \ 6 s n r \ + \ 8 n r^2 \right ]  \ ,
		\end{aligned}
	\end{equation}
with the implicit restriction that the multiplicities be non-negative.
For this family of models, the anomaly polynomial factorises as
	\begin{equation} \label{anpolsundiag}
		I_8 = \frac{1}{8} \left \{ \left ( 3 \text{tr} R^2 + r \text{tr} F^2  \right )^2 -  \left (  \text{tr} R^2 + (s+3r) \text{tr} F^2 \right  )^2 \right \}.	
	\end{equation}
From this expression one can read off the corresponding Bianchi identities,
	\begin{equation}
		\begin{aligned}
			&d H_1= \tfrac{3}{4} p_1 - r c_2 \, ,
			\\
			&d H_2= \tfrac{1}{4} p_1 - (s+3r) c_2 \, .
		\end{aligned}
	\end{equation}
Studying this family in full generality is quite cumbersome, since one can trivialise Bianchi classes in various ways and, similarly, quadratic refinements can cancel anomalies in various ways. The story would be considerably more interesting if we could understand which quadratic refinements, if any, are singled out by string theory. Hints of this phenomenon were laid out in \cite{Dierigl:2022zll}, but at present a complete understanding is hindered by the difficulties in treating global aspects with a top-down approach.

\noindent Because of this, we complement this analysis of allowable $SU(n)$ families for diagonal lattice embeddings with a computation of anomalies for the specific quadratic refinements defined in \cite{Hsieh:2020jpj}, but we leave a comprehensive analysis of all quadratic refinements for future work. Models with zero tensor multiplets are simpler, and we discuss them in \cref{sec:abelian_no_tensors} and \cref{sec:gepner}. In particular, in \cref{sec:gepner} we show that the Gepner orientifold without tensor multiplets is anomaly-free for a unique choice of quadratic refinement defined in \cite{Hsieh:2020jpj}, compatibly with the considerations in \cite{Dierigl:2022zll}.

\noindent For Lens spaces, the quadratic refinement defined in \cite{Hsieh:2020jpj} evaluates to
\begin{equation}
    \widetilde{Q}( \check{\mathbf{A}})=- 12 \left ( \eta_1 - \eta_0 \right ) \, , 
\end{equation}
thus giving rise in general to
\begin{equation} \label{quaddef}
	\begin{aligned}
		\widetilde{Q}(b \check{\mathbf{A}}) &= b \, \widetilde{Q}(\check{\mathbf{A}}) + \frac{b (b-1)}{2} \left ( \check{\mathbf{A}}, \check{\mathbf{A}} \right )
		\\
		&= -b \, \frac{p^2-1}{2p} - \frac{ b (b-1)}{2p}  \, ,
	\end{aligned}
\end{equation}
where the differential character  $\check{\mathbf{A}}$ generates the integer cohomology group of the Lens space. Since the anomaly theory is now fully specified, one can seek anomalous models within the family parametrised by \cref{sundiag}. As an example, the model with $r=-8$ and $s=20$, corresponding to
\begin{eqaed}
    d_F = 160-48n \, , \quad d_S = 15 \, , \quad d_A = 35
\end{eqaed}
is anomalous. Indeed, the Bianchi classes are $-5y$ and $-3y$, with $y$ a generator of degree-four cohomology, and the anomaly on $L_5^7$ evaluates to $\tfrac{1}{5}$. Hence, the $SU(3)$ model with sixteen fundamental, fifteen symmetric and thirty-five antisymmetric hypermultiplets is anomalous. As another example, the $SU(3)$ model with $r=-7$ and $s=16$ has both Bianchi classes equal to $-4y$, and the anomaly on $L_4^7$ evaluates to $\tfrac{1}{2}$.

\subsubsection{Off-diagonal lattice}

Similarly to the preceding case, in order to have a consistent embedding into the off-diagonal lattice, one can express the multipliticies in terms of two integers $r \, , \, s$ according to
\begin{equation}
		\begin{aligned} \label{sunoff}
			&d_S=1 \ - \ s \ - \ r  s \ - \ r^2 \ ,
			\\
			&d_A=1 \ + \ s \ - \ r  s \ - \ r^2 \ ,
			\\
			& d_F= 2 n r \  ( n r \ + \ s) \ + \ 16 s   \ ,
			\\
			& d_T = 243 \ - \ n r \ ( n r \ + \ n s ) \ + \ 15 n s  \ ,
		\end{aligned}
	\end{equation}
once again with the implicit restriction that the multiplicities be non-negative. The resulting anomaly polynomial factorises as
	\begin{equation} \label{anpolsunoff}
		I_8 = \frac{1}{4} \left ( 2 \text{tr} R^2 + r \text{tr} F^2  \right )  \left ( 2 \text{tr} R^2 - (r+s) \text{tr} F^2 \right  ).	
	\end{equation}
As a result, the Bianchi identity encoding the twisted string structure are
	\begin{equation}
		\begin{aligned}
			&dH_1= \tfrac{1}{2} p_1 - r c_2 \, ,
			\\
			&d H_2= \tfrac{1}{2} p_1 + (r+s) c_2 \, .
		\end{aligned}
	\end{equation} 
For $SU(n)$, the bundles described in \cref{appendix:systematics} allow $2k \leq n$, with second Chern class $c_2 = - \, k y$ where $y$ is a generator of degree-four cohomology as before. Thus, the Bianchi classes on Lens backgrounds evaluate to $(2 + k r)y$ and $(2 - k r - k s)y$. One can trivialise either Bianchi class choosing $ks = 2 - kr - mp$ (for the first class) or $kr = mp - 2$ (for the second class) for some integer $m$. The resulting anomaly simplifies to
\begin{eqaed}
    \mathcal{A}(L_p^7) & = \frac{(1-p)(p^2+p+6+3kr+3m(p+kr+2))}{3} \\
    & \equiv_1 - \, \frac{(p-1) \, p \, (p+1)}{3} \equiv_1 0
\end{eqaed}
in the former case, and
\begin{eqaed}
    \mathcal{A}(L_p^7) & = \frac{(1-p)(p^2+p+12-3m^2p-3m(ks-4)-3ks)}{3} \\
    & \equiv_1 - \, \frac{(p-1) \, p \, (p+1)}{3} \equiv_1 0
\end{eqaed}
in the latter case. Thus the anomaly vanishes whenever the twisted string structure exists.

\noindent All in all, for $SU(n)$ no novel constraints emerged for the off-diagonal models, which is a non-trivial result in its own right. Diagonal families may some anomalies where, perhaps, no choice of quadratic refinements can cancel them, but at least for particular choices such as the one of \cite{Hsieh:2020jpj} there are anomalous examples.

\section{Dai-Freed anomalies for Spin groups}\label{sec:anomaly_Spin2n}

We now move on to Spin groups, in particular the $D_n$ series $\text{Spin}(2n)$. We include once again trivial, vector, symmetric and antisymmetric (adjoint) representations. The conditions to cancel irreducible gravitational and gauge anomalies now read
\begin{equation}
	d_T + 2n d_F+ n(2n-1)( d_{\text{Ad}}-1) + n(2n+1) d_S= 244
\end{equation}
for gravitational anomalies, and
\begin{equation}
	d_F+ (2n-8)( d_{\text{Ad}}-1) +(2n+8) d_{S}=0 \, .
\end{equation}
for gauge anomalies. Here we denote the antisymmetric multiplicity with $d_\text{Ad}$, since it corresponds to the adjoint representation in this case. The complete anomaly polynomial is then given by
\begin{equation} \label{anpolso2n}
	I_8=\left (\text{tr} R^2 \right )^2-\frac{1}{4} \text{tr} R^2 \ \text{tr} F^2 \left (  d_{\text{Ad}}-1  -  d_S \right ) +	\frac{1}{8}  \left (  \text{tr} F^2 \right )^2 \left (  d_{\text{Ad}}-1+ d_S \right ).
\end{equation}
As before, we now seek families of models whose anomaly lattice admit consistent embeddings into a self-dual lattice following \cite{Seiberg:2011dr}. Once again, the multiplicities of each family can be expressed in terms of two integers $r \, , \, s$ with the implicit restriction that the multiplicities be non-negative.

\subsection{Diagonal lattice}

For the diagonal lattice, there is a single family, which is described by
\begin{equation}
		\label{famso2ndiag}
		\begin{aligned}
			&d_F= r (8 + 2 r n) + 6 r n s + 4 n s^2 \, ,
			\\
			&d_{\text{Ad}}= 1 +  \frac{r - r^2 - 3 r s - 2 s^2}{2} \, ,
			\\
			&d_S=\frac{-r - r^2 - 3 r s - 2 s^2}{2} \, ,
			\\
			&d_T= 244 - \frac{ 2n (2 r^2 n + 4 n s^2 + 3 r (5 + 2 n s))}{2} \, .
		\end{aligned},
	\end{equation}
As a result, the anomaly polynomial simplifies to the factorised form
	\begin{equation}
		I_8 = \frac{1}{8} \left \{ \left ( 3 \text{tr} R^2 + \frac{s}{2} \text{tr} F^2  \right )^2 -  \left (  \text{tr} R^2 + \frac{3s+2r}{2} \text{tr} F^2 \right  )^2 \right \},
	\end{equation} 
from which one can read off the corresponding Bianchi identities
	\begin{equation}
		\begin{aligned}
			&d H_1= \tfrac{3}{4} p_1 - \frac{s}{2} c_2 \, ,
			\\
			&d H_2= \tfrac{1}{4} p_1 - \tfrac{2r +3 s}{2} c_2 \, .
		\end{aligned}	
	\end{equation}
As discussed in the preceding section on unitary groups, we postpone a detailed analysis of the anomalies for this kind of models, due to the difficulties in handling two potentially different quadratic refinements.  As an example, the $\text{Spin}(4)$ model with $r=8$ and $s=-6$, corresponding to
\begin{eqaed}
    d_F = 32 \, , \quad d_\text{Ad} = 9 \, , \quad d_S = 0 \, ,
\end{eqaed}
has Bianchi classes equal to $\pm 3 y$, with $y$ a generator of degree-four cohomology. The fermionic anomaly on $L_3^7$ evaluates to $\tfrac{1}{3}$, and thus admits a combination of quadratic refinements that allows to cancel such contribution. Therefore, as for the unitary case, the choice of quadratic refinement plays a crucial role, and indeed, if we use for example the definition of \cite{Hsieh:2020jpj} for \cref{quaddef} the model is anomalous.

\subsection{Off-diagonal lattice}

For the off-diagonal lattice, two families are found. The first one is parametrised according to
\begin{equation} \label{famso2noff1}
		\begin{aligned}
			&d_F= 4 (n s^2 + r (4 + n s)) \, ,
			\\
			&d_{\text{Ad}}= (1-s)(1+r+s) \, ,
			\\
			&d_S=-s^2-r(1+s) \, ,
			\\
			&d_T=244 - 30 n r - 4 n^2 s (r + s) \, ,
		\end{aligned}
	\end{equation}
which yields the factorised anomaly polynomial
	\begin{equation} \label{anpolso2noff1}
		I_8= \frac{1}{4}  \left ( 2 \text{tr} R^2 + \frac{2s}{2} \text{tr} F^2  \right )  \left ( 2 \text{tr} R^2 - \frac{2r+2s}{2} \text{tr} F^2 \right  ).
	\end{equation}
 The second family is described instead by
 \begin{equation} \label{famso2noff2}
		\begin{aligned}
			&d_F= 2 (4 + 8 r + n (1 + r + s) (1 + 2 s)) \, ,
			\\
			&d_{\text{Ad}}= (1-2s)\frac{2+r+s}{2} \, ,
			\\
			&d_S=-\frac{2+(r+s)(3+2s)}{2} \, ,
			\\
			&d_T=244 - n (15 + 30 r + 2 n (1 + r + s) (1 + 2 s)) \, ,
		\end{aligned}
	\end{equation}
leading to the factorised anomaly polynomial
	\begin{equation} \label{anpolso2noff2}
		I_8= \frac{1}{4}  \left ( 2 \text{tr} R^2 + \frac{2s+1}{2} \text{tr} F^2  \right )  \left ( 2 \text{tr} R^2 - \frac{2r+2+2s}{2} \text{tr} F^2 \right  ).
	\end{equation}
The resulting Bianchi identities can be actually encompassed into a single case for both families, namely
	\begin{equation}
		\begin{aligned}
			&d H_1= \tfrac{1}{2} p_1 - \tfrac{m}{2} c_2 \, ,
			\\
			&d H_2= \tfrac{1}{2} p_1 + \tfrac{m + w}{2} c_2 \, ,
		\end{aligned}
	\end{equation}   
where $m=2s+\sigma$ and $w=2r+\sigma$, for which $\sigma=0$ identifies \cref{famso2noff1} whereas $\sigma=1$ identifies \cref{famso2noff2}. For these models, the Chern class of $\text{Spin}(2n)$ bundles evaluates to $-2k \, y$ according to the results in \cref{appendix:systematics}.

\noindent The two families do not admit any anomalous model as for the $SU(n)$ off-diagonal case. Indeed, imposing the trivialisation of the first Bianchi identity implies $2+k m =0$ mod $p$ and it can be shown after some trivial algebra that the anomaly recasts as
\begin{equation} \label{anomalyoffdiagspin2n}
    \begin{aligned}
        \mathcal{A}(L_p^7) &=- \frac{p(p^2-1)}{3}
        \\
        &\equiv_1 0 \, ,
    \end{aligned}
\end{equation}
while trivialising the second Bianchi identity implies $2-k (m+w)=0$ mod $p$ which gives rise to the same result contained in \eqref{anomalyoffdiagspin2n}.

\noindent As a concrete example, we can focus on the $\text{Spin}(28)$ model in the second family, with $s = r = -1$. This corresponds to the heterotic orbifold model ``2a'' in \cite{Honecker:2006qz} (at least at the level of the Lie algebra), with
\begin{eqaed}
    d_F = 20 \, , \quad d_\text{Ad} = 0 \, , \quad d_S = 0 \, .
\end{eqaed}
The Bianchi classes are $(2-k)y$ and $2(1+k)y$. Thus, choosing $k = 2+mp$ or $k = mp - 1$ for some integer $m$ trivialises one of them, and in either case the anomaly simplifies to
\begin{eqaed}
    \mathcal{A}(L_p^7) \equiv_1 - \, \frac{(p-1) \, p \, (p+1)}{3} \equiv_1 0.
\end{eqaed}
It would be interesting to perform a complete analysis that can lead us to the knowledge of the bordism group, that would allow us to understand whether the bordism groups associated to these string structures are trivial, as this result seems to suggest, or if it admits a non trivial generator on which these models can be tested.

\section{Dai-Freed anomalies for exceptional groups}\label{sec:anomaly_E7E8}

For exceptional groups the classification of the landscape, including only fundamental and adjoint representations for the charged hypermultiplets, is quite simpler. The price to pay is that computing Dai-Freed anomalies is more difficult at the technical level, because the bundles described in \cref{appendix:systematics} require decomposing spinorial representations of $\text{Spin}(12)$ and $\text{Spin}(16)$ for $E_7$ and $E_8$ respectively. Once again, we present a systematic description of allowable models also for the diagonal lattice embedding, showing an example of an anomalous model for the choice of quadratic refinement of \cite{Hsieh:2020jpj}, and we leave a comprehensive classification of (pairs of) quadratic refinements for future work.

\subsection{The case of \texorpdfstring{$E_7$}{E7}}

Let us begin with $E_7$. The multiplicities of uncharged matter and fundamental and adjoint charged matter are restricted to
\begin{equation}
	d_T + 56 d_F+ 133( d_{\text{Ad}}-1) = 244
\end{equation} 
by cancellation of irreducible anomalies. Making use of the helpful trace identities
\begin{equation}
	\begin{aligned}
		&\text{tr}_{\text{Ad}} F^2= 3 \,\text{tr} F^2 
		\, , \\
		& \text{tr}_{\text{Ad}} F^4=  \frac{1}{6} \left ( \text{tr} F^2 \right )^2 , \\
        & \text{tr} F^4=  \frac{1}{24} \left ( \text{tr} F^2 \right )^2 ,
	\end{aligned}
\end{equation}
it is possible to simplify the anomaly polynomial to
\begin{equation}
	I_8=\left ( \text{tr} R^2 \right )^2 - \frac{1}{4} \text{tr} R^2  \ \text{tr} F^2 \left ( d_F +  3 (d_{\text{Ad}}-1) \right ) + \frac{1}{4} \left ( \text{tr} F^2 \right )^2 \left ( \frac{ d_F}{4} +   (d_{\text{Ad}}-1) \right ) ,
\end{equation}
where for convenience we have rescaled $\text{tr} F^2 \to 6 \, \text{tr} F^2$. As in the preceding sections, we seek families of models with consistent embeddings according to the conditions in \cite{Seiberg:2011dr}. Once more, the multiplicities are fixes in terms of two integers $r \, , \, s$ as before.

\subsubsection{Diagonal lattice for \texorpdfstring{$E_7$}{E7}}

Embedding the anomaly lattice into the diagonal lattice fixes
\begin{equation}
		\begin{aligned}
			&d_T=244-91 s - 70 (2r+s) (r+s) \, ,
			\\
			&d_F=4 s + 6(r+s) (2r+s) \, ,
			\\
			& d_{\text{Ad}}= 1-s - 4 r^2- 6 s r-2s^2 \, .
		\end{aligned}
	\end{equation} 
The resulting anomaly polynomial factorises in the form
	\begin{equation}
		I_8 = \frac{1}{8} \left \{ \left ( 3 \text{tr} R^2 + \frac{r}{2} \text{tr} F^2  \right )^2 -  \left (  \text{tr} R^2 + \frac{3r+2s}{2} \text{tr} F^2 \right  )^2 \right \}.
	\end{equation}
From this expression one can read off the Bianchi identities
	\begin{equation}
		\begin{aligned}
			&d H_1= \tfrac{3}{4} p_1 - \frac{r}{2} c_2 \, ,
			\\
			&d H_2= \tfrac{1}{4} p_1 - \frac{3r+2s}{2} c_2 \, .
		\end{aligned}
	\end{equation}   

\noindent As an example of anomalous model with the choice of quadratic refinement in \cref{quaddef}, picking $r=-3$ and $s=4$ leads to
\begin{eqaed}
    d_F = 4 \, , \quad d_\text{Ad} = 1 \, .
\end{eqaed}
For such model we can choose to trivialise the first Bianchi with the $E_7$ bundle identified by $k=2$, which admits $L_3^7$ as a background, leading to
\begin{equation} \label{andiagE7}
    \mathcal{A}(L_3^7)= \frac{4}{9} \, . 
\end{equation}
The second Bianchi is instead trivialised by choosing $p=1+k$ with $k=1,2$ giving rise to a vanishing anomaly for $k=1$, while \eqref{andiagE7} is reproduced for $k=2$.

In the case for $k=2$ there is no possible combination of quadratic refinement allowing to cancel the fermion anomaly and thus such models do not admit a UV completion in quantum gravity. However, in such a case it is possible to introduce only probe strings on which the theory is non-unitary, thus violating the KSV constraints. This means that, although the bordism group is non-trivial and there are anomalous models, the latter are already excluded by the analysis performed in \cite{Kim:2019vuc}.

\subsubsection{Off-diagonal lattice for \texorpdfstring{$E_7$}{E7}}

Embedding the anomaly lattice into the off-diagonal lattice fixes 
\begin{equation}
		\begin{aligned}
			&d_T=244-91 s - 35 r (r+s) \, ,
			\\
			&d_F=4 s + 3 r(r+s) \, ,
			\\
			& d_{\text{Ad}}= 1-s - r^2-s r \, ,
		\end{aligned}
	\end{equation} 
	for which the anomaly polynomial factorises as
	\begin{equation}
		I_8= \frac{1}{4} \left ( 2 \text{tr} R^2 + \frac{r}{2} \text{tr} F^2  \right )  \left ( 2 \text{tr} R^2 - \frac{r+s}{2}\text{tr} F^2 \right  ).
	\end{equation}
The corresponding Bianchi identities then read
	\begin{equation}
		\begin{aligned}
			&d H_1= \tfrac{1}{2} p_1 - \frac{r}{2} c_2 \, ,
			\\
			&d H_2= \tfrac{1}{2} p_1 + \frac{r+s}{2} c_2 \, .
		\end{aligned}
	\end{equation}

\noindent As for the $SU(n)$ and $\text{Spin}(2n)$ cases, it can be shown that such models yield no anomalies on Lens spaces trivialising at least one of the Bianchi identity with the bundle identified by $k=1$. Indeed, the first Bianchi is trivialised if $2+r=0$ mod $p$ and the second one whenever $2-(r+s)=0$ mod $p$. In the former case the anomaly can be written as
\begin{equation}
\begin{aligned}
    \mathcal{A}(L_p^7) &= \frac{6}{p} - \frac{20 p}{3} - \frac{p^3}{3} - \frac{
 2 ( - 3 - s + 2 s)}{p} - \frac{ 4 s +6 (2 - s)}{p}
 \\
 & \equiv_1 -\frac{p(p^2-1)}{3}
 \\
 & \equiv_1 0 \, ,
 \end{aligned}
\end{equation}
while the it reads 
\begin{equation}
\begin{aligned}
    \mathcal{A}(L_p^7) &= \frac{6}{p} - \frac{20 p}{3} - \frac{p^3}{3} -\frac{6 (2 - s) + 4 s}{p} - \frac{2 (1 - (2 - s)^2 - s - (2 - s) s)}{p}
 \\
 & \equiv_1 0 \, ,
 \end{aligned}
\end{equation}
in the second case.

\noindent For $k=2$ instead there exists anomalous models, as for example the case in which $r=-9$ and $s=10$ admitting $L_{16}^7$ as an allowed background, but on which the anomaly theory is given by
\begin{equation}
    \mathcal{A}(L_{16}^7)= \frac{9}{32} \, .
\end{equation}
However, as it happens for the diagonal case, these models turn out to be already excluded by imposing the KSV constraints and thus, in such a case, Dai-Freed anomalies on Lens spaces cannot provide any additional conditions through which we can bind the set of low energy theories.

\subsection{The case of \texorpdfstring{$E_8$}{E8}}

Let us now discuss $E_8$ as the final example of non-abelian gauge group. The situation here is much simpler than the ones analysed previously, and indeed the number of cases is manifestly finite. The cancellation of the irreducible gravitational anomaly constrains the number of trivial and adjoint representations that can appear according to
\begin{equation} \label{irrgrave8}
	d_T+ 248(d_{\text{Ad}}-1)=244 \, ,
\end{equation}
from which it immediately follows that no such theory with just $E_8$ as gauge group is allowed. Nevertheless, we can relax \cref{irrgrave8}, allowing the presence of other gauge group factors whose vector multiplets appear as hypermultiplets with the wrong chirality. With this subtlety in mind, the anomaly polynomial is
\begin{equation}
	I_8=\left ( \text{tr} R^2 \right )^2 - \frac{30}{24} \text{tr} R^2  \ \text{tr} F^2 \left ( d_{\text{Ad}}-1 \right ) + \frac{9}{24} \left ( \text{tr} F^2 \right )^2 \left ( d_{\text{Ad}}-1 \right ) ,
\end{equation}
where we have made use of the identity
\begin{equation}
	\text{tr} F^4= \frac{1}{100} \left ( \text{tr} F^2 \right )^2
\end{equation}
as well as rescaled $F^2 \to 30 \, F^2$, once again to take into account the minimal instanton numbers. Once more, the Seiberg-Taylor conditions lead to consistent embeddings of the anomaly lattice into a self-dual lattice in terms of integers $r \, , \, s$ as follows.

\subsubsection{Diagonal lattice for \texorpdfstring{$E_8$}{E8}}

For the diagonal lattice, there is only solution
\begin{equation}
		d_{\text{Ad}}=2 \, ,
	\end{equation}
which leads to the Bianchi identities
	\begin{equation}
		\begin{aligned}
			&d H_1= \tfrac{3}{4} p_1 + \tfrac{4}{2}  c_2 \, ,
			\\
			&d H_2= \tfrac{1}{4} p_1 + \tfrac{2}{2} c_2 \, .
		\end{aligned}
	\end{equation}
The uniqueness of the factorisation once the diagonal embedding is imposed allows to carry out the analysis case by case. Since the second Chern class for $E_8$ bundles on Lens spaces is given by $c_2=-2 k y$, the first Bianchi identity is trivialised by choosing $p$ to be a divisor of $4k-3$ and of $2k-1$ when instead we choose to trivialise the second Bianchi identity. Indeed, when $k=2$\footnote{The case $k=1$ does not give rise to meaningful Lens spaces and thus cannot be considered in the following.} the only available Lens space associated to the first Bianchi identity corresponds to $p=5$, yielding 
\begin{eqaed}
    \mathcal{A}(L_{5}^7) \equiv_1 \frac{3}{5} \, ,
\end{eqaed}
while trivialising the second class requires $p=3$ and the anomaly reads
\begin{eqaed}
    \mathcal{A}(L_{3}^7) \equiv_1 \frac{2}{3} \, .
\end{eqaed}
If instead $k=3$, the available Lens spaces trivialising the first Bianchi identity are given by $p=3,9$, yielding
\begin{eqaed}
    &\mathcal{A}(L_{3}^7) \equiv_1 \frac{1}{3} \, ,
    \\
    & \mathcal{A}(L_{9}^7) \equiv_1 \frac{1}{9} \, ,
\end{eqaed}
whereas the second one is trivialised by choosing $p=5$, giving rise to the anomaly
\begin{eqaed}
    \mathcal{A}(L_{5}^7) \equiv_1 \frac{3}{5} \, .
\end{eqaed}
All in all, if only one Bianchi identity is required for the twisted string structure, this model is anomalous for the choice of quadratic refinement in \cite{Hsieh:2020jpj}. Since anomaly-free models, and in particular string constructions, seem to be indifferent to this choice, this seems to be a reasonable criterion to exclude models when it is impossible to trivialise both Bianchi classes.

\subsubsection{Off-diagonal lattice for \texorpdfstring{$E_8$}{E8}}

For the off-diagonal lattice we find a similar result for the parametrisation of the allowed models. Namely the multiplicity of adjoint hypermultiplets takes the form
\begin{equation}
		d_{\text{Ad}}=2 \, ,
	\end{equation}
and the corresponding Bianchi identities are
	\begin{equation}
		\begin{aligned}
			&d H_1= \tfrac{1}{2} p_1 + \tfrac{3}{2} c_2 \, ,
			\\
			&d H_2= \tfrac{1}{2} p_1 +  \tfrac{2}{2} c_2 \, .
		\end{aligned}
	\end{equation}
Once again, since there is only one model to be considered, one can simply look for an anomaly on a case-by-case basis. By the previous considerations, the choices trivialising the first Bianchi identity for $k=2$ correspond to $p=4$, yielding no anomaly, and $p=2$, which trivialises simultaneously also the second Bianchi, for which, again, the anomaly cancels. The other possible $E_8$ bundle is given by $k=3$, that identifies the Lens space with $p=7$ as the result of the trivialisation of the first Bianchi identity, for which no anomaly is found. Trivialising the second Bianchi implies $p=2,4$ and again the anomaly cancels on the corresponding Lens spaces $L_{2,4}^7$. 

\section{Dai-freed anomalies for abelian charges}\label{sec:anomaly_U1}

In this section we study theories with matter charged under the gauge group $U(1)$. Swampland bounds on the number $r$ of abelian factors have been obtained \cite{Taylor:2018khc,Park:2011wv,Lee:2019skh}, depending to the presence of tensor multiplets. In particular, it has been shown \cite{Park:2011wv} that the cancellation of local anomalies, when other non-abelian factors are present, implies the upper bound
\begin{eqaed}
    r \leq (n_T+2)(n_T+ 7/2 + (n_T^2 - 51 n_T + 2225/4)^{1/2})
\end{eqaed}
on the rank. If only abelian groups are considered such constraints can be further refined. Indeed, in \cite{Lee:2019skh} it has been found out that $r \leq 20 \, (22)$ for $n_T \geq 1$ in presence (absence) of charged matter, while $r \leq 32$ if there are no tensor multiplets. 

\noindent Although these constraints are great steps toward finiteness of the landscape, they turn out to be ineffective in limiting the possible highest charge of the hypermultiplets. Indeed, restricting to $r=1$ and $n_T \leq 1$, a previous analysis has been carried out in \cite{Taylor:2018khc}, where infinite families with unbounded $U(1)$ charges were shown to satisfy all the swampland criteria without a known string/F-theory realisation\footnote{In \cite{Taylor:2018khc} the term {\em swampland} is used to identify these kind of theories, rather than EFTs that do not couple consistently to gravity.} with the exception of a finite number of cases\footnote{Some examples with charges bounded by $5$ admit an F-theory construction either directly or as a result of an Higgs mechanism breaking larger non-abelian groups.}. In this section, we shall discuss the families in \cite{Taylor:2018khc} with $n_T \leq 1$ in order to see whether they can be ruled out by Dai-Freed anomalies. In order to do so, we shall first repeat the analysis performed in sections \ref{sec:anomaly_SUn}, \ref{sec:anomaly_Spin2n} and \ref{sec:anomaly_E7E8} in the presence of one tensor multiplet, by imposing $|q|\leq 2$. This choice allows to recover the families with $n_T=1$ described in \cite{Taylor:2018khc} as a testbed. The analysis will be carried out from scratch in order to set notations and conventions. Afterwards, we will consider the infinite families with $n_T=0$ and arbitrary large charges found in \cite{Taylor:2018khc}. Given this setup, we will discuss the role played by Dai-Freed anomalies, aiming to rule out at least some of the models contained therein.

\subsection{Families with one tensor multiplet} 

With one tensor multiplet the anomaly polynomial reads 
\begin{equation}\label{I8u1}
	I_8= \left (\sum_{q=0}^2 d_q- 245 \right ) \frac{1}{360} \text{tr} R^4 \, + \, \left (\text{tr} R^2 \right )^2-\frac{1}{24} \text{tr} R^2 \sum_{q=1}^2 q^2 d_q F^2 +	\frac{1}{24} \sum_{q=1}^2 q^4 d_q F^4 \, .
\end{equation}
The conditions stemming from cancellation of (ir)reducible local gravitational and gauge anomalies constrain the number of degeneracies in the spectrum. In \cref{I8u1}, the degeneracies are summed over positive $q$, since each hypermultiplet contains fields with opposite charges. As described in the previous sections, with one tensor multiplet there are only two ways for the anomaly lattice to be embedded in a self-dual lattice of string charges of signature $(1,1)$, namely those with diagonal or off-diagonal bilinear forms.

\subsubsection{Diagonal lattice}

In the former case, the anomaly polynomial $I_8$ factorises as 
\begin{eqaed} 
	I_8&= \frac{1}{8} \left \{ \left ( 3 \text{tr} R^2 + b^{(1)} F^2 \right )^2 -\left ( \text{tr} R^2 + b^{(2)} F^2 \right )^2\right \}
 \\
 &= \frac{1}{2} \left \{ \left ( \tfrac{3}{4} p_1 + \tfrac{b^{(1)}}{2} c_1^2 \right )^2 -\left ( \tfrac{1}{4} p_1 + \tfrac{ b^{(2)}}{2} c_1^2 \right  )^2 \right \} \, ,
\end{eqaed}
where now the second Chern class is replaced by $-c_1^2/2$ as explained in \cref{appendix:anomaly_formulae}. The coefficients $b^{(i)}$ are obtained imposing the usual factorisation conditions, and for the abelian case one finds \cref{I8u1}
\begin{equation}
	\begin{aligned}
	&{b^{(1)}}^2- 	{b^{(2)}}^2= \frac{1}{3} \sum_{q=1}^2 q^4 d_q = h\, ,
	\\
	&3 b^{(1)}- 	b^{(2)}= -\frac{1}{6} \sum_{q=1}^2 q^2 d_q =-r\, ,
	\end{aligned}
\end{equation}
in which we have introduced two integer parameters $h \, , \, r$.

\noindent This system of equations is solved by $b^{(1)}=s$ and $r+3s$, where $r,s$ are integer parameters constrained to lie on the surface $h=h(r,s)=- r^2 - 6 r s - 8 s^2$. In terms of these quantities we can conveniently write the degeneracies as 
\begin{equation} \label{u1degdiag}
	\begin{aligned}
		& d_0= 245-(30 r - 3 h(r,s))/4 \, , \qquad d_1= 8 r -  h(r,s) \, , \qquad d_2= (h(r,s)-2r)/4 \,  .
	\end{aligned}
\end{equation}
In order for these expressions to be meaningful, these multiplicities are required to be non-negative integers, which entails $2r \leq h \leq 8 r$ and $h > 10 r-327$. This reduces the number of charged hypermultiplets to a smaller set of degeneracies.

\noindent We now study Dai-Freed anomalies for these models. The anomaly theory in terms of $r \, , \, s$ reads
\begin{equation} \label{anu1diag}
	\begin{aligned}
		\mathcal{A}\big ( Y \big )= \frac{8 r -  h}{2} \, \widetilde{\eta}_1 + \frac{h-2r}{8} \, \widetilde{\eta}_2 + \widetilde{\mathcal{Q}}_+\left (\tfrac{s}{2} \mathbf{\check{c}} \right ) - \widetilde{\mathcal{Q}}_-\left ( \tfrac{r+3s}{2} \mathbf{\check{c}} \right ) + \mathcal{A}_{\text{grav}} \, ,
	\end{aligned}
\end{equation}     
where now $\mathbf{\check{c}}$ corresponds to the generator of the cohomology group given by $c_1^2$ and we isolated the purely gravitational contribution
\begin{equation} \label{angravu1}
\mathcal{A}_{\text{grav}}= 245 \eta_0 - \, \eta_\text{gravitino} + \mathcal{Q}_+(0)-\mathcal{Q}_-(0) = 0 \, . 
\end{equation}
In eqs. \eqref{anu1diag} and \eqref{angravu1} the index of the $\eta$-invariants denotes $U(1)$ charge. On the Lens space $L_p^7$, one obtains
\begin{equation} \label{andiagu1}
	\begin{aligned}
		\mathcal{A}\big (L_p^7 \big ) &\equiv_1  \frac{p r^2+4s(p^2-2)+2r(2p^2+3ps+p-4)}{8 p} \, .
	\end{aligned}
\end{equation} 
As we have explained in the preceding sections, studying these anomalies for generic pairs of quadratic refinements is difficult. Thus, once again in \cref{andiagu1} we chose the one in \cref{quaddef} derived in \cite{Hsieh:2020jpj}.

\noindent We are now in the position to compute the anomaly on those backgrounds in which at least one of the Bianchi identities is satisfied. The only surviving models correspond to those in which the polynomial in \cref{andiagu1} is integer-valued on all allowed backgrounds. We performed a scan of the $123$ available models where $\abs{r} \, , \, \abs{s} \leq 500$ where the Bianchi classes are integral, as required by \cite{Seiberg:2011dr}. Using the quadratic refinement defined in \cite{Hsieh:2020jpj}, only $21$ of them turn out to be anomaly-free. These models are reported in \cref{u1familydiag}.
\begin{table}[ht!] \centering
	\begin{tabular}{ |p{1.8 cm}||p{0.5cm} | p{0.5cm}| p{0.5cm}|} 
		\hline
		$(r,s)$ & $d_0$ &  $d_1$ &  $d_2$
		\\
		\hline
$(24, -10)$ & 113 & 128 & 4 \\
$(26, -10)$ & 113 & 124 & 8 \\
$(26, -8)$ & 95 & 148 & 2 \\
$(28, -10)$ & 107 & 128 & 10 \\
$(30, -12)$ & 101 & 132 & 12 \\
$(30, -10)$ & 95 & 140 & 10 \\
$(32, -10)$ & 77 & 160 & 8 \\
$(34, -10)$ & 53 & 188 & 4 \\
$(36, -16)$ & 59 & 176 & 19 \\
$(36, -12)$ & 83 & 144 & 18 \\
$(38, -14)$ & 95 & 124 & 26 \\
$(42, -14)$ & 77 & 140 & 28 \\
$(46, -16)$ & 89 & 116 & 40 \\
$(48, -18)$ & 101 & 96 & 48 \\
$(54, -18)$ & 83 & 108 & 54 \\
$(62, -22)$ & 131 & 28 & 86 \\
$(62, -20)$ & 77 & 100 & 68 \\
$(70, -22)$ & 71 & 92 & 82 \\
$(86, -26)$ & 59 & 76 & 110 \\
$(94, -28)$ & 53 & 68 & 124 \\
$(126, -36)$ & 29 & 36 & 180 \\
		\hline
	\end{tabular}
	\caption{Anomaly-free families of degeneracies of $U(1)$ charges in terms of the integer parameters $r \, , \, s$ with a diagonal lattice embedding.}
	\label{u1familydiag}
\end{table}

\newpage

\subsubsection{Off-diagonal lattice}

The case in which the lattice bilinear form is off-diagonal induces the factorisation
\begin{eqaed}
	I_8&= \frac{1}{4} \left ( 2 \text{tr} R^2 + b^{(1)} F^2 \right )\left ( 2 \text{tr} R^2 + b^{(2)} F^2 \right )
 \\
 &=    \left ( \tfrac{1}{2} p_1 +\tfrac{b^{(1)}}{2} c_1^2 \right ) \left ( \tfrac{1}{2} p_1 + \tfrac{ b^{(2)}}{2} c_1^2 \right  )  \, ,
\end{eqaed} 
where now the previous conditions for $b^{(1,2)}$ are modified and read
\begin{equation}
	\begin{aligned}
		& b^{(1)} \, b^{(2)} = \frac{1}{6} \sum_{q=1}^2 q^4 d_q = h\, ,
		\\
		& b^{(1)} +	b^{(2)}= -\frac{1}{12} \sum_{q=1}^2 q^2 d_q =-r\, .
	\end{aligned}
\end{equation}
Solving this system of equations $b^{(1)}=s$ and $ b^{(2)}=-r-s$, with $r \, , \, s$ integer parameters lying on the surface $h=h(r,s)=-r s-s^2$. The corresponding degeneracies are
\begin{equation} \label{u1degoff}
	\begin{aligned}
		& d_0= 245-15 r + 3 h(r,s)/2 \, , \quad d_1= 16 r - 2 h(r,s) \, , \quad d_2= h(r,s)/2-r \,  .
	\end{aligned}
\end{equation}
As before, these solutions are restricted to the cases in which $d_q$'s are non-negative integers, and thus satisfy $ 2r \leq h \leq 8 r$ and $h> 164-10 r$. Given these degeneracies we can straightforwardly compute the anomaly theory on $L_p^7$, finding
\begin{equation} \label{anu1off}
	\begin{aligned}
		\mathcal{A}\big ( L_p^7 \big ) &=  (8 r -  h) \widetilde{\eta}_1 + \frac{h-2r}{4} \widetilde{\eta}_2 + \mathcal{A}_{\text{grav}} 
		\\
    & = - \, \frac{p-1}{12p} \, \left(4p^3+4p^2-6p(r-8)-3(s+4)(r+s-4)\right)
	\end{aligned}
\end{equation} 
where now the gravitational anomaly is dictated only by the fermionic piece
\begin{equation}
    \mathcal{A}_{\text{grav}}= 245 \eta_0 - \, \eta_\text{gravitino} \, . 
\end{equation}
Since the $2$-form field lifts in integer cohomology it is non-chiral, and thus its electric-magnetic anomaly \cite{Hsieh:2020jpj} on Lens spaces vanishes, as discussed in the previous sections. We can now evaluate this expression on the allowed $L_p^7$ backgrounds.

\noindent To begin with, we impose the condition in \cite{Seiberg:2011dr}. Bianchi classes are integral if and only if both $r$ and $s$ are even. Thus, we replace $r \to 2r$ and $s \to 2s$. The Bianchi classes are then $(2+s)y$ and $(2-r-s)y$. Choosing $s = lp-2$ for some integer $l$ trivialises the former, while $r = 2-s-lp$ trivialises the latter. In both cases, the anomaly simplifies to
\begin{eqaed}
    \mathcal{A}(L_p^7) \equiv_1 - \, \frac{(p-1) \, p \, (p+1)}{3} \equiv_1 0 \, .
\end{eqaed}
Hence, none of these off-diagonal models is anomalous.

\subsection{A family with no tensor multiplets}\label{sec:abelian_no_tensors} 

Different infinite families of models with no tensor multiplets, which satisfy the swampland criteria discussed in the literature but for which no string/F-theory realisation is known, have been found in \cite{Taylor:2018khc}. In this section we attempt to find out whether Dai-Freed anomalies can exclude some cases. It turns out that for a class of quadratic refinements the anomaly is eventually non-vanishing for large charges, thus excluding all but a finite number of these models. For other choices of quadratic refinements, only models with even charges survive. Unless one can choose the parameter $m$ \emph{ad hoc} for each background, or unless there is an underlying theoretical reason to do so, at least three quarters of these models are excluded. Generically, as we will discuss, all but a finite number are anomalous if $m \neq 6$ is bounded in the large charge limit.

\noindent Before analysing anomalies, let us provide some intuition for why large charges may problematic, beyond merely spoiling the desireable feature of finiteness. Recall that in the non-abelian case representations of unbounded dimension tend to violate unitarity constraints on (BPS) string probes \cite{Tarazi:2021duw}. In the abelian case, the presence of a large charge $Q$ in the spectrum of elementary fields does not increase the number of massless modes, but it does increase the effective gauge coupling $g_\text{eff} = gQ$, which in $6d$ has dimensions of length. One can immediately observe that perturbativity $gQ \ll \ell_\text{Pl}$ is in conflict with the (upper bound to the) quantum gravity cutoff $\Lambda_\text{QG}$ determined by the magnetic weak gravity conjecture \cite{Arkani-Hamed:2006emk, Harlow:2022ich},
\begin{eqaed}
    \Lambda_\text{QG} \lesssim g \, M_\text{Pl}^{2} \ll \frac{M_\text{Pl}}{Q} \, .
\end{eqaed}
Intuitively, this means that large charges are obstructed in weakly coupled gauge theories when quantum gravity is involved. We now study in more detail whether this potential obstruction exists and is detected by Dai-Freed anomalies in a family of EFTs found in \cite{Taylor:2018khc}.

\noindent Without tensor multiplets there is only one possible lattice embedding, dictated by $a=3$, and the factorised anomaly polynomial is
\begin{eqaed} \label{0tensorI8}
	I_8&=  \tfrac{1}{8} \left ( 3 \text{tr} R^2 + b F^2  \right )^2
 \\
 &= \tfrac{1}{2} \left ( \tfrac{3}{4} p_1 + \tfrac{b}{2} c_1^2  \right )^2 \, .
\end{eqaed}    
In \cite{Taylor:2018khc} a family of charges $q,r,q+r$ admitting such factorisation was found, with fixed multiplicities
\begin{equation}\label{eq:u1_no_tensor_fam}
	\begin{aligned}
		d_q= 54 \, , \qquad d_r= 54 \, , \qquad d_{q+r}= 54 \, .
	\end{aligned}
\end{equation}
Here $q,r$ are integers determining the anomaly coefficient $b=- 6( q^2 +r q + r^2)$.

\noindent The anomaly theory for these models is described by
\begin{equation} \label{an0T}
	\mathcal{A} \big (Y \big ) = 27 ( \widetilde{\eta}_q +  \widetilde{\eta}_r + \widetilde{\eta}_{q+r} ) - \frac{b(b+2m)}{8p} \, ,
\end{equation}
where we have considered the general form of the quadratic refinement in terms of an integer parameter $m$. Using the expressions contained in \cref{appendix:systematics}, we can compute the value of \cref{an0T} for Lens spaces $L_p^7$ with $p$ a divisor of $| 3-3( q^2 +r q + r^2)|$, which trivialises the Bianchi class. For general $m$ and $p$, one has
\begin{eqaed}\label{eq:anom_abelian_no_tensors}
    \mathcal{A}(L_p^7) \equiv_1 \frac{\beta (p^2+3m-18)}{2p} \, ,
\end{eqaed}
where $\beta \equiv -b/6 = q^2+q r+r^2$ is the only combination of charges that appears. This is also true for the Bianchi class, which is $3(1-\beta)y$. For $p=3$ one has 
\begin{eqaed}
    \mathcal{A}(L_3^7) \equiv_1 \frac{\beta (m-3)}{2} \, ,
\end{eqaed}
which vanishes either for even $\beta$ ($q$ and $r$ even) or $m = 3$ (mod $2$). For general $\beta$, $p=\beta-1$ remains a valid choice, and we now study this case for large charge. We restrict to choices of quadratic refinement such that $m$ which are bounded (mod $2p$), and show that all but finitely many theories are excluded with this assumption unless $m=6$. For this family of models the choice of \cite{Hsieh:2020jpj} amounts to $m = p^2-2$, which for $p$ even is bounded mod $2p$ for $p = \beta-1 \gg 1$. For $p$ odd, $m \equiv_{2p} p-2 = \beta - 3$ with $\beta$ even, so for this choice $m$ is not bounded and a separate analysis is needed.

\noindent To begin with, notice if $(q,r) \to \infty$ in $\mathbb{R}^2$ then $\beta \to +\infty.$ Indeed,
\begin{eqaed}
    2\beta - \abs{(q,r)}^2 = (q+r)^2 \geq 0 \; \Longrightarrow \; \beta \geq \frac{1}{2} \, \abs{(q,r)}^2 \to +\infty \, .
\end{eqaed}
For $m=6$ the anomaly always vanishes. Assuming that $m \neq 6$ and bounded (taken between $0$ and $2p-1$), letting $p = \beta-1$ one has
\begin{eqaed}
    \mathcal{A}(L_{\beta-1}^7) & \overset{\beta \gg 1}{\sim} \frac{\beta \, \left(\beta^2-\beta+3(m-6)\right) + 3(m-6)}{2\beta} + \frac{3(m-6)}{2\beta^2} \\
    & \equiv \frac{\beta \, N(\beta) + N_0}{2\beta} + \frac{N_0}{2\beta^2} \, ,
\end{eqaed}
where for our purposes we only need to use that $N(\beta) \in \mathbb{Z}[\beta]$ is a polynomial with integer coefficients and that $N_0 \in \mathbb{Z}$ is an integer. The first term cannot be integer for large enough $\beta$, since the numerator cannot be a multiple of $\beta$ for $\beta > N_0$. The second term is subleading for large $\beta$, and thus cannot compensate the fractional part since the series is asymptotic. Furthermore, having packaged the divergent part in the first term, all other terms resum to a finite result, namely the limit of \cref{eq:anom_abelian_no_tensors} (as a rational number rather than mod 1) with its divergent contribution subtracted. All in all, in order to cancel the anomaly on $L_{\beta-1}^7$ one necessarily needs $m=6$, whereas for $p=3$ only even charges survive unless $m=3$ mod 2. If $m$ does not depend on the choice of background, one can save at most even charges, if any.

\noindent In order to conclude our analysis for the choice of quadratic refinement of \cite{Hsieh:2020jpj}, for $\beta \equiv 2z$ even and $p=\beta-1$, the parameter $m = p^2-2 \equiv_{2p} p-2 = 2z-3$ is not bounded for $z \gg 1$. The anomaly simplifies to
\begin{eqaed}
    \mathcal{A}(L_{2z-1}^7) \equiv_1 \frac{2z^2+z-13}{2z-1} \equiv_1 - \, \frac{12}{2z-1} \, ,
\end{eqaed}
which once more shows that the anomaly is eventually non-vanishing.

\noindent The result just described is confirmed by a numerical scan. Indeed, fixing the quadratic refinement to the choice $m=p^2-2$, which corresponds to the definition in \cite{Hsieh:2020jpj}, only $6$ models out of $440$ are anomaly-free\footnote{There are additional models for which the analysis cannot be performed, since they would formally yield $p=0$ as valid backgrounds. These correspond to $(q,r)=\{ (-1,0), (0,-1), (1,0), (0,1), (1,-1), (-1,1) \}$.}. The charge is allowed to vary $|q|\leq 10$ and $|r|\leq 10$, but the non-anomalous models are bounded by $|q|=2$ and $|r|=2$, as listed in \cref{u1family0TY}.
 \begin{table}[ht!] \centering
 	\begin{tabular}{ |p{1.4 cm}|} 
 		\hline
 		$(r,s)$
 		\\
 		\hline
        $(-2, 1)$
        \\
        $(-1, -1)$
        \\
        $(-1, 2)$
        \\
        $(1, -2)$
        \\
        $(2, -1)$
        \\
        $(1,1)$
 		\\
 		\hline
 	\end{tabular}
 	\caption{Anomaly-free families of the $U(1)$ models in \cref{eq:u1_no_tensor_fam} with no tensor multiplets in terms of $q,r$ with $n_T=0$ and $\abs{q} \, , \, \abs{r} \leq 10$, choosing $m=p^2-2$.}
 	\label{u1family0TY}
 \end{table}
 
\noindent To summarise our findings, unless we were to find a mechanism indicating which specific quadratic refinements are selected in string/F-theory, we cannot place these theories in the swampland with certainty. A genericity argument would tend to exclude most of them, since only a very specific choice for each backround cancels all the anomalies we have investigated. It would be interesting to further explore this issue, trying to deduce a top-down criterion to select the allowed quadratic refinement(s). Nonetheless, we now know that unless $m = 6$, no choice with $m$ bounded can cancel the anomaly on $L_{\beta-1}^7$ for infinitely many charges, while the anomaly on $L_3^7$ only saves even charges unless $m=3$ mod 2.

\section{Non-supersymmetric heterotic models}\label{sec:non-susy_models}
	
In the previous Sections we have discussed the role played by Dai-Freed anomalies as a consistency swampland criterion for six-dimensional $\mathcal{N}=(1,0)$ supergravity theories with simply laced and abelian gauge groups. The reason why one could call this a swampland condition lies in the origin of Dai-Freed anomalies, which requires spacetime topology change. It is natural to think of this as an intrinsically quantum-gravitational effect, whereby cancellation of such anomalies may be unnecessary if the theory is not coupled to gravity. 

\noindent The analysis carried out so far is twofold: on the one hand, we have discussed the consistency of six-dimensional supergravity theories on the allowed Lens backgrounds, from which some anomalous EFTs have been discarded up to the choice of the quadratic refinement characterising the global dynamics of the chiral fields. In this sense, including this choice as part of the specification of the EFT, many of them are excluded. On the other hand, all the examples arising from the string landscape, namely perturbative heterotic or F-theory constructions, are devoid of such anomalies. This means that Dai-Freed anomaly cancellation is not an empty requirement, and it adds extra constraints to the ones considered in the literature up to now. It can provide further evidence for the non-perturbative consistency of string theory, constrain gravitational EFTs and provide an additional tool to investigate string universality. Among the examples that we have checked are models obtained through a $K3$ compactification in the orbifold limit of the heterotic $SO(32)$ \cite{Honecker:2006qz} and $E_8 \times E_8$ \cite{Walton, Honecker:2006qz} theories, which fit in the restricted setting of six-dimensional supersymmetric theories. This result is neither new nor surprising, since \cite{Tachikawa:2021mby} has shown that anomalies of this kind are always absent in heterotic constructions when supersymmetry is present. 

\noindent However, an analogous general result is not available for non-supersymmetric vacua\footnote{To our knowledge, neither for supersymmetric and non-supersymmetric orientifold vacua a similar result is available.} and thus there is no \emph{a priori} guarantee that Dai-Freed anomalies would cancel for these theories as well. As explained in \cref{sec:anomaly_intro}, the cancellation of Dai-freed anomalies is connected to the possibility of writing unambiguously the partition functions of chiral fields \cite{Witten:2019bou, Yonekura:2016wuc}, and makes no reference to supersymmetry. Hence, it is important to check their absence in non-supersymmetric settings. This program has been initiated in \cite{Basile:2023knk}, where ten-dimensional tachyon-free models have been shown to be free of Dai-Freed anomalies. This also implies that any smooth geometric compactification thereof is anomaly-free, but in principle lower-dimensional vacua may arise from different constructions. Furthermore, it is not \emph{a priori} obvious that singular (\emph{e.g.} orbifold) points in spaces of deformation parameters remain anomaly-free, since in the absence of supersymmetry these parameters are not \emph{bona fide} moduli of the theory. In light of these considerations, the purpose of this section is to verify that Dai-Freed anomalies on Lens spaces cancel for certain non-supersymmetric heterotic orbifolds in six dimensions. For concreteness, we shall discuss $SU(2)$ anomalies on Lens spaces for the $SO(16)\times SO(16)$ heterotic model\footnote{The global form of the gauge group is not in fact $SO(16) \times SO(16)$, but this subtlety will be immaterial for our considerations.} compactified on $K3$ in its orbifold limits. The specific expressions for the anomaly are slightly different, since now chiral spectra do not arrange themselves into supermultiplets, but the overall methodology is the same as in the rest of the paper.

\subsection{The \texorpdfstring{$SO(16) \times SO(16)$}{SO(16) x SO(16)} model}

For simplicity we shall study orbifolds \cite{Dixon:1985jw,Dixon:1986jc} whose point group $P$ is a discrete subgroup of the $SU(2)$ holonomy of $K3$, thus corresponding to $\mathbb{T}^4/\mathbb{Z}_N$ with $N=2,3,4,6$ used to exploit the supersymmetric heterotic landscape in \cite{Walton, Honecker:2006qz}. In the following, we shall also restrict to the standard embedding in which the gauge connection on one of the $SO(16)$ factors is identified with the spin connection, so that the orbifold action on the gauge worldsheet fermions is identical to those on spacetime. Since in the settings at stake supersymmetry is broken already at the string scale, these restrictions are not imposed by general principles. Rather, they are chosen for convenience and it is easy to relax these requirements to investigate other corners of the landscape. With this setup, the point group action on the worldsheet fermions is dictated by the vectorial shifts
\begin{equation} \label{vectshifts}
    v_{\text{st}}=\frac{1}{N} (1,-1) \, , \qquad	v_{\text{gauge}}= \frac{1}{N} (0^6,1,-1) \otimes (0^8) \, ,
\end{equation}        
for spacetime and gauge degrees of freedom respectively, acting on worldsheet fermions and bosons via
\begin{equation} \label{pointgroupaction}
	g \cdot \psi^i_R= e^{2 \pi i v_{\text{gauge}}^i} \psi^i_R \, , \quad g \cdot \psi^i_L = e^{2 \pi i v_{\text{st}}^i} \psi^i_L \, ,  \qquad \text{and} \qquad 	g \cdot z^i= e^{2 \pi i v_{\text{st}}^i} z^i \, , 
\end{equation}
where $\psi^i_{L,R}$ denotes the complex combinations of Majorana spinors determined by the complex structure on the $i$-th torus $\mathbb{T}^2$ parametrised by $z_i$. With this choice of shift vectors, the expression for the torus amplitude of the $SO(16)\times SO(16)$ vacuum compactified on such $K3$ orbifolds can be straightforwardly obtained from the partition function in ten dimensions \cite{Dixon:1986iz,Alvarez-Gaume:1986ghj}, and reads
\begin{equation} \label{torusso16xso16}
	\mathcal{T}= \frac{1}{N} \sum_{\alpha, \beta=0}^{N-1} \ d_{\alpha, \beta} \ \modZ{\alpha}{\beta} \Lambda_{\alpha,\beta} \, .
\end{equation}
Here we have packaged the contributions from the Hilbert space of oscillators for non-compact bosons and worldsheet fermions in the blocks $\modZ{\alpha}{\beta}$, and for the compact bosons in $\Lambda_{\alpha,\beta}$. The explicit expressions are presented in \cref{K3 orbifolds partition functions}.
The multiplicities $d_{\alpha,\beta}$ in \cref{torusso16xso16} reflect the number of fixed points under the action of the orbifold group, and are required to obtain a modular invariant result. Moreover, when the point group $\mathbb{Z}_N$ admits $\mathbb{Z}_k$ as a non-trivial subgroup, {\em i.e.} when $k$ is a non-trivial divisor of $N$, the multiplicities for the $N/k$-twisted sectors encode the number of fixed points for $\mathbb{Z}_k$ organised into  suitable multiplets of the orbifold group. For such $K3$ orbifolds this occurs for $N=4,6$, where fixed points for the subgroups $\mathbb{Z}_2$ and $\mathbb{Z}_{2,3}$ are organised into multiplets of $\mathbb{Z}_4$ and $\mathbb{Z}_6$ respectively. Instead, for $N=2,3$ such subtleties do not appear, and thus we shall begin discussing these constructions. 

\subsubsection{The \texorpdfstring{$\mathbb{Z}_2$}{Z2} orbifold }
 
 In light of the action of the point group on the torus in \cref{pointgroupaction}, the $N=2$ case dictates a simple sign flip on worldsheet fermions and bosons. As a consequence, the action on spacetime coordinates implies the presence of $16$ fixed points 
 \begin{equation}
 	\zeta_{ab}= (0,\tfrac{1}{2}, \tfrac{1}{\sqrt{2}}e^{\frac{ \pi i}{4}},\tfrac{1}{2} i ) \times (0,\tfrac{1}{2}, \tfrac{1}{\sqrt{2}}e^{\frac{ \pi i}{4}},\tfrac{1}{2}i)
 \end{equation}
where $a,b=1,\ldots, 4$ label the $\mathbb{Z}_2$ fixed points for each torus. The induced twisted multiplicities
 \begin{equation}
 	d_{\alpha, \beta}=  \begin{pmatrix}
 		1 & 1 \\ 16 & 16
 	\end{pmatrix}
 \end{equation}
are necessary to make the partition function modular invariant. Moreover, the orbifold action on the $SO(4)$ characters is diagonal,
 \begin{equation} \label{Z2 action}
 	\begin{aligned}
 		&O_4 \to O_4 \, , \qquad V_4 \to - V_4 \, ,
 		\\
 		&S_4 \to S_4 \, , \qquad C_4 \to - C_4 \, .
 	\end{aligned}
 \end{equation}
This allows us to write the contributions from internal directions in the modular blocks $\modZ{\alpha}{\beta}$ in terms of a combination of the conjugacy classes of $SO(4)$. As a result, the gauge group induced by the $\mathbb{T}^4/\mathbb{Z}_2$ compactification is obtained via the breaking
\begin{equation}
	SO(16) \times SO(16) \to SO(12) \times SU(2)^2 \times SO(16) \, .
\end{equation} 
We now have all the ingredients to write the twisted and untwisted sectors of the partition function describing this model. The untwisted unprojected sector reads
\begin{equation}
	\begin{aligned}
		\modZ{0}{0}= \frac{1}{\left ( \eta \bar \eta \right )^4} 
		& \bigg \{ \left (V_4 O_4 + O_4 V_4 \right ) \left [ \left ( \overline O_{12} \overline O_4 + \overline V_{12} \overline V_4 \right )  \overline O_{16} + \left ( \overline S_{12} \overline S_4 + \overline C_{12} \overline C_4 \right )  \overline S_{16} \right ]
		\\
		& + \left (O_4 O_4 + V_4 V_4 \right ) \left [ \left ( \overline O_{12} \overline V_4 + \overline V_{12} \overline O_4 \right )  \overline C_{16} + \left ( \overline S_{12} \overline C_4 + \overline C_{12} \overline S_4 \right )  \overline V_{16} \right ]
		\\
		&- \left (S_4 S_4 + C_4 C_4 \right ) \left [ \left ( \overline O_{12} \overline O_4 + \overline V_{12} \overline V_4 \right )  \overline S_{16} + \left ( \overline S_{12} \overline S_4 + \overline C_{12} \overline C_4 \right )  \overline O_{16} \right ]
		\\
		& - \left (S_4 C_4 + C_4 S_4 \right ) \left [ \left ( \overline O_{12} \overline V_4 + \overline V_{12} \overline O_4 \right )  \overline V_{16} + \left ( \overline S_{12} \overline C_4 + \overline C_{12} \overline S_4 \right )  \overline C_{16} \right ] \bigg \},
	\end{aligned}
\end{equation}
 where the compact bosons are described by the Narain lattice sum $\Lambda_{0,0}= \Lambda_{4,4}$. The untwisted projected contribution is encoded in 
\begin{equation}
	\begin{aligned}
		\modZ{0}{1}=  \frac{1}{\left ( \eta \bar \eta \right )^4} & \bigg \{ \left (V_4 O_4 - O_4 V_4 \right ) \left [ \left ( \overline O_{12} \overline O_4 - \overline V_{12} \overline V_4 \right )  \overline O_{16} + \left ( \overline S_{12} \overline S_4 - \overline C_{12} \overline C_4 \right )  \overline S_{16} \right ]
		\\
		& + \left (O_4 O_4 - V_4 V_4 \right ) \left [ \left (- \overline O_{12} \overline V_4 + \overline V_{12} \overline O_4 \right )  \overline C_{16} + \left (- \overline S_{12} \overline C_4 + \overline C_{12} \overline S_4 \right )  \overline V_{16} \right ]
		\\
		&- \left (S_4 S_4 - C_4 C_4 \right ) \left [ \left ( \overline O_{12} \overline O_4 - \overline V_{12} \overline V_4 \right )  \overline S_{16} + \left ( \overline S_{12} \overline S_4 - \overline C_{12} \overline C_4 \right )  \overline O_{16} \right ]
		\\
		& - \left (- S_4 C_4 + C_4 S_4 \right ) \left [ \left (- \overline O_{12} \overline V_4 + \overline V_{12} \overline O_4 \right )  \overline V_{16} + \left ( - \overline S_{12} \overline C_4 + \overline C_{12} \overline S_4 \right )  \overline C_{16} \right ] \bigg \} \, ,
	\end{aligned}
\end{equation}
whose compact bosons are described by
\begin{equation}
	\Lambda_{0,1}=  \prod_{i=1}^2 \frac{ (2 \sin{\pi v_i})^2 \eta \bar \eta }{ \Jtheta{1/2}{1/2+v_i}{0}{\tau}  \Jbartheta{1/2}{1/2-v_i}{0}{\overline \tau} } \, .
\end{equation}
As anticipated, modular invariance of the partition function introduces {\em twisted sectors} via the images under $S$ and $TS$ of $\modZ{0}{1}$ required to complete the modular orbit. Thus, the $g$-twisted unprojected piece reads
\begin{equation}
	\begin{aligned}
		&\modZ{1}{0}=  \frac{1}{\left ( \eta \bar \eta \right )^4}  \bigg \{ \left (O_4 S_4 + V_4 C_4 \right )  \left [ \left ( \overline O_{12} \overline C_4 + \overline V_{12} \overline S_4 \right )  \overline O_{16} + \left ( \overline C_{12} \overline O_4 + \overline S_{12} \overline V_4 \right )  \overline S_{16} \right ]
		\\
		& \qquad \quad  + \left (O_4 C_4 + V_4 S_4 \right ) \left [ \left ( \overline O_{12} \overline S_4 + \overline V_{12} \overline C_4 \right )  \overline C_{16} + \left ( \overline S_{12} \overline O_4 + \overline C_{12} \overline V_4 \right )  \overline V_{16} \right ]
		\\
		& \qquad \quad - \left (S_4 O_4 + C_4 V_4 \right )  \left [ \left ( \overline S_{12} \overline O_4 + \overline C_{12} \overline V_4 \right )  \overline C_{16} + \left ( \overline O_{12} \overline S_4 + \overline V_{12} \overline C_4 \right )  \overline V_{16} \right ]
		\\
		& \qquad \quad  - \left ( S_4 V_4 + C_4 O_4 \right )   \left [ \left ( \overline O_{12} \overline C_4 + \overline V_{12} \overline S_4 \right )  \overline S_{16} + \left ( \overline S_{12} \overline V_4 + \overline C_{12} \overline O_4 \right )  \overline O_{16} \right ]\bigg \} \, ,
		\\
		& \Lambda_{1,0}=\prod_{i=1}^2 \frac{  \eta \bar \eta }{ \Jtheta{1/2+v_i}{1/2}{0}{\tau}  \Jbartheta{1/2-v_i}{1/2}{0}{\overline \tau} } \, ,
		\end{aligned}
\end{equation}
while the $g$-twisted projected part is given by
\begin{equation}
	\begin{aligned}
 &\modZ{1}{1}=  \frac{1}{\left ( \eta \bar \eta \right )^4}  \bigg \{ \left (O_4 S_4 - V_4 C_4 \right )  \left [ \left (- \overline O_{12} \overline C_4 + \overline V_{12} \overline S_4 \right )  \overline O_{16} + \left ( \overline C_{12} \overline O_4 - \overline S_{12} \overline V_4 \right )  \overline S_{16} \right ]
 \\
 & \qquad \quad + \left (-O_4 C_4 + V_4 S_4 \right ) \left [ \left ( \overline O_{12} \overline S_4 - \overline V_{12} \overline C_4 \right )  \overline C_{16} + \left ( \overline S_{12} \overline O_4 - \overline C_{12} \overline V_4 \right )  \overline V_{16} \right ]
 \\
 & \qquad \quad  - \left (S_4 O_4 - C_4 V_4 \right )  \left [ \left ( \overline S_{12} \overline O_4 - \overline C_{12} \overline V_4 \right )  \overline C_{16} + \left ( \overline O_{12} \overline S_4 - \overline V_{12} \overline C_4 \right )  \overline V_{16} \right ]
 \\
 & \qquad \quad  - \left (- S_4 V_4 + C_4 O_4 \right )   \left [ \left (- \overline O_{12} \overline C_4 + \overline V_{12} \overline S_4 \right )  \overline S_{16} + \left ( -\overline S_{12} \overline V_4 + \overline C_{12} \overline O_4 \right )  \overline O_{16} \right ]\bigg \} \, , 
 \\
 & \Lambda_{1,1}= \prod_{i=1}^2 \frac{  \eta \bar \eta }{ \Jtheta{1/2+v_i}{1/2+v_i}{0}{\tau}  \Jbartheta{1/2-v_i}{1/2-v_i}{0}{\overline \tau} } \, .
\end{aligned}
\end{equation}
The particle content of this model for each mass level can be directly read from the $q$-expansion of these characters, once the level matching condition is imposed. In particular, the massless contribution to the spectrum comprises the graviton, the non-chiral Kalb-Ramond field and the dilaton in the singlet of the gauge group, the gauge boson in the adjoint of $SO(12) \times SU(2)^2 \times SO(16)$, four scalars in $(12,2,2,1) \oplus 4 (1,1,1,1)$, a doublet of left fermions in $(1,1,1,128_s) \oplus (32_s,2,1,1) \oplus (1,2,2,16)$ and a doublet of right fermions in $(32_c,2,2,1) \oplus (12,1,1,16)$ from the untwisted sector. The contribution from the twisted sector follows similarly and comprises thirty-two scalars in $(12,2,1,1)$, one hundred and twenty-eight scalars in $(1,1,2,1)$, eight doublets of right fermions in $(32_c,1,1,1)$ and eight doublets of left fermions in $(1,2,1,16)$. 

Given the massless spectrum, we can compute the anomaly theory associated to this vacuum. As discussed in \cref{sec:anomaly_intro}, the contribution from the fermions is encoded into the $\eta$ invariant of the Dirac operator with a sign depending on chirality of fermions in spacetime. This leads to the contribution
\begin{equation} \label{feranso16xso16Z2}
	\begin{aligned}
		\mathcal{A}_{\text{fermions}} \big ( Y \big )&= \eta^D_{(32_c,1,2,1)} \big ( Y \big )+ \eta^D_{(12,1,1,16)} \big ( Y \big ) + 8 \, \eta^D_{(32_c,1,1,1)} \big ( Y \big )
		\\
		&- \eta^D_{(1,1,1,128_s)} \big ( Y \big )- \eta^D_{(32_s,2,1,1)} \big ( Y \big )- \eta^D_{(1,2,2,16)} \big ( Y \big )- 8 \, \eta^D_{(1,2,1,16)} \big ( Y \big ).
	\end{aligned}
\end{equation}  
When $Y$ is a boundary of an eight dimensional manifold $Z$, \cref{feranso16xso16Z2} can be expressed via the APS index theorem \cite{Atiyah:1975jf} by
\begin{equation} \label{fermloc}
   \mathcal{A}_{\text{fermions}} \big ( \partial Z \big ) \equiv_1 - \int_Z I_8 \, ,
\end{equation}
where $I_8$ reproduces precisely the perturbative contribution
\begin{eqaed}
	I_8&= \frac{1}{2}\left ( 2 \ \text{tr} R^2  -  \tfrac{1}{2} \  \text{tr} F_1^2  -  \text{tr} F_2^2 - \text{tr} F_3^2 - \tfrac{1}{2} \ \text{tr} F_4^2   \right ) \left ( \text{tr} F_4^2 - 2 \  \text{tr} F_1^2  + 8 \  \text{tr} F_2^2  \right )\\
    & =  X_4^1 X_4^2 \, .
\end{eqaed}
The result suitably factorises in order for a generalised Green Schwarz mechanism \cite{Green:1984sg,Green:1984bx,Sagnotti:1992qw} to take place. In addition, the factorisation above tells us that the non-chiral B-fields lift to Cheeger-Simons character $\check{\mathbf{A}}^i=(  N^i, A^i, X_4^i )$, where $ [N^i]= [ X_4^i ]_{\mathbb{Z}}$  and $ [X_4^i]= [ X_4^i ]_{\text{dR}}$ \cite{Hsieh:2020jpj,Freed:2000ta,Freed:2006ya,Freed:2006yc}. As discussed in \cite{Hsieh:2020jpj}, the anomaly associated to the non-chiral field is then given by the cohomology pairing\footnote{As explained in \cref{sec:anomaly_intro} we could have chosen to switch the role to the factors coupled ``electrically'' and ``magnetically'', which are expected to yield equivalent results.}
\begin{equation} \label{bfieldsan}
	\mathcal{A}_{\text{B-fields}} \big ( Y \big )= ( \check{\mathbf{A}}^1, \check{\mathbf{A}}^2)_Y \, ,
\end{equation} 
thus providing the complete expression for the anomaly theory on a general background,
\begin{equation} \label{anomaly}
	\mathcal{A} \big (Y \big )= \mathcal{A}_{\text{fermions}} \big ( Y \big ) -  ( \check{\mathbf{A}}^1, \check{\mathbf{A}}^2)_Y \, .
\end{equation}
If the $i=1$ character is topologically trivial, {\em i.e.} $ [X_4^1]_{\mathbb{Z}} =0$, the coupling reduces to the known Chern-Simons coupling $A^1 \wedge X_4^2$. When $Y=\partial Z$, this contribution reproduces the Green-Schwarz term $X_4^1 \wedge X_4^2$. Requiring that the differential characters $\check{\mathbf{A}}^i$ be topologically trivial entails a twisted string structure on $Y$, since at the level of de Rham cohomology one obtains the Bianchi identities
\begin{equation} \label{bianchiz2}
	\begin{aligned}
		&d H_1= X_4^1= \tfrac{1}{2} p_1 + \tfrac{1}{2} c_2(F_1)+ c_2(F_2)  + c_2(F_3) + \tfrac{1}{2} c_2(F_4) \, ,
		\\
		&d H_2= X_4^2= 4 \,  c_2(F_1) - 16 \, c_2(F_2)  -2 \, c_2(F_4) \, .
	\end{aligned}
\end{equation}
One can choose to trivialise either class at the integral level, but of course the consistency of the model should not depend on this choice. The corresponding twisted string bordism groups ought to classify equivalent anomaly backgrounds.

\noindent We now study anomalies on Lens spaces, turning off all the groups beside the first $SU(2)$\footnote{Turning on only the second $SU(2)$ makes the second Bianchi identity trivial. However, this forbids the Lens space as a valid background for any $p$, as will be explained in the following.}. In this case $L_p^7$ trivialises $X_4^2$ for $p=2,4,8,16$, since the Pontryagin and Chern classes are $ \frac{1}{2} p_1= 2y$ and $c_2=-y$ with $y$ a generator of degree-four cohomology. On these backgrounds, the contribution from \cref{bfieldsan} vanishes, as in the off-diagonal supersymmetric models, and thus the anomaly is simply given in terms of the net number of fundamentals for the first $SU(2)$, $d_F=12 \cdot 16$ as in \cref{ansu2Lp7off}. As a result,
\begin{equation} \label{anlensz2}
	\begin{aligned}
		\mathcal{A} \big ( L_p^7 \big ) = 12 \cdot 16 \, \tilde \eta_1 + \mathcal{A}_{\text{grav}} = 12 \cdot 16 \,  \frac{p^2-1}{12 p} \equiv_1 0 \, .
	\end{aligned}
\end{equation}
It is worth noticing that the available backgrounds are exactly the only ones for which the anomaly vanishes, whereas it would have not been the case for any other value of $p$.
  
\subsubsection{The  \texorpdfstring{$\mathbb{Z}_3$}{Z3} orbifold}

Generalising the $\mathbb{Z}_2$ orbifold to higher-order abelian groups brings along a complex orbifold action on worldsheet fermions and bosons. This prevents a gauge group enhancement $U(1) \to SU(2)$ from taking place, as in the case of $\mathbb{Z}_2$. This means that the breaking of the gauge group is maximal and reflects a simple $K3$ compactification \cite{Walton}, 
\begin{equation} \label{gaugegroupzN}
	SO(16) \times SO(16) \to SO(12) \times SU(2) \times U(1) \times SO(16) \, .
\end{equation} 
Furthermore, the $SO(4)$ characters are no longer eigenvectors of the orbifold action. The corresponding partition function can only be expressed in terms of $SU(2)$ and $U(1)$ characters. However, there is no ``algorithmic'' procedure to determine the level of the affine $U(1)$ algebra, and thus the characters appearing in the partition function in \cref{torusso16xso16}. Concretely, the $q$-expansion of the modular blocks contains information about the $SU(2)$ representations at the massless level, leaving the $U(1)$ charges undetermined\footnote{Alternatively, one can look at the (tree-level) spectrum obtained compactifying the $SO(16) \times SO(16)$ model on $\mathbb{T}^4$, select states that are invariant under the orbifold action, and complete them with suitable twisted sectors. This allows to determine charges asides from an overall normalisation, but for our purpose these technicalities can be neglected.}. Since we are interested in $SU(2)$ anomalies, we simply turn off the $U(1)$ in \cref{gaugegroupzN}. At any rate, a more complete discussion would require computing the relevant twisted string bordism group.

\noindent The complex action of the point group also affects the complex structure of the torus, since gauging $\mathbb{Z}_N$ requires that the torus lattice be invariant. For the $\mathbb{Z}_3$ case, the complex structure modulus $U$ of the torus is fixed to $U=e^{\frac{2 \pi i}{3}}$.  The vectorial shifts in \cref{vectshifts} on the worldsheet bosons and fermions in \cref{pointgroupaction} determine nine fixed points,
\begin{equation}
	\zeta_{ab}= (0,\tfrac{1}{\sqrt{3}} e^{\frac{i \pi }{6}}, \tfrac{1}{\sqrt{3}} i ) \times (0,\tfrac{1}{\sqrt{3}} e^{\frac{i \pi }{6}}, \tfrac{1}{\sqrt{3}} i ) \,  ,
\end{equation}
where $a,b=1,2,3$ label the fixed points for each torus. This is reflected in the matrix of multiplicities
\begin{equation}
	d_{\alpha, \beta}=	\begin{pmatrix}
		1 & 1 & 1 \\
		9 & 9 & 9 \\
		9 & 9 & 9
	\end{pmatrix} .
\end{equation}  
With this data one can perform a $q$-expansion of the modular blocks in \cref{torusso16xso16}. The resulting massless spectrum containing the graviton, the non-chiral Kalb-Ramond field and the dilaton in the singlet of the gauge group. There are in addition twelve scalars in the $(1,1,1)$, four scalars in $(12,2,1)$, the gauge boson in the adjoint of $SO(12) \times SU(2) \times U(1) \times SO(16)$, a doublet of left fermions in $(32_s,2,1) \oplus (1,1,128_s) \oplus (1,2,16) $ and a doublet of right fermions in $(32_c,1,1) \oplus (12,1,16)$. The twisted sectors comprise instead two hundred and fifty-two scalars in $(1,1,1)$, thirty-six scalars in $(12,2,1)$, nine doublets of left fermions in $(1,2,16)$ and nine doublets of right fermions in $(32_c,1,1)$. 

\noindent From the massless spectrum one can straightforwardly compute the anomaly theory. The fermionic contribution is
\begin{equation} \label{feranso16xso16Z3}
	\begin{aligned}
		\mathcal{A}_{\text{fermions}} \big ( Y \big ) & = \eta^D_{(32_c,1,1)} \big ( Y \big )+ \eta^D_{(12,1,16)} \big ( Y \big ) + 9 \, \eta^D_{(32_c,1,1)} \big ( Y \big )
		\\
		&- \eta^D_{(1,1,128_s)} \big ( Y \big )- \eta^D_{(32_s,2,1)} \big ( Y \big )- \eta^D_{(1,2,16)} \big ( Y \big )- 9 \, \eta^D_{(1,2,16)} \big ( Y \big ) \, .
	\end{aligned}
\end{equation}  
The corresponding anomaly polynomial factorises as
\begin{equation} \label{polso16xso16z3}
	I_8= \frac{1}{2}\left ( 2 \ \text{tr} R^2  -  \tfrac{1}{2} \  \text{tr} F_1^2  -  \text{tr} F_2^2 - \tfrac{1}{2} \ \text{tr} F_3^2   \right ) \left ( \text{tr} F_3^2 - 2 \  \text{tr} F_1^2  + 8 \  \text{tr} F_2^2  \right ) \, ,
\end{equation}
as required by the Green-Schwarz mechanism. The corresponding Bianchi classes are thus
\begin{equation} \label{z3pol}
	\begin{aligned}
		& X_4^1= \tfrac{1}{2} p_1 + \tfrac{1}{2} c_2(F_1)+ c_2(F_2) + \tfrac{1}{2} c_2(F_3) \, ,
		\\
		&X_4^2= 4 \,  c_2(F_1) - 16 \, c_2(F_2)  -2 \, c_2(F_3) \, .
	\end{aligned}
\end{equation}

\noindent As before, turning off all the groups aside from $SU(2)$, the Lens spaces $L_p^7$ are valid twisted string backgrounds trivialising the second Bianchi class when $p=2,4,8,16$. The anomaly theory that depends only on the number of fundamentals $d_F=12 \cdot 16$, thus yielding the same result as \cref{anlensz2} and the anomaly vanishes.

\subsubsection{The \texorpdfstring{$\mathbb{Z}_4$}{Z4} orbifold}

As in the $\mathbb{Z}_3$ case, the orbifold action is complex and induces the maximal breaking of the $SO(16) \times SO(16)$ gauge group in \cref{gaugegroupzN}. In addition, as before, one introduces the $\widehat{su(2)}_1$ and $\widehat{u(1)}_k$ characters whose level cannot be determined directly from the partition function. Gauging $\mathbb{Z}_4$ fixes the modulus $U=i$, and determines the fixed points
\begin{equation}
	\zeta_{ab}= (0, \tfrac{1}{\sqrt{2}} e^{ \frac{i\pi}{4}}) \times (0, \tfrac{1}{\sqrt{2}} e^{ \frac{i\pi}{4}}) \, ,
\end{equation} 
with $a,b=1,2$ labelling the fixed points on each torus. Although the situation looks similar to the $\mathbb{Z}_3$ case, the main novelty resides in the non-trivial subgroup $\mathbb{Z}_2$ whose exclusive fixed points are interchanged under $\mathbb{Z}_4$ transformations. This implies that for each torus $\zeta_a=0,\frac{1}{\sqrt{2}} e^{ \frac{i\pi}{4}}$ are $\mathbb{Z}_4$ fixed points but $\tilde \zeta_a=\frac{1}{2},\frac{i}{2}$ form a doublet under the $\mathbb{Z}_4$ action. The structure of the matrix of the degeneracies is then given by
\begin{equation}
	d_{\alpha, \beta}=	\begin{pmatrix}
		1 & 1 & 1 & 1\\
		4 & 4 & 4 & 4 \\
		16 & 4 & 16 & 4 \\
		4 & 4 & 4 & 4
	\end{pmatrix} \, ,
\end{equation} 
where as explained the $16$ appearing in the $g^2$ should be interpreted as $4+ 6 \cdot 2$, in which the $4$ counts the number of $\mathbb{Z}_4$ fixed points and the $6 \cdot 2$ encodes six doublets of $\mathbb{Z}_2$ ones. Taking these considerations into account, we can read the massless spectrum from the $q$-expansion and the level matching condition. It comprises the graviton, the non-chiral Kalb-Ramond field and the dilaton in the singlet of the gauge group, eight scalars in $(1,1,1)$, four scalars in $(12,2,1)$, the gauge boson in the adjoint of $SO(12) \times SU(2) \times U(1) \times SO(16)$, a doublet of left fermions in $(32_s,2,1) \oplus (1,1,128_s) \oplus (1,2,16) $ and a doublet of right fermions in $(32_c,1,1) \oplus (12,1,16)$ from the untwisted sector. From the $\mathbb{Z}_4$ twisted sectors we have instead a hundred and sixty scalars in $(1,1,1)$, thirty-two scalars in $(12,2,1)$, six doublets of left fermions in $(1,2,16)$ and six doublets of right fermions in $(32_c,1,1)$. Moreover, from the $\mathbb{Z}_2$ twisted sector we have ninety-six scalars in $(1,1,1)$, twelve scalars in $(12,2,1)$, three doublets of left fermions in $(1,2,16)$ and three doublets of right fermions in $(32_c,1,1)$  organised into $\mathbb{Z}_4$ doublets.

\noindent All in all, one thus finds the fermion anomaly theory
\begin{equation} \label{feranso16xso16Z4}
	\begin{aligned}
		\mathcal{A}_{\text{fermions}} \big ( Y \big ) &= \eta^D_{(32_c,1,1)} \big ( Y \big )+ \eta^D_{(12,1,16)} \big ( Y \big ) + 9 \, \eta^D_{(32_c,1,1)} \big ( Y \big )
		\\
		&- \eta^D_{(1,1,128_s)} \big ( Y \big )- \eta^D_{(32_s,2,1)} \big ( Y \big )- \eta^D_{(1,2,16)} \big ( Y \big )- 9 \, \eta^D_{(1,2,16)} \big ( Y \big ) \, ,
	\end{aligned}
\end{equation}     
reproducing exactly the contribution of the preceding case\footnote{The difference lies in the $U(1)$ charges, which we are not considering.}. The anomaly polynomial is that of \cref{polso16xso16z3} and the Bianchi classes those of \cref{z3pol}. The anomaly on $L_p^7$ thus vanishes on account of the preceding computation.

\subsubsection{The \texorpdfstring{$\mathbb{Z}_6$}{Z6} orbifold}

The $\mathbb{Z}_6$ orbifold exhausts the classification but does not present any further conceptual novelties with respect to the $\mathbb{Z}_4$ case. Indeed, also in this case the action of the point group is complex, entailing the maximal breaking of \cref{gaugegroupzN} as well as the impossibility to express the partition function in terms of $SO(4)$ characters. Once more the complex structure fixes $U=e^{\frac{2 \pi i}{6}}$. Now there is only one fixed point invariant under $\mathbb{Z}_6$, namely $\zeta=(0,0)$. However, the presence of the subgroups $\mathbb{Z}_2$ and $\mathbb{Z}_3$ induces corresponding fixed points organised in triplets and doublets under the $\mathbb{Z}_6$ action. This is reflected by the degeneracy matrix 
	\begin{equation}
	d_{\alpha, \beta}=	\begin{pmatrix}
		1 & 1 & 1 & 1 & 1 & 1 \\
		1 & 1 & 1 & 1 & 1 & 1 \\
		9 & 1 & 9 & 1 & 9 & 1 \\
		16 & 1 & 1 & 16 & 1 & 1 \\
		9 & 1 & 9 & 1 & 9 & 1 \\
		1 & 1 & 1 & 1 & 1 & 1
	\end{pmatrix} \, ,
\end{equation}
where now the $9$ $\mathbb{Z}_3$ fixed points should be interpreted as $1+ 4 \cdot 2$ and the $16$ $\mathbb{Z}_2$ fixed points as interpreted as $1+5 \cdot 3$. With these subtleties addressed, we have all the ingredients to write the massless spectrum. One finds that it comprises the graviton, the non-chiral Kalb-Ramond field and the dilaton in the singlet of the gauge group, eight scalars in $(1,1,1)$, four scalars in $(12,2,1)$, the gauge boson in the adjoint of $SO(12) \times SU(2) \times U(1) \times SO(16)$, a doublet of left fermions in $(32_s,2,1) \oplus (1,1,128_s) \oplus (1,2,16) $ and a doublet of right fermions in $(32_c,1,1) \oplus (12,1,16)$. From the twisted sectors we have instead two hundred and sixteen scalars in $(1,1,1)$, thirty-six scalars in $(12,2,1)$, nine doublets of left fermions in $(1,2,16)$ and nine doublets of right fermions in $(32_c,1,1)$.

\noindent As a result, the fermion anomaly theory reads
\begin{equation} \label{feranso16xso16Z6}
	\begin{aligned}
		\mathcal{A}_{\text{fermions}} \big ( Y \big )&= \eta^D_{(32_c,1,1)} \big ( Y \big )+ \eta^D_{(12,1,16)} \big ( Y \big ) + 9 \, \eta^D_{(32_c,1,1)} \big ( Y \big )
		\\
		&- \eta^D_{(1,1,128_s)} \big ( Y \big )- \eta^D_{(32_s,2,1)} \big ( Y \big )- \eta^D_{(1,2,16)} \big ( Y \big )- 9 \, \eta^D_{(1,2,16)} \big ( Y \big ) \, ,
	\end{aligned}
\end{equation}     
which precisely matches that of the $\mathbb{Z}_3$ and $\mathbb{Z}_4$ cases, since the $U(1)$ gauge group is turned off. The Bianchi classes are those of \cref{z3pol}, and thus the only available Lens backgrounds have $p=2,4,8,16$. Once again, for all these the anomaly vanishes, since it is given by \cref{anlensz2}.

\section{The Gepner orientifold with no tensor multiplets}\label{sec:gepner}

Gepner \cite{Gepner:1987qi} has shown that consistent superstring worldsheet theories in a background with $SU(n)$ holonomy can be realised in terms of tensor products of $\mathcal{N}=2$ superconformal minimal models. These models correspond to special points of the moduli space of a compactification of the type IIB superstring on Calabi-Yau manifolds. The chiral closed-string spectrum comprises the gravity multiplet and twenty-one tensor multiplets with $\mathcal{N}=(2,0)$ spacetime supersymmetry. The number of supercharges can be halved to $\mathcal{N}=(1,0)$ within the open descendants \cite{Angelantonj:1996mw}, whose spectra depends on the point of the moduli space in which the model lies. In the following, we will focus on the orientifold projection of the Gepner model with $81$ characters, whose (unoriented) closed-string spectrum comprises ten hypermultiplets in the antisymmetric representation of the Chan-Paton gauge group $SO(8)$ and twenty-one uncharged ones. Our interest in this model is motivated by the absence of tensor multiplets, which highlights the single chiral $2$-form contained in the gravity multiplet.

\noindent The full anomaly theory for a generic closed manifold $Y$ contains a single quadratic refinement, and takes the form
\begin{equation} \label{angepnerY}
		\mathcal{A} \big ( Y \big )= (10-1) \, \eta_{\mathbf{28}} (Y) + 21 \, \eta_{\mathbf{0}} (Y) - \eta_{\text{grav}} (Y) + \mathcal{Q}(Y) \, .
\end{equation}
On boundaries $Y=\partial Z$, \eqref{angepnerY} reproduces the anomaly polynomial
	\begin{equation} \label{angepner}
		\begin{aligned}
			\mathcal{A} \big ( \partial Z \big ) &\equiv_1 - \int_Z I_8 = - \frac{9}{8} \int_Z \left ( \text{tr} R^2 - \text{tr} F^2\right )^2 \, ,
			\end{aligned}
	\end{equation}
from which one reads off the Bianchi identity
\begin{equation} \label{bianchigepner}
		d H= 3 \left ( \tfrac{1}{4} p_1 + c_2 \right ) = X_4 \, .
	\end{equation}
The Green-Schwarz term cancelling \cref{angepner} arises from the quadratic refinement
\begin{equation} \label{sepquadgepner}
    \mathcal{Q}_{Y} ( 3 \mathbf{\check{c}})  = \widetilde{ \mathcal{Q}}_{Y} ( 3 \mathbf{\check{c}}) +  \mathcal{Q}_{Y} ( 0) \, .
\end{equation}
As discussed above \cref{eq:quadratic_refinement_0}, we write the gravitational contribution
\begin{equation}
    \mathcal{Q}_{Y} ( 0)= \int_Z \left \{ \frac{1}{2} \left ( \frac{a}{4} p_1  \right )^2-\, \frac{1}{8} L \right \} = - \, \frac{7(35 a^2-3) }{8} \, \eta_D(Y) +\frac{(a^2-1) }{8} \, \eta_{\text{grav}}(Y) \, ,
\end{equation}
where $a=3$ in the present case. Thus the only unknown contribution is $\widetilde{\mathcal{Q}}_Y( 3 \mathbf{\check{c}})$, which solves \cref{eq:quadrefcharacteristiceq} and can be parametrised as in \cref{eq:tildeQ_lens} on Lens spaces. The Bianchi identity in \cref{bianchigepner} singles out Lens spaces $L_p^7$ with  $p=3$, $p = 1 - 2 k$ or $p = 3(1 - 2k)$, where $k= 1 \, , \, 2$ parametrises the bundles in \cref{appendix:systematics}. Thus, $p=3 \, , 9$ are allowed depending on the value of $k$. In these cases, the anomaly in \cref{angepnerY} evaluates to 
\begin{equation}\label{eq:an_gepner_lens}
\begin{aligned}
    \mathcal{A} (L_p^7) &= 9 (k (2 k - 1) \Tilde{\eta}_2 + 4 k (4 - 2 k)\Tilde{\eta}_1) + 28 \cdot 9 \eta_0 + 21 \eta_0- 7 \cdot 39 \eta_0 + \widetilde{\mathcal{Q}}_Y( 3 \mathbf{\check{c}}) 
    \\
    &= 9 \left(k (2 k - 1) \frac{(p-1)(p-2)}{3p}+ 4 k (4 - 2 k)\frac{p^2-1}{12p}\right) 
    \\
    &- (28 \cdot 9  + 21 - 7 \cdot 39) \, \frac{(p^2-1)(p^2+11)}{720p}- \frac{-6 k(-6k+m)}{2p} \\
    & = \frac{3k}{p} \left( m + 3 (p^2 - 2kp + p - 2) \right) \equiv_1 \frac{3k}{p} \, (m - 6) \, ,
\end{aligned}
\end{equation}
where the quadratic refinement is parametrised as in \cref{eq:tildeQ_lens},
\begin{equation}\label{eq:tildeQ_gepner}\widetilde{\mathcal{Q}}_Y( 3 \mathbf{\check{c}})= - \frac{-6 k(-6k+m)}{2p} \, , \qquad \text{with} \ \ m=0,\ldots, 2p-1 \ \ \text{mod} \ \ 2p \, .
\end{equation}
Choosing $p=3$, valid for all $k$, manifestly trivialises \cref{eq:an_gepner_lens}, while for $k=2$ and $p=9$ one finds
\begin{equation}
   \mathcal{A} (L_9^7) \equiv _1 \frac{2}{3} \, m \, ,
\end{equation}
which cancels only for $m=0$ mod 3. This is not the value of $m$ one finds for the choice in \cite{Hsieh:2020jpj}, which is $m=2$. All in all, we have shown that the anomaly vanishes in this model for a specific choice of quadratic refinement for all the allowed $k$ and $p$, although the rationale behind this choice remains obscure.
 
\section{Conclusions}\label{sec:conclusions}

In this paper we have extensively studied (a subset of) Dai-Freed anomalies in some corners of the six-dimensional supergravity landscape. Unless a specific quadratic refinement describing the anomaly theory of chiral fields is chosen to cancel the anomaly in each allowed background, we have ruled out some models with abelian and simply laced gauge groups. The analysis is particularly effective in the case of large abelian charges, where the known results on finiteness, bounds on ranks and so on do not apply. From the point of view of the swampland program, this means that consistent global specifications of EFTs coupled to quantum gravity are ``highly unnatural'', in the sense that randomly choosing a quadratic refinement likely leads to anomalies. On the other hand, the same class of models is anomaly-free if the $2$-form fields uplifted in differential cohomology are non-chiral, and therefore no quadratic refinement enters the picture. Furthermore, we have shown that the same result holds non-trivially for all string theory constructions that we have tested, including non-supersymmetric and non-geometric ones. This outcome is yet another testament to how all the (necessary) ingredients in the theory interweave in precisely such a way as to guarantee consistency. The rigidity and interconnectedness of the framework mean that it could have been easily compromised by any inconsistency. As we have discussed, we have not found any anomalies on Lens spaces in the string constructions that we have studied, but it is worth noting that, were such an inconsistency to be found, they may be remedied by a discrete form of the Green-Schwarz mechanism \cite{Debray:2021vob, Dierigl:2022zll}. In order to fully establish the absence of all possible anomalies, one would have to compute the complete bordism groups associated to the twisted string structure, an interesting mathematical task in its own right which may unveil new physical insights. This type of analysis can also be extended to theories with more tensor multiplets. Understanding the details of the twisted string structure, for instance whether the bordism groups depend on the choice of Bianchi identity to trivialise, is an additional interesting challenge.

\noindent From a bottom-up perspective, the implications of unitarity for the consistency of EFTs of gravity have already driven remarkable progress in understanding the string landscape and uncovering physical explanations of its observed patterns. The main point that we have emphasised in this work is that the full extent of the consistency conditions arising from anomaly cancellation has not yet been harnessed. Combining it with supersymmetry may eventually lead to a completion of the ``swampland task'' in six dimensions. Even if this ambitious goal were to be achieved, it seems doubtful that such results could be straightforwardly extended to settings in lower dimensions and/or lower amounts of supersymmetry. In the former case, the absence of purely gravitational anomalies significantly decreases the constraining power of these considerations, and furthermore bordism groups tend to be less rich (if not altogether trivial). In the latter case, the field content need not be arranged in a rigid handful of supermultiplets with tightly constrained interactions, and in particular there is no \emph{a priori} connection between chiral and non-chiral degrees of freedom, leading to a much larger pool of \emph{a priori} allowable models. Together with the technical difficulties in handling multiple potentially independent quadratic refinements, one can foresee how this strategy may not be as fruitful when multiple genuinely chiral form fields are present.

\noindent At present, it is unclear how to proceed to make comparable progress in these more complicated settings. A natural development of the ideas that we have presented is investigating anomaly inflow on defects at the Dai-Freed level. Indeed, even in a bottom-up approach, exploiting the completeness principle \cite{Polchinski:2003bq, Banks:2010zn, Heidenreich:2020pkc, Heidenreich:2021xpr} arising from cobordism triviality \cite{McNamara:2019rup, McNamara:2021cuo} and holography \cite{Harlow:2018tng, McNamaraThesis} can predict the existence of novel non-perturbative defects in the theory (see \emph{e.g.} \cite{Debray:2021vob, Debray:2023yrs} for a detailed analysis in type IIB supergravity). In turn, the consistency of anomaly inflow on these defects can restrict the theory even with lower amounts of supersymmetry, as exemplified in \cite{Martucci:2022krl}. Perhaps one can achieve more mileage complementing such kinematical considerations with dynamical ones, for instance relying on S-matrix bootstrap techniques along the lines of \cite{Guerrieri:2021ivu, Guerrieri:2022sod}. Another promising avenue is the study of equivariant topological modular forms \cite{douglas2014topological, hopkins, Gukov:2018iiq}, a generalised cohomology theory believed to encode deformation classes of $2d$ SQFTs (and thus of heterotic worldsheets) \cite{Gaiotto:2019asa} according to the Stolz-Teichner conjecture \cite{Stolz:2011zj}. As we have already mentioned, this approach was fruitfully employed in \cite{Tachikawa:2021mvw, Tachikawa:2021mby} to exclude all anomalies in supersymmetric heterotic strings, but an equivariant version \cite{Chua_2022, gepner2023equivariant} could apply to non-supersymmetric settings (where the GSO projection is implemented with an additional $\mathbb{Z}_2$ gauging) and to refined invariants taking into account gauge charges. This approach could exclude further models from the (perturbative heterotic) landscape. The road ahead looks daunting, but the remarkable and rapid progress witnessed by the last few years of research are certainly encouraging, and motivate further efforts in the pursuit of a global understanding of the landscape.

\section*{Acknowledgements}\label{sec:acknowledgements}

We are grateful to Arun Debray, Matilda Delgado, Markus Dierigl, Cameron Krulewski, Miguel Montero, Paul-Konstantin Oehlmann, Natalia Pacheco-Tallaj and Michelangelo Tartaglia for insightful discussions and/or collaboration on related projects. We thank Carlo Angelantonj for helpful suggestions and insights on string constructions. I.B. thanks Monica Jinwoo Kang and Ethan Torres for discussions. G.L. is grateful to the Max Planck Institute for Physics, the Ecole Polytechnique of Paris and CERN for the hospitality while this project was underway.

\appendix

\section{Some formulae and conventions for anomalies}\label{appendix:anomaly_formulae}

Throughout the paper, we have used somewhat interchangeably Chern and Pontryagin classes and their representation in terms of invariant polynomials in the curvature of the connection of the corresponding bundle via the Chern-Weil isomorphism. This appendix is meant to explicitly spell out our conventions, in order to properly connect the two descriptions. We follow the conventions of \cite{Angelantonj:2020pyr}, where the curvatures are normalised with respect to \cite{Bilal:2008qx} as
\begin{eqaed}
    F \to \frac{F}{2 \pi} \qquad \text{and} \qquad R \to \frac{R}{4 \pi} \, .
\end{eqaed}
Concretely, this means that the APS index theorem for $\eta$-invariants of the Dirac operator, corresponding to a Weyl fermion in representation $\mathbf{R}$, is
\begin{equation}
    \text{Index}_D= \eta^D_{\mathbf{R}} ( \partial Z ) + \int_Z \hat{A}(R) \, \text{ch}_{\mathbf{R}}(F)  \, ,
\end{equation}
where the $A$-roof genus takes the form
\begin{eqaed}
    \hat{A}(R)&= 1 + \frac{1}{12} \text{tr} R^2 + \left [ \frac{1}{360} \text{tr} R^4 + \frac{1}{288} \left ( \text{tr} R^2 \right )^2  \right ] 
    \\
    & \qquad \qquad  + \left [ \frac{1}{5670} \text{tr} R^6 + \frac{1}{4320} \text{tr} R^4 \, \text{tr} R^2 + \frac{1}{10368} \left ( \text{tr} R^2 \right )^3  \right ] + \ldots
    \\
    &\equiv \hat{A}_0 + \hat{A}_1 + \hat{A}_2 + \hat{A}_3 \ldots
\end{eqaed}
Similarly, the Chern character
\begin{eqaed}
    \text{ch}_{\mathbf{R}}(F)= \text{tr}_{\mathbf{R}} e^{i F} &= \text{dim}\mathbf{R} + i \, \text{tr}_{\mathbf{R}} F -\frac{1}{2} \text{tr}_{\mathbf{R}} F^2 + \ldots
    \\
    &= \text{ch}_{\mathbf{R},0}(F) + \text{ch}_{\mathbf{R},1}(F) + \text{ch}_{\mathbf{R},2}(F) +  \ldots
\end{eqaed}
When dealing with gravitini, we use the APS index theorem for the $\eta$-invariant of the Rarita-Schwinger operator, which reads
\begin{equation}
    \text{Index}_{RS} = \eta^{RS}( \partial Z ) + \int_Z \hat{A}(R) \left ( \text{tr} \, e^{ 2 i R} - 1 \right )\, ,
\end{equation}
where now the integrand can be expanded as
\begin{eqaed}
    \hat{A}(R) \left ( \text{tr} \, e^{ 2 i R} - 1 \right ) & = (d-1) + \frac{d-25}{12} \text{tr} R^2 + \left [ \frac{d+239}{360} \text{tr} R^4 + \frac{d-49}{288} \left ( \text{tr} R^2 \right )^2  \right ] 
    \\
    & + \left [ \frac{d-505}{5670} \text{tr} R^6 + \frac{d+215}{4320} \text{tr} R^4 \, \text{tr} R^2 + \frac{d-73}{10368} \left ( \text{tr} R^2 \right )^3  \right ] + \ldots
\end{eqaed}
Finally, similarly to the previous cases, the APS theorem expresses the signature of the manifold in terms of the Hirzebruch genus and the $\eta$-invariant of the operator in \cref{eq:signature_operator} as
\begin{equation}
    \frac{\sigma(Z)}{8}= \frac{1}{8} 2 \eta(\Tilde{D}_{\partial Z}^{\text{Sig}}) + \int_Z \frac{1}{8} L \, ,
\end{equation}
where the Hirzebruch genus is given by
\begin{eqaed}
    L(R)&= 1- \frac{2}{3} \text{tr} R^2 + 16 \left [ -\frac{7}{180} \text{tr} R^4 + \frac{1}{72} \left ( \text{tr} R^2 \right )^2 \right ] + 
    \\
    & + 64 \left [ -\frac{31}{2835} \text{tr} R^6 + \frac{7}{1080} \text{tr} R^4 \, \text{tr} R^2 - \frac{1}{1296} \left ( \text{tr} R^2 \right )^3  \right ] + \ldots
\end{eqaed}
It is often convenient to express the invariant polynomials in terms of characteristic classes, namely the total Chern class
\begin{eqaed}
    \sum_i c_{\mathbf{R},i} t^i &= \text{det} \left ( e^{i F t} + 1 \right )
    \\
    &= 1+ i \, \text{tr}_{\mathbf{R}} F \, t + \frac{\text{tr}_{\mathbf{R}} F^2 - \left (\text{tr}_{\mathbf{R}} F \right )^2 }{2} t^2 
    \\
    & \qquad -i  \frac{   2\text{tr}_{\mathbf{R}} F^3 - 3 \left ( \text{tr}_{\mathbf{R}} F^2 \right ) \left ( \text{tr}_{\mathbf{R}} F \right ) + \left ( \text{tr}_{\mathbf{R}} F \right )^3}{6} t^3 + \ldots
\end{eqaed}
and the total Pontryagin class 
\begin{eqaed}
    \sum_i p_i t^i &= \text{det} \left ( e^{2 i R} + 1 \right ) 
    \\
    &= 1- 2 \, \text{tr} R^2 \, t + 2 \left [ \left (  \text{tr} R^2 \right )^2 - \text{tr} R^4 \right ]  t^2 
    \\
    & \qquad - \frac{4}{3} \left [ \left (  \text{tr} R^2 \right )^3 - 6 \left ( \text{tr} R^4 \right ) \left ( \text{tr} R^2 \right ) + 8 \left ( \text{tr} R^6 \right )\right ] + \ldots
\end{eqaed}
Therefore, in our conventions 
\begin{eqaed}
    &p_1=- 2 \, \text{tr} R^2 \, ,
    \\
    &p_2= 2 \left [ \left (  \text{tr} R^2 \right )^2 - \text{tr} R^4 \right ] \, ,
    \\
    &p_3= \frac{4}{3} \left [ \left (  \text{tr} R^2 \right )^3 - 6 \left ( \text{tr} R^4 \right ) \left ( \text{tr} R^2 \right ) + 8 \left ( \text{tr} R^6 \right )\right ] \, ,
\end{eqaed}
so that the $A$-roof genus can be recast as
\begin{eqaed}
    &\hat{A}_1= - \frac{p_1}{24} \, ,
    \\
    & \hat{A}_2= \frac{7 p_1^2- 4 p_2}{5760} \, ,
    \\
    & \hat{A}_3= \frac{-31 p_1^3+ 44 p_1^2 p_2- 16 p_3}{967680} \, .
\end{eqaed}
Similarly, the Hirzebruch genus is 
\begin{eqaed}
    &L_1=  \frac{p_1}{3} \, ,
    \\
    & L_2= \frac{7 p_2-  p_1^2}{45} \, ,
    \\
    & L_3= \frac{2 p_1^3-13 p_1^2 p_2+ 62 p_3}{945} \, .
\end{eqaed}
It is also convenient to write the Chern classes in terms of the components of Chern characters as
\begin{eqaed}
    &c_{\mathbf{R},1}= \text{ch}_{\mathbf{R},1}( F) \, ,
    \\
    &c_{\mathbf{R},2}= \tfrac{1}{2}\text{ch}_{\mathbf{R},1}( F)^2 - \text{ch}_{\mathbf{R},2}( F) \, ,
    \\
    &c_{\mathbf{R},3}= 2 \text{ch}_{\mathbf{R},3}( F) -\text{ch}_{\mathbf{R},1}( F)\text{ch}_{\mathbf{R},2}( F) + \tfrac{1}{6 }\text{ch}_{\mathbf{R},1}( F)^3 \, .
\end{eqaed}
For our purposes, it is convenient to organise the above results into the contributions to the anomaly polynomial of $6d$ $\mathcal{N}=(1,0)$ multiplets. The gravity multiplet, comprising the graviton, an antiself-dual $2$-form field and a doublet of left-handed gravitinos, gives
\begin{equation}
    I_{G}= -\frac{273}{360} \, \text{tr} R^4 + \frac{51}{288} \left ( \text{tr} R^2 \right )^2 \, .
\end{equation}
A tensor multiplet comprises a scalar, a self-dual $2$-form and a doublet of fermions with opposite chirality, yielding
\begin{equation}
    I_T=\frac{29}{360} \text{tr} R^4 - \frac{7}{288} \left ( \text{tr} R^2 \right )^2 \, .
\end{equation}
A vector multiplet of a Lie algebra $\mathfrak{g}$ comprises the gauge boson and a doublet of left-handed fermions, which give
\begin{equation}
    I_V=- \frac{\text{dim} \mathfrak{g}}{360} \, \text{tr} R^4 - \frac{\text{dim} \mathfrak{g}}{288} \left ( \text{tr} R^2 \right )^2 - \frac{1}{24} \text{tr}_{\textbf{adj}} F^4 + \frac{1}{24} \text{tr}_{\textbf{adj}} F^2 \, \text{tr} R^2 \, .
\end{equation}
Finally, hypermultiplets comprise four scalars and a doublet of right-handed fermions, and their anomaly polynomial reads
\begin{equation}
    I_H= \frac{\text{dim} \textbf{R}}{360} \, \text{tr} R^4 + \frac{\text{dim} \textbf{R}}{288} \left ( \text{tr} R^2 \right )^2 + \frac{1}{24} \text{tr}_{\textbf{R}} F^4 - \frac{1}{24} \text{tr}_{\textbf{R}} F^2 \, \text{tr} R^2 \, .
\end{equation}
Finally, in \cref{tab:lambda_factors} we list the group-theoretic factors $\lambda_i$ which normalise the instanton number.
\begin{table}[ht!] \centering
 	\begin{tabular}{ |p{1.5 cm}|| p{1.35cm}| p{1.35 cm}| p{1.35 cm}| p{1.35 cm}| p{1.35 cm}|} 
 		\hline
 		$G$ & $SU(N)$ & $SO(2N)$ & $E_7$ & $E_8$ & $U(1)$
 		\\
 		\hline
 		$\lambda$ & $1$ & $2$ & $12$ & $60$ & $1$
 		\\
 		\hline
 	\end{tabular}
 	\caption{The group-theoretic factors entering the perturbative local anomalies of \cref{eq:local_anomaly_reproduced}.}
 	\label{tab:lambda_factors}
 \end{table}

\section{Systematics for computing anomalies}\label{appendix:systematics}

In this appendix we collect useful formulae to systematically compute anomalies on Lens spaces. These expressions can be also used for more general backgrounds of the type $L^{2k-1}_p \times X$.

\noindent The Pontryagin class of Lens spaces can be computed using its relation to the Chern class of the complexified tangent bundle, which is stably trivial since
\begin{eqaed}\label{eq:lens_tangent}
    (TL^{2n-1}_p(j_i) \otimes \mathbb{C}) \oplus \mathbb{C} = \bigoplus_{i=1}^n L^{j_i} \oplus L^{-j_i} \, .
\end{eqaed}
Therefore, one finds
\begin{eqaed}\label{eq:pontryagin_lens}
    p = \prod_{i=1}^n \left(1 + j_i^2 x^2\right) \qquad \longrightarrow \qquad \frac{p_1}{2} = \frac{1}{2} \sum_{i=1}^n j_i^2 \, x^2 \, .
\end{eqaed}
In order to discuss the role of the gauge group $G$, we build an inclusion $\mathbb{Z}_p \to U(1) \to G$. Here we focus on $SU(N)$, $\text{Spin}(2N)$ and exceptional groups. For the first and second cases, one can include $U(1) \to SU(N) \to \text{Spin}(2N)$ in the standard way, according to which the (complexified) vector representation of $\text{Spin}(2N)$ splits as
\begin{eqaed}
    \mathbf{2N} \to \mathbf{N} \oplus \mathbf{N}^* \, .
\end{eqaed}
Placing the $U(1)$ fundamental representation $\mathcal{L}$ in $k$ pairs $L \equiv \mathcal{L} \oplus \mathcal{L}^{-1}$ inside the fundamental of $SU(N)$ in $k$ diagonal blocks, and the rest in the trivial representation $\mathcal{L}^0$, the vector representation of $\text{Spin}(2N)$ further splits into
\begin{eqaed}
    \mathbf{2N} \to \mathbf{N} \oplus \mathbf{N}^* \to [kL \oplus (N-2k)\mathcal{L}^0] \oplus [kL \oplus (N-2k)\mathcal{L}^0] \, .
\end{eqaed}
This describes a family of inclusions, parametrised by $k = 0 \, , \, \dots \, , \, [N/2]$ for $SU(N)$ and $\text{Spin}(2N)$. In order to find the branching rules for other representations, it is convenient to use Chern characters. Letting $x \equiv c_1(\mathcal{L})$, the Chern character of $\mathbf{N}$ (and $\mathbf{N}^*$) is
\begin{eqaed}
    \text{ch}(\mathbf{N}) = k \left(e^x + e^{-x}\right) + (N-2k) \, .
\end{eqaed}
Then we can build the characters for adjoint, symmetric and antisymmetric $SU(N)$ representations, from which we can reconstruct the characters for $\text{Spin}(2N)$ representations of interest, such as the adjoint (antisymmetric) and spinorial. The latter is composed of antisymmetric $SU(N)$ representations, which can be computed using the graded Chern character of the exterior algebra $\Lambda(V) = \oplus_n \Lambda^n(V)$ of a bundle $V$, defined by
\begin{eqaed}
    \text{ch}(\Lambda(V)) \equiv \sum_n t^n \, \text{ch}(\Lambda^n(V))
\end{eqaed}
where the coefficients can be extracted noticing that $\Lambda(U \oplus V) \simeq \Lambda(U) \otimes \Lambda(V)$ and that, for line bundles $\mathcal{L}$,
\begin{eqaed}
    \text{ch}(\Lambda(\mathcal{L})) = 1 + t \, e^{c_1(\mathcal{L})} \, .
\end{eqaed}
For the cases of interest, we will only need antisymmetric powers $\Lambda^n$ with $n \leq 4$, so that we can build the spinorial representations which play a role in $E_7$ and $E_8$.

\noindent In detail, we obtain the following decompositions, where we omit the trivial representation whose degeneracy is fixed by the total dimension of the original representation. In the following expressions, we denote $L^n \equiv \mathcal{L}^n \oplus \mathcal{L}^{-n}$.

\begin{itemize}
    \item Decompositions for $SU(N)$:
\end{itemize}
\begin{eqaed}
    \mathbf{N} & \quad \longrightarrow \quad k L \, , \\
    \mathbf{adj} & \quad \longrightarrow \quad k^2 L^2 \oplus 2k(N-2k)L \, , \\
    \mathbf{\frac{N(N-1)}{2}} & \quad \longrightarrow \quad \frac{k(k-1)}{2} L^2 \oplus k(N-2k) L \, , \\
    \mathbf{\frac{N(N+1)}{2}} & \quad \longrightarrow \quad \frac{k(k+1)}{2} L^2 \oplus k(N-2k) L \, . \\
\end{eqaed}
For $\text{Sp}(N)$, the inclusion we employ is $\mathbb{Z}_p \hookrightarrow U(1) \hookrightarrow \text{Sp}(1) \simeq SU(2) \overset{k}{\hookrightarrow} \text{Sp}(N)$, under which the $\mathbf{2N}$ representation branches according to
\begin{eqaed}
    \mathbf{2N} \quad \longrightarrow \quad kL \oplus (2N - 2k)\mathcal{L}^0 \, .
\end{eqaed}
We shall not make use of this fact in this paper, but it may be useful for future applications.

\begin{itemize}
    \item Decompositions for $\text{Spin}(2N)$:
\end{itemize}
\begin{eqaed}
    \mathbf{2N} & \quad \longrightarrow \quad 2 k L \, , \\
    \mathbf{adj} & \quad \longrightarrow \quad k(2k-1) L^2 \oplus 4k(N-2k) L \, , \\
    \mathbf{sym} & \quad \longrightarrow \quad k(2k+1) L^2 \oplus 4k(N-2k) L \, . \\
\end{eqaed}

\noindent In order to study anomalies for exceptional groups with inclusions of this type, we need the branching rules for spinorial representations of $\text{Spin}(12)$ and $\text{Spin}(16)$. These can be found by summing antisymmetric representations of various ranks, but the resulting decompositions are very complicated to express for general $k$. Since only a few values of $k$ are allowed, one can compute the branching rules for each value, finding the following.

\begin{itemize}
    \item Decomposition of spinorial representations for $\text{Spin}(16)$:
\end{itemize}
\begin{eqaed}
    \mathbf{128}^\pm & \quad \overset{k = 1}{\longrightarrow} \quad 32L \, , \\
    \mathbf{128}^\pm & \quad \overset{k = 2}{\longrightarrow} \quad 8L^2 \oplus 32 L \, , \\
    \mathbf{128}^\pm & \quad \overset{k = 3}{\longrightarrow} \quad 2L^3 \oplus 12L^2 \oplus 30L \, , \\
    \mathbf{128}^+ & \quad \overset{k = 4}{\longrightarrow} \quad  8L^3 \oplus 56L \, , \\
    \mathbf{128}^- & \quad \overset{k = 4}{\longrightarrow} \quad  L^4 \oplus 28L^2 \, , \\
    \mathbf{adj}(E_8) & \quad \longrightarrow \quad \mathbf{adj} \oplus \mathbf{128} \, .
\end{eqaed}

\begin{itemize}
    \item Decomposition of spinorial representations for $\text{Spin}(12)$:
\end{itemize}
\begin{eqaed}
    \mathbf{32}^\pm & \quad \overset{k = 1}{\longrightarrow} \quad 8L \, , \\
    \mathbf{32}^\pm & \quad \overset{k = 2}{\longrightarrow} \quad 2L^2 \oplus 8 L \, , \\
    \mathbf{32}^+ & \quad \overset{k = 3}{\longrightarrow} \quad  L^3 \oplus 15L \, , \\
    \mathbf{32}^- & \quad \overset{k = 3}{\longrightarrow} \quad  6L^2 \, , \\
    \mathbf{fund}(E_7) & \quad \longrightarrow \quad 2 \times \mathbf{12} \oplus \mathbf{32}^+ \, , \\
    \mathbf{adj}(E_7) & \quad \longrightarrow \quad \mathbf{adj} \oplus \mathbf{32}^+ \oplus \mathbf{32}^- \, .
\end{eqaed}

\noindent Having found the relevant branching rules, one needs to compute Chern classes and eta invariants on various spaces. The above expressions spell out the decomposition of the gauge bundle into $\mathbb{Z}_p$ charges $q = 0 \, , \, \dots \, , \, p-1$. By the splitting principle, one can then write the total Chern class
\begin{eqaed}
    c\left(\bigoplus_m a_m L^m\right) = \prod_m (1-m^2 x^2)^{a_m} \, ,
\end{eqaed}
so that the second Chern class is given by
\begin{eqaed}
    c_2 = - \sum_m m^2 \, a_m \, x^2 \, .
\end{eqaed}
Since eta invariants are additive under direct sum, we provide explicit formulae for $\eta_q$ for charged spinor fields and $\eta_0$ for Rarita-Schwinger fields, since in the settings at stake gravitini are always uncharged. For future convenience we provide these invariants for Lens spaces of dimension three, seven and eleven, where the spin structure is the simplest.

\begin{itemize}
    \item Charged eta invariants for (bulk) Dirac fermions on Lens spaces:
\end{itemize}
\begin{eqaed}
    \eta^\text{D}_q(L^{3}_p) & = \frac{p^2 - 1 - 6pq + 6q^2}{12p} \, , \\
    \eta^\text{D}_q(L^{7}_p) & = - \, \frac{p^4 + 10p^2 - 11 - 30 p^2 q^2 - 60 pq + 60 pq^3 - 30 q^4 + 60 q^2}{720p} \, , \\
    \eta^\text{D}_q(L^{11}_p) & = \frac{2 p^6 + 21 p^4 + 168 p^2 - 191 - 42 p^4 q^2 + 210 p^2 q^4 - 630 p^2 q^2}{60480p} \, , \\
    & + \frac{- 252 p q^5 + 1260 p q^3 - 1008 pq + 84 q^6 - 630 q^4 + 1008 q^2}{60480p} \, .
\end{eqaed}
\begin{itemize}
    \item Neutral Rarita-Schwinger invariants on Lens spaces:
\end{itemize}
\begin{eqaed}
    \eta^\text{RS}_0(L^{3}_p) & = \frac{p^2 - 8p + 7}{4p} \, , \\
    \eta^\text{RS}_0(L^{7}_p) & = - \, \frac{7p^4 - 170p^2 + 163}{720p} \, , \\
    \eta^\text{RS}_0(L^{11}_p) & = \frac{22p^6 - 273p^4 - 3192p^2 + 3443}{60480p} \, .
\end{eqaed}
For the reader's convenience, we also provide useful formulae for trace decompositions, which we extensively use in the main text to write anomaly polynomials in a factorised form. In the following expressions we denote the fundamental traces with $\text{tr}$ without any suffix.

\begin{itemize}
    \item Trace decompositions for $SU(N)$ ($\mathbf[2]$ denotes the rank-two antisymmetric representation):
\end{itemize}
\begin{eqaed}
    \tr_\mathbf{adj} F^2 & = 2N \tr F^2 \, , \\
    \tr_\mathbf{adj} F^4 & = 2N \tr F^4 + 6 (\tr F^2)^2 \, , \\
    \tr_\mathbf{[2]} F^2 & = (N-2) \tr F^2 \, , \\
    \tr_\mathbf{[2]} F^4 & = (N-8) \tr F^4 + 3 (\tr F^2)^2 \, , \\
    \tr_\mathbf{sym} F^2 & = (N+2) \tr F^2 \, , \\
    \tr_\mathbf{sym} F^4 & = (N+8) \tr F^4 + 3 (\tr F^2)^2 \, .
\end{eqaed}

\begin{itemize}
    \item Trace decompositions for $\text{Spin}(2N)$:
\end{itemize}
\begin{eqaed}
    \tr_\mathbf{adj} F^2 & = (2N-2) \tr F^2 \, , \\
    \tr_\mathbf{adj} F^4 & = (2N-8) \tr F^4 + 3 (\tr F^2)^2 \, , \\
    \tr_\mathbf{sym} F^2 & = (2N+2) \tr F^2 \, , \\
    \tr_\mathbf{sym} F^4 & = (2N+8) \tr F^4 + 3 (\tr F^2)^2 \, .
\end{eqaed}

\begin{itemize}
    \item Trace decompositions for spinors of $\text{Spin}(16)$ and $\text{Spin}(12)$:
\end{itemize}
\begin{eqaed}
    \tr_\mathbf{128} F^2 & = 16 \tr F^2 \, , \\
    \tr_\mathbf{128} F^4 & = -8 \tr F^4 + 6 (\tr F^2)^2 \, , \\
    \tr_\mathbf{32} F^2 & = 4 \tr F^2 \, , \\
    \tr_\mathbf{32} F^4 & = -2 \tr F^4 + \frac{3}{2} (\tr F^2)^2 \, .
\end{eqaed}

\section{Partition functions for \texorpdfstring{$SO(16) \times SO(16)$}{SO(16) x SO(16)} \texorpdfstring{$K3$}{K3} orbifolds } \label{K3 orbifolds partition functions}
     
     This appendix is devoted to a brief discussion on the structure of the partition functions of the $SO(16) \times SO(16)$ theory compactified on the orbifolds $\mathbb{T}^4 / \mathbb{Z}_N$ for $N=2,3,4,6$. The general strategy is to compute the trace on the Hilbert space of oscillators of the Virasoro zero modes for the $SO(16) \times SO(16)$ vacuum compactified on $\mathbb{T}^4$, and afterwards performing the projection on $P_N=\frac{1}{N} \sum_{\alpha=0}^{N-1} g^\beta$. The result is not modular invariant anymore, and one has to include suitable images of the modular orbit. This leads to the presence of twisted sectors and multiplicities $d_{\alpha,\beta}$. The partition function can be expressed in terms of
     traces on Hilbert space of oscillators in each sector,
     \begin{equation}
     	\text{tr}_{\alpha} \, g^\beta q^{L_0-c/24} \, \bar{q}^{\bar{L}_0-\bar{c}/24}= \modZ{\alpha}{\beta} \Lambda_{\alpha,\beta} \, ,
     \end{equation} 
   where the contributions of worldsheet fermions and non compact bosons are given by     
     \begin{equation}
     	\begin{aligned}
     		&\modZ{\alpha}{\beta}=  \frac{1}{\left ( \eta \bar \eta \right )^4} \times \\
     		& \bigg \{ \left (V_4 O_4^{(\alpha,\beta)} + O_4 V_4^{(\alpha,\beta)} \right ) \left [ \left ( \overline O_{12} \overline O_4^{(\alpha,\beta)} + \overline V_{12} \overline V_4^{(\alpha,\beta)} \right )  \overline O_{16} + \left ( \overline S_{12} \overline S_4^{(\alpha,\beta)} + \overline C_{12} \overline C_4^{(\alpha,\beta)} \right )  \overline S_{16} \right ]
     		\\
     		& + \left (O_4 O_4^{(\alpha,\beta)} + V_4 V_4^{(\alpha,\beta)} \right ) \left [ \left ( \overline O_{12} \overline V_4^{(\alpha,\beta)} + \overline V_{12} \overline O_4^{(\alpha,\beta)} \right )  \overline C_{16} + \left ( \overline S_{12} \overline C_4^{(\alpha,\beta)} + \overline C_{12} \overline S_4^{(\alpha,\beta)} \right )  \overline V_{16} \right ]
     		\\
     		&- \left (S_4 S_4^{(\alpha,\beta)} + C_4 C_4^{(\alpha,\beta)} \right ) \left [ \left ( \overline O_{12} \overline O_4^{(\alpha,\beta)} + \overline V_{12} \overline V_4^{(\alpha,\beta)} \right )  \overline S_{16} + \left ( \overline S_{12} \overline S_4^{(\alpha,\beta)} + \overline C_{12} \overline C_4^{(\alpha,\beta)} \right )  \overline O_{16} \right ]
     		\\
     		& - \left (S_4 C_4^{(\alpha,\beta)} + C_4 S_4^{(\alpha,\beta)} \right ) \left [ \left ( \overline O_{12} \overline V_4^{(\alpha,\beta)} + \overline V_{12} \overline O_4^{(\alpha,\beta)} \right )  \overline V_{16} + \left ( \overline S_{12} \overline C_4^{(\alpha,\beta)} + \overline C_{12} \overline S_4^{(\alpha,\beta)} \right )  \overline C_{16} \right ] \bigg \}
     	\end{aligned}
     \end{equation}
     where the various affine characters of the level-one orthogonal algebras are given by
     \begin{equation} \label{orbifoldschar}
     	\begin{aligned}
     		& O_4^{(\alpha,\beta)}=\frac{1}{2 \eta(\tau)^2}  \bigg ( \prod_{i=1}^2 \Jtheta{\alpha v_i}{\beta v_i }{0}{\tau}  + \prod_{i=1}^2 \Jtheta{\alpha v_i}{1/2+\beta v_i }{0}{\tau}  \bigg ) 	\, ,
     		\\
     		&V_4^{(\alpha,\beta)}=\frac{1}{2 \eta(\tau)^2}  \bigg ( \prod_{i=1}^2 \Jtheta{\alpha v_i}{\beta v_i }{0}{\tau}  - \prod_{i=1}^2 \Jtheta{\alpha v_i}{1/2+\beta v_i }{0}{\tau}  \bigg ) 	\, ,
     		\\
     		&S_4^{(\alpha,\beta)}= \frac{1}{2 \eta(\tau)^2}  \bigg ( \prod_{i=1}^2 \Jtheta{1/2+\alpha v_i}{\beta v_i }{0}{\tau}  - \prod_{i=1}^2 \Jtheta{1/2+\alpha v_i}{1/2+\beta v_i }{0}{\tau}  \bigg ) \, ,
     		\\
     		&C_4^{(\alpha,\beta)}= \frac{1}{2 \eta(\tau)^2} \bigg ( \prod_{i=1}^2 \Jtheta{1/2+\alpha v_i}{\beta v_i }{0}{\tau}  + \prod_{i=1}^2 \Jtheta{1/2+\alpha v_i}{1/2+\beta v_i }{0}{\tau}  \bigg ) \, .
     	\end{aligned}
     \end{equation}
     In the above expressions, $v_i$ stands for the shift vectors described in \cref{vectshifts}, which should be properly inserted when dealing with the left or right moving sectors. Furthermore, such expressions are valid whenever the spin connection is identified with the gauge connection (namely when the Bianchi identity is satisfied with the standard embedding), and should be modified to incorporate other cases. Finally, the contributions from compact bosons read 
     \begin{equation}
     	\Lambda_{\alpha,\beta}= \begin{cases}
     		\Lambda_{4,4}= \frac{1}{\left (\eta \bar \eta \right )^4} \displaystyle{\sum_{m,n}} q^{\frac{\alpha'}{4} \left ( \frac{m}{R}- \frac{n R}{\alpha'} \right ) } \bar q^{\frac{\alpha'}{4} \left ( \frac{m}{R} +  \frac{n R}{\alpha'} \right ) } \ , & \ \ (\alpha,\beta)=(0,0) \, ,
     		\\
     		\displaystyle{\prod_{i=1}^2}  (2 \sin{\pi v_i})^{2 \delta_{\alpha,0}} \frac{  \eta \bar \eta }{ \Jtheta{1/2+ \alpha v_i }{1/2+ \beta v_i}{0}{\tau} \,  \Jbartheta{1/2 - \alpha v_i}{1/2-\beta v_i}{0}{\overline \tau} } \ , & \ \ \text{otherwise} \, .
     	\end{cases}
     \end{equation} 
    The expressions in \cref{orbifoldschar} contain characters of the $\widehat{su(2)}_1$ and $\widehat{u(1)}_k$ algebras, in which however the level $k$ and the $U(1)$ characters themselves are not easy to extract. Thus these ``characters'' can only give us explicit information about $SU(2)$ representations encoded into the coefficients of their $q$-expansion. In particular, it is possible to see that $O_4^{(\alpha,\beta)}$ and $C_4^{(\alpha,\beta)}$ contain the trivial representation of $SU(2)$, whereas $V_4^{(\alpha,\beta)}$ and $S_4^{(\alpha,\beta)}$ contain the fundamental, for every value of $\alpha$ and $\beta$.

\printbibliography

@article{Freed:1986hv,
    author = "Freed, D. S.",
    title = "{ON DETERMINANT LINE BUNDLES}",
    journal = "Conf. Proc. C",
    volume = "8607214",
    pages = "189--238",
    year = "1986"
}

@article{Freed:2016rqq,
    author = "Freed, Daniel S. and Hopkins, Michael J.",
    title = "{Reflection positivity and invertible topological phases}",
    eprint = "1604.06527",
    archivePrefix = "arXiv",
    primaryClass = "hep-th",
    doi = "10.2140/gt.2021.25.1165",
    journal = "Geom. Topol.",
    volume = "25",
    pages = "1165--1330",
    year = "2021"
}

@article{Szabo:2012hc,
    author = "Szabo, Richard J.",
    editor = "Bonora, Loriano and Bytsenko, A. A. and Guimaraes, M. E. X. and Helayel-Neto, J. A.",
    title = "{Quantization of Higher Abelian Gauge Theory in Generalized Differential Cohomology}",
    eprint = "1209.2530",
    archivePrefix = "arXiv",
    primaryClass = "hep-th",
    reportNumber = "HWM-12-13, EMPG-12-20, ESI-2381",
    doi = "10.22323/1.175.0009",
    journal = "PoS",
    volume = "ICMP2012",
    pages = "009",
    year = "2012"
}

@article{PhysRevD.12.3845,
  title = {Concept of nonintegrable phase factors and global formulation of gauge fields},
  author = {Wu, Tai Tsun and Yang, Chen Ning},
  journal = {Phys. Rev. D},
  volume = {12},
  issue = {12},
  pages = {3845--3857},
  numpages = {0},
  year = {1975},
  month = {12},
  publisher = {American Physical Society},
  doi = {10.1103/PhysRevD.12.3845},
  url = {https://link.aps.org/doi/10.1103/PhysRevD.12.3845}
}

@article{PhysRev.176.1489,
  title = {Quantum Field Theory of Particles with Both Electric and Magnetic Charges},
  author = {Zwanziger, Daniel},
  journal = {Phys. Rev.},
  volume = {176},
  issue = {5},
  pages = {1489--1495},
  numpages = {0},
  year = {1968},
  month = {12},
  publisher = {American Physical Society},
  doi = {10.1103/PhysRev.176.1489},
  url = {https://link.aps.org/doi/10.1103/PhysRev.176.1489}
}

@article{PhysRev.144.1087,
  title = {Magnetic Charge and Quantum Field Theory},
  author = {Schwinger, Julian},
  journal = {Phys. Rev.},
  volume = {144},
  issue = {4},
  pages = {1087--1093},
  numpages = {0},
  year = {1966},
  month = {4},
  publisher = {American Physical Society},
  doi = {10.1103/PhysRev.144.1087},
  url = {https://link.aps.org/doi/10.1103/PhysRev.144.1087}
}

@article{PhysRev.74.817,
  title = {The Theory of Magnetic Poles},
  author = {Dirac, P. A. M.},
  journal = {Phys. Rev.},
  volume = {74},
  issue = {7},
  pages = {817--830},
  numpages = {0},
  year = {1948},
  month = {10},
  publisher = {American Physical Society},
  doi = {10.1103/PhysRev.74.817},
  url = {https://link.aps.org/doi/10.1103/PhysRev.74.817}
}

@article{Angelantonj:1996mw,
    author = "Angelantonj, Carlo and Bianchi, Massimo and Pradisi, Gianfranco and Sagnotti, Augusto and Stanev, Yassen S.",
    title = "{Comments on Gepner models and type I vacua in string theory}",
    eprint = "hep-th/9607229",
    archivePrefix = "arXiv",
    reportNumber = "ROM2F-96-35",
    doi = "10.1016/0370-2693(96)01124-0",
    journal = "Phys. Lett. B",
    volume = "387",
    pages = "743--749",
    year = "1996"
}

@article{Gepner:1987qi,
    author = "Gepner, Doron",
    editor = "Schellekens, B.",
    title = "{Space-Time Supersymmetry in Compactified String Theory and Superconformal Models}",
    reportNumber = "Print-87-0370 (PRINCETON), PUPT-1056",
    doi = "10.1016/0550-3213(88)90397-5",
    journal = "Nucl. Phys. B",
    volume = "296",
    pages = "757",
    year = "1988"
}

@article{Schommer-Pries:2017sdd,
    author = "Schommer-Pries, Christopher",
    title = "{Invertible Topological Field Theories}",
    eprint = "1712.08029",
    archivePrefix = "arXiv",
    primaryClass = "math.AT",
    month = "12",
    year = "2017"
}

@article{Lurie:2009keu,
    author = "Lurie, Jacob",
    title = "{On the Classification of Topological Field Theories}",
    eprint = "0905.0465",
    archivePrefix = "arXiv",
    primaryClass = "math.CT",
    month = "5",
    year = "2009"
}

@article{Baez:1995xq,
    author = "Baez, J. C. and Dolan, J.",
    title = "{Higher dimensional algebra and topological quantum field theory}",
    eprint = "q-alg/9503002",
    archivePrefix = "arXiv",
    doi = "10.1063/1.531236",
    journal = "J. Math. Phys.",
    volume = "36",
    pages = "6073--6105",
    year = "1995"
}

@article{Yonekura:2018ufj,
    author = "Yonekura, Kazuya",
    title = "{On the cobordism classification of symmetry protected topological phases}",
    eprint = "1803.10796",
    archivePrefix = "arXiv",
    primaryClass = "hep-th",
    reportNumber = "IPMU-18-0040",
    doi = "10.1007/s00220-019-03439-y",
    journal = "Commun. Math. Phys.",
    volume = "368",
    number = "3",
    pages = "1121--1173",
    year = "2019"
}

@article{Freed:2007vy,
    author = "Freed, Daniel S. and Hopkins, Michael J. and Teleman, Constantin",
    title = "{Consistent orientation of moduli spaces}",
    eprint = "0711.1909",
    archivePrefix = "arXiv",
    primaryClass = "math.AT",
    month = "11",
    year = "2007"
}

@article{Honecker:2006qz,
    author = "Honecker, Gabriele and Trapletti, Michele",
    title = "{Merging Heterotic Orbifolds and K3 Compactifications with Line Bundles}",
    eprint = "hep-th/0612030",
    archivePrefix = "arXiv",
    reportNumber = "CERN-PH-TH-2006-249, HD-THEP-06-31",
    doi = "10.1088/1126-6708/2007/01/051",
    journal = "JHEP",
    volume = "01",
    pages = "051",
    year = "2007"
}

@inproceedings{Witten:1985mj,
    author = "Witten, Edward",
    title = "{GLOBAL ANOMALIES IN STRING THEORY}",
    booktitle = "{Symposium on Anomalies, Geometry, Topology}",
    reportNumber = "Print-85-0620 (PRINCETON)",
    month = "6",
    year = "1985"
}

@article{Walton,
  title = {Heterotic string on the simplest Calabi-Yau manifold and its orbifold limits},
  author = {Walton, Mark A.},
  journal = {Phys. Rev. D},
  volume = {37},
  issue = {2},
  pages = {377--390},
  numpages = {0},
  year = {1988},
  month = {1},
  publisher = {American Physical Society},
  doi = {10.1103/PhysRevD.37.377},
  url = {https://link.aps.org/doi/10.1103/PhysRevD.37.377}
}

@article{Hsieh:2020jpj,
    author = "Hsieh, Chang-Tse and Tachikawa, Yuji and Yonekura, Kazuya",
    title = "{Anomaly Inflow and p-Form Gauge Theories}",
    eprint = "2003.11550",
    archivePrefix = "arXiv",
    primaryClass = "hep-th",
    reportNumber = "IPMU-20-0028, TU-1098",
    doi = "10.1007/s00220-022-04333-w",
    journal = "Commun. Math. Phys.",
    volume = "391",
    number = "2",
    pages = "495--608",
    year = "2022"
}

@article{Seiberg:2011dr,
    author = "Seiberg, Nathan and , Washington",
    title = "{Charge Lattices and Consistency of 6D Supergravity}",
    eprint = "1103.0019",
    archivePrefix = "arXiv",
    primaryClass = "hep-th",
    reportNumber = "MIT-CTP-4216",
    doi = "10.1007/JHEP06(2011)001",
    journal = "JHEP",
    volume = "06",
    pages = "001",
    year = "2011"
}

@article{Dierigl:2022zll,
    author = "Dierigl, Markus and Oehlmann, Paul-Konstantin and Schimannek, Thorsten",
    title = "{The discrete Green-Schwarz mechanism in 6D F-theory and elliptic genera of non-critical strings}",
    eprint = "2212.04503",
    archivePrefix = "arXiv",
    primaryClass = "hep-th",
    reportNumber = "LMU-ASC 33/22",
    doi = "10.1007/JHEP03(2023)090",
    journal = "JHEP",
    volume = "03",
    pages = "090",
    year = "2023"
}

@article{Kim:2019vuc,
    author = "Kim, Hee-Cheol and Shiu, Gary and Vafa, Cumrun",
    title = "{Branes and the Swampland}",
    eprint = "1905.08261",
    archivePrefix = "arXiv",
    primaryClass = "hep-th",
    doi = "10.1103/PhysRevD.100.066006",
    journal = "Phys. Rev. D",
    volume = "100",
    number = "6",
    pages = "066006",
    year = "2019"
}

@article{Angelantonj:2020pyr,
    author = "Angelantonj, Carlo and Bonnefoy, Quentin and Condeescu, Cezar and Dudas, Emilian",
    title = "{String Defects, Supersymmetry and the Swampland}",
    eprint = "2007.12722",
    archivePrefix = "arXiv",
    primaryClass = "hep-th",
    reportNumber = "CPHT-RR046.072020, DESY 20-123, DESY-20-123",
    doi = "10.1007/JHEP11(2020)125",
    journal = "JHEP",
    volume = "11",
    pages = "125",
    year = "2020"
}

@article{Debray:2023yrs,
    author = "Debray, Arun and Dierigl, Markus and Heckman, Jonathan J. and Montero, Miguel",
    title = "{The Chronicles of IIBordia: Dualities, Bordisms, and the Swampland}",
    eprint = "2302.00007",
    archivePrefix = "arXiv",
    primaryClass = "hep-th",
    reportNumber = "LMU-ASC 06/23, IFT-UAM/CSIC-23-7",
    month = "1",
    year = "2023"
}

@article{Kumar:2010ru,
    author = "Kumar, Vijay and Morrison, David R. and Taylor, Washington",
    title = "{Global aspects of the space of 6D N = 1 supergravities}",
    eprint = "1008.1062",
    archivePrefix = "arXiv",
    primaryClass = "hep-th",
    reportNumber = "MIT-CTP-4132, NSF-KITP-10-096, UCSB-MATH-2010-19",
    doi = "10.1007/JHEP11(2010)118",
    journal = "JHEP",
    volume = "11",
    pages = "118",
    year = "2010"
}

@article{Park:2011wv,
    author = "Park, Daniel S. and Taylor, Washington",
    title = "{Constraints on 6D Supergravity Theories with Abelian Gauge Symmetry}",
    eprint = "1110.5916",
    archivePrefix = "arXiv",
    primaryClass = "hep-th",
    reportNumber = "MIT-CTP-4228",
    doi = "10.1007/JHEP01(2012)141",
    journal = "JHEP",
    volume = "01",
    pages = "141",
    year = "2012"
}

@article{Taylor:2018khc,
    author = "Taylor, Washington and Turner, Andrew P.",
    title = "{An infinite swampland of U(1) charge spectra in 6D supergravity theories}",
    eprint = "1803.04447",
    archivePrefix = "arXiv",
    primaryClass = "hep-th",
    reportNumber = "MIT-CTP/4992, MIT-CTP-4992",
    doi = "10.1007/JHEP06(2018)010",
    journal = "JHEP",
    volume = "06",
    pages = "010",
    year = "2018"
}

@article{Tachikawa:2021mby,
    author = "Tachikawa, Yuji and Yamashita, Mayuko",
    title = "{Topological Modular Forms and the Absence of All Heterotic Global Anomalies}",
    eprint = "2108.13542",
    archivePrefix = "arXiv",
    primaryClass = "hep-th",
    doi = "10.1007/s00220-023-04761-2",
    journal = "Commun. Math. Phys.",
    volume = "402",
    number = "2",
    pages = "1585--1620",
    year = "2023",
    note = "[Erratum: Commun.Math.Phys. 402, 2131 (2023)]"
}

@article{Tachikawa:2021mvw,
    author = "Tachikawa, Yuji",
    title = "{Topological modular forms and the absence of a heterotic global anomaly}",
    eprint = "2103.12211",
    archivePrefix = "arXiv",
    primaryClass = "hep-th",
    doi = "10.1093/ptep/ptab060",
    journal = "PTEP",
    volume = "2022",
    number = "4",
    pages = "04A107",
    year = "2022"
}

@article{Debray:2021vob,
    author = "Debray, Arun and Dierigl, Markus and Heckman, Jonathan J. and Montero, Miguel",
    title = "{The anomaly that was not meant IIB}",
    eprint = "2107.14227",
    archivePrefix = "arXiv",
    primaryClass = "hep-th",
    reportNumber = "LMU-ASC 24/21",
    doi = "10.1002/prop.202100168",
    month = "7",
    year = "2021"
}

@article{Alvarez-Gaume:1986ghj,
    author = "Alvarez-Gaume, Luis and Ginsparg, Paul H. and Moore, Gregory W. and Vafa, C.",
    title = "{An O(16) x O(16) Heterotic String}",
    reportNumber = "HUTP-86/A013",
    doi = "10.1016/0370-2693(86)91524-8",
    journal = "Phys. Lett. B",
    volume = "171",
    pages = "155--162",
    year = "1986"
}

@article{Dixon:1986iz,
    author = "Dixon, Lance J. and Harvey, Jeffrey A.",
    editor = "Schellekens, B.",
    title = "{String Theories in Ten-Dimensions Without Space-Time Supersymmetry}",
    reportNumber = "PRINT-86-0244 (PRINCETON)",
    doi = "10.1016/0550-3213(86)90619-X",
    journal = "Nucl. Phys. B",
    volume = "274",
    pages = "93--105",
    year = "1986"
}

@article{Dixon:1985jw,
    author = "Dixon, Lance J. and Harvey, Jeffrey A. and Vafa, C. and Witten, Edward",
    editor = "Schellekens, B.",
    title = "{Strings on Orbifolds}",
    reportNumber = "PRINT-85-0616 (PRINCETON)",
    doi = "10.1016/0550-3213(85)90593-0",
    journal = "Nucl. Phys. B",
    volume = "261",
    pages = "678--686",
    year = "1985"
}

@article{Dixon:1986jc,
    author = "Dixon, Lance J. and Harvey, Jeffrey A. and Vafa, C. and Witten, Edward",
    title = "{Strings on Orbifolds. 2.}",
    reportNumber = "PRINT-86-0246 (PRINCETON)",
    doi = "10.1016/0550-3213(86)90287-7",
    journal = "Nucl. Phys. B",
    volume = "274",
    pages = "285--314",
    year = "1986"
}

@article{Green:1984sg,
    author = "Green, Michael B. and Schwarz, John H.",
    title = "{Anomaly Cancellation in Supersymmetric D=10 Gauge Theory and Superstring Theory}",
    reportNumber = "CALT-68-1182",
    doi = "10.1016/0370-2693(84)91565-X",
    journal = "Phys. Lett. B",
    volume = "149",
    pages = "117--122",
    year = "1984"
}

@article{Green:1984bx,
    author = "Green, Michael B. and Schwarz, John H. and West, Peter C.",
    title = "{Anomaly Free Chiral Theories in Six-Dimensions}",
    reportNumber = "CALT-68-1210",
    doi = "10.1016/0550-3213(85)90222-6",
    journal = "Nucl. Phys. B",
    volume = "254",
    pages = "327--348",
    year = "1985"
}

@article{Sagnotti:1992qw,
    author = "Sagnotti, Augusto",
    title = "{A Note on the Green-Schwarz mechanism in open string theories}",
    eprint = "hep-th/9210127",
    archivePrefix = "arXiv",
    reportNumber = "ROM2F-92-49",
    doi = "10.1016/0370-2693(92)90682-T",
    journal = "Phys. Lett. B",
    volume = "294",
    pages = "196--203",
    year = "1992"
}

@article{Atiyah:1975jf,
    author = "Atiyah, M. F. and Patodi, V. K. and Singer, I. M.",
    title = "{Spectral asymmetry and Riemannian Geometry 1}",
    doi = "10.1017/S0305004100049410",
    journal = "Math. Proc. Cambridge Phil. Soc.",
    volume = "77",
    pages = "43",
    year = "1975"
}

@inproceedings{Freed:2000ta,
    author = "Freed, Daniel S.",
    title = "{Dirac charge quantization and generalized differential cohomology}",
    eprint = "hep-th/0011220",
    archivePrefix = "arXiv",
    month = "11",
    year = "2000"
}

@article{Freed:2006ya,
    author = "Freed, Daniel S. and Moore, Gregory W. and Segal, Graeme",
    title = "{The Uncertainty of Fluxes}",
    eprint = "hep-th/0605198",
    archivePrefix = "arXiv",
    reportNumber = "NSF-KITP-05-119",
    doi = "10.1007/s00220-006-0181-3",
    journal = "Commun. Math. Phys.",
    volume = "271",
    pages = "247--274",
    year = "2007"
}

@article{Freed:2006yc,
    author = "Freed, Daniel S. and Moore, Gregory W. and Segal, Graeme",
    title = "{Heisenberg Groups and Noncommutative Fluxes}",
    eprint = "hep-th/0605200",
    archivePrefix = "arXiv",
    reportNumber = "NSF-KITP-05-118",
    doi = "10.1016/j.aop.2006.07.014",
    journal = "Annals Phys.",
    volume = "322",
    pages = "236--285",
    year = "2007"
}

@inproceedings{Witten:2019bou,
    author = "Witten, Edward and Yonekura, Kazuya",
    title = "{Anomaly Inflow and the $\eta$-Invariant}",
    booktitle = "{The Shoucheng Zhang Memorial Workshop}",
    eprint = "1909.08775",
    archivePrefix = "arXiv",
    primaryClass = "hep-th",
    month = "9",
    year = "2019"
}

@article{Garcia-Etxebarria:2018ajm,
    author = "Garc\'\i{}a-Etxebarria, I\~naki and Montero, Miguel",
    title = "{Dai-Freed anomalies in particle physics}",
    eprint = "1808.00009",
    archivePrefix = "arXiv",
    primaryClass = "hep-th",
    reportNumber = "MPP-2018-188",
    doi = "10.1007/JHEP08(2019)003",
    journal = "JHEP",
    volume = "08",
    pages = "003",
    year = "2019"
}

@article{Yonekura:2016wuc,
    author = "Yonekura, Kazuya",
    title = "{Dai-Freed theorem and topological phases of matter}",
    eprint = "1607.01873",
    archivePrefix = "arXiv",
    primaryClass = "hep-th",
    reportNumber = "IPMU-16-0094",
    doi = "10.1007/JHEP09(2016)022",
    journal = "JHEP",
    volume = "09",
    pages = "022",
    year = "2016"
}

@article{Dai:1994kq,
    author = "Dai, Xian-zhe and Freed, Daniel S.",
    title = "{eta invariants and determinant lines}",
    eprint = "hep-th/9405012",
    archivePrefix = "arXiv",
    doi = "10.1063/1.530747",
    journal = "J. Math. Phys.",
    volume = "35",
    pages = "5155--5194",
    year = "1994",
    note = "[Erratum: J.Math.Phys. 42, 2343--2344 (2001)]"
}

@article{Lee:2019skh,
    author = "Lee, Seung-Joo and Weigand, Timo",
    title = "{Swampland Bounds on the Abelian Gauge Sector}",
    eprint = "1905.13213",
    archivePrefix = "arXiv",
    primaryClass = "hep-th",
    reportNumber = "CERN-TH-2019-082",
    doi = "10.1103/PhysRevD.100.026015",
    journal = "Phys. Rev. D",
    volume = "100",
    number = "2",
    pages = "026015",
    year = "2019"
}

@mastersthesis{Michelangelo,
   author  = "Tartaglia, Michelangelo",
   title   = " {Self-dual fields in 6D Supergravity}",
   school  = {Ludwig-Maximilians-Universitaet M\"unchen},
   type    = "MSc Thesis",
   year    = "2023",
}

@inproceedings{Taylor2022GaussSI,
  title={Gauss Sums in Algebra and Topology},
  author={Laurence R. Taylor},
  year={2022},
  url={https://api.semanticscholar.org/CorpusID:12686252}
}

@article{DELOUP2005105,
title = {Quadratic functions on torsion groups},
journal = {Journal of Pure and Applied Algebra},
volume = {198},
number = {1},
pages = {105-121},
year = {2005},
issn = {0022-4049},
doi = {https://doi.org/10.1016/j.jpaa.2004.10.011},
url = {https://www.sciencedirect.com/science/article/pii/S0022404904002166},
author = {Florian Deloup and Gwénaël Massuyeau}}

@article{Witten:2018lgb,
    author = "Witten, Edward",
    title = "{A note on boundary conditions in Euclidean gravity}",
    eprint = "1805.11559",
    archivePrefix = "arXiv",
    primaryClass = "hep-th",
    doi = "10.1142/S0129055X21400043",
    journal = "Rev. Math. Phys.",
    volume = "33",
    number = "10",
    pages = "2140004",
    year = "2021"
}

@InProceedings{10.1007/BFb0075216,
author="Cheeger, Jeff and Simons, James",
title="Differential characters and geometric invariants",
booktitle="Geometry and Topology",
year="1985",
publisher="Springer Berlin Heidelberg",
address="Berlin, Heidelberg",
pages="50--80",
isbn=" "
}

@article{Deligne,
author="Deligne, Pierre",
title="Théorie de Hodge, II",
journal="Publications Mathématiques de l'Institut des Hautes Études Scientifiques",
volume="40",
pages="5-57",
year="1971"
}

@article{Hopkins:2002rd,
    author = "Hopkins, M. J. and Singer, I. M.",
    title = "{Quadratic functions in geometry, topology, and M theory}",
    eprint = "math/0211216",
    archivePrefix = "arXiv",
    journal = "J. Diff. Geom.",
    volume = "70",
    number = "3",
    pages = "329--452",
    year = "2005"
}

@article{Adler:1969gk,
    author = "Adler, Stephen L.",
    title = "{Axial vector vertex in spinor electrodynamics}",
    doi = "10.1103/PhysRev.177.2426",
    journal = "Phys. Rev.",
    volume = "177",
    pages = "2426--2438",
    year = "1969"
}

@article{Bell:1969ts,
    author = "Bell, J. S. and Jackiw, R.",
    title = "{A PCAC puzzle: $\pi^0 \to \gamma \gamma$ in the $\sigma$ model}",
    doi = "10.1007/BF02823296",
    journal = "Nuovo Cim. A",
    volume = "60",
    pages = "47--61",
    year = "1969"
}

@book{Fujikawa:2004cx,
    author = "Fujikawa, K. and Suzuki, H.",
    title = "{Path integrals and quantum anomalies}",
    doi = "10.1093/acprof:oso/9780198529132.001.0001",
    year = "2004"
}

@inproceedings{Harvey:2005it,
    author = "Harvey, Jeffrey A.",
    title = "{TASI 2003 lectures on anomalies}",
    eprint = "hep-th/0509097",
    archivePrefix = "arXiv",
    reportNumber = "EFI-05-16",
    month = "9",
    year = "2005"
}

@article{Alvarez-Gaume:1983ict,
    author = "Alvarez-Gaume, Luis and Ginsparg, Paul H.",
    title = "{The Topological Meaning of Nonabelian Anomalies}",
    reportNumber = "HUTP-83/A081",
    doi = "10.1016/0550-3213(84)90487-5",
    journal = "Nucl. Phys. B",
    volume = "243",
    pages = "449--474",
    year = "1984"
}

@article{Alvarez-Gaume:1984zlq,
    author = "Alvarez-Gaume, Luis and Ginsparg, Paul H.",
    editor = "Salam, A. and Sezgin, E.",
    title = "{The Structure of Gauge and Gravitational Anomalies}",
    reportNumber = "HUTP-84/A016",
    doi = "10.1016/0003-4916(85)90087-9",
    journal = "Annals Phys.",
    volume = "161",
    pages = "423",
    year = "1985",
    note = "[Erratum: Annals Phys. 171, 233 (1986)]"
}

@article{Alvarez-Gaume:1983ihn,
    author = "Alvarez-Gaume, Luis and Witten, Edward",
    editor = "Salam, A. and Sezgin, E.",
    title = "{Gravitational Anomalies}",
    reportNumber = "HUTP-83/A039",
    doi = "10.1016/0550-3213(84)90066-X",
    journal = "Nucl. Phys. B",
    volume = "234",
    pages = "269",
    year = "1984"
}

@article{Alvarez-Gaume:1984zst,
    author = "Alvarez-Gaume, Luis and Della Pietra, S. and Moore, Gregory W.",
    title = "{Anomalies and Odd Dimensions}",
    reportNumber = "HUTP-84-A028",
    doi = "10.1016/0003-4916(85)90383-5",
    journal = "Annals Phys.",
    volume = "163",
    pages = "288",
    year = "1985"
}

@article{Bilal:2008qx,
    author = "Bilal, Adel",
    title = "{Lectures on Anomalies}",
    eprint = "0802.0634",
    archivePrefix = "arXiv",
    primaryClass = "hep-th",
    reportNumber = "LPTENS-08-05",
    month = "2",
    year = "2008"
}

@article{Alvarez-Gaume:2022aak,
    author = "Alvarez-Gaume, Luis and Vazquez-Mozo, Miguel A.",
    title = "{Anomalies and the Green-Schwarz Mechanism}",
    eprint = "2211.06467",
    archivePrefix = "arXiv",
    primaryClass = "hep-th",
    month = "11",
    year = "2022"
}

@incollection{10.1093/acprof:oso/9780198507628.003.0004,
    author = {Bertlmann, Reinhold A.},
    title = "{177Anomalies in QFT}",
    booktitle = "{Anomalies in Quantum Field Theory}",
    publisher = {Oxford University Press},
    year = {2000},
    month = {11},
}

@article{Fukaya:2019qlf,
    author = "Fukaya, Hidenori and Furuta, Mikio and Matsuo, Shinichiroh and Onogi, Tetsuya and Yamaguchi, Satoshi and Yamashita, Mayuko",
    title = "{The Atiyah\textendash{}Patodi\textendash{}Singer Index and Domain-Wall Fermion Dirac Operators}",
    eprint = "1910.01987",
    archivePrefix = "arXiv",
    primaryClass = "math.DG",
    reportNumber = "OU-HET-1026",
    doi = "10.1007/s00220-020-03806-0",
    journal = "Commun. Math. Phys.",
    volume = "380",
    number = "3",
    pages = "1295--1311",
    year = "2020"
}

@article{Gaiotto:2014kfa,
    author = "Gaiotto, Davide and Kapustin, Anton and Seiberg, Nathan and Willett, Brian",
    title = "{Generalized Global Symmetries}",
    eprint = "1412.5148",
    archivePrefix = "arXiv",
    primaryClass = "hep-th",
    doi = "10.1007/JHEP02(2015)172",
    journal = "JHEP",
    volume = "02",
    pages = "172",
    year = "2015"
}

@article{Gross:1987kza,
    author = "Gross, David J. and Mende, Paul F.",
    title = "{The High-Energy Behavior of String Scattering Amplitudes}",
    reportNumber = "PUPT-1062",
    doi = "10.1016/0370-2693(87)90355-8",
    journal = "Phys. Lett. B",
    volume = "197",
    pages = "129--134",
    year = "1987"
}

@article{Gross:1987ar,
    author = "Gross, David J. and Mende, Paul F.",
    title = "{String Theory Beyond the Planck Scale}",
    reportNumber = "PUPT-1067",
    doi = "10.1016/0550-3213(88)90390-2",
    journal = "Nucl. Phys. B",
    volume = "303",
    pages = "407--454",
    year = "1988"
}

@article{Mende:1989wt,
    author = "Mende, Paul F. and Ooguri, Hirosi",
    title = "{Borel Summation of String Theory for Planck Scale Scattering}",
    reportNumber = "MIT-CTP-1800, EFI-89-62",
    doi = "10.1016/0550-3213(90)90202-O",
    journal = "Nucl. Phys. B",
    volume = "339",
    pages = "641--662",
    year = "1990"
}

@article{Witten:1995ex,
    author = "Witten, Edward",
    title = "{String theory dynamics in various dimensions}",
    eprint = "hep-th/9503124",
    archivePrefix = "arXiv",
    reportNumber = "IASSNS-HEP-95-18",
    doi = "10.1016/0550-3213(95)00158-O",
    journal = "Nucl. Phys. B",
    volume = "443",
    pages = "85--126",
    year = "1995"
}

@article{Vafa:2005ui,
	title        = {{The String landscape and the swampland}},
	author       = {Vafa, Cumrun},
	year         = 2005,
	month        = 9,
	eprint       = {hep-th/0509212},
	archiveprefix = {arXiv},
	reportnumber = {HUTP-05-A043}
}

@article{Brennan:2017rbf,
	title        = {{The String Landscape, the Swampland, and the Missing Corner}},
	author       = {Brennan, T. Daniel and Carta, Federico and Vafa, Cumrun},
	year         = 2017,
	journal      = {PoS},
	booktitle    = {{Proceedings, Theoretical Advanced Study Institute in Elementary Particle Physics: Physics at the Fundamental Frontier (TASI 2017): Boulder, CO, USA, June 5-30, 2017}},
	volume       = {TASI2017},
	pages        = {015},
	doi          = {10.22323/1.305.0015},
	eprint       = {1711.00864},
	archiveprefix = {arXiv},
	primaryclass = {hep-th},
	reportnumber = {IFT-UAM-CSIC-17-105},
	slaccitation = {%%CITATION = ARXIV:1711.00864;%%}
}

@article{Palti:2019pca,
	title        = {{The Swampland: Introduction and Review}},
	author       = {Palti, Eran},
	year         = 2019,
	journal      = {Fortsch. Phys.},
	volume       = 67,
	number       = 6,
	pages        = 1900037,
	doi          = {10.1002/prop.201900037},
	eprint       = {1903.06239},
	archiveprefix = {arXiv},
	primaryclass = {hep-th},
	reportnumber = {MPP-2019-53},
	slaccitation = {%%CITATION = ARXIV:1903.06239;%%}
}

@article{vanBeest:2021lhn,
    author = "van Beest, Marieke and Calder\'on-Infante, Jos\'e and Mirfendereski, Delaram and Valenzuela, Irene",
    title = "{Lectures on the Swampland Program in String Compactifications}",
    eprint = "2102.01111",
    archivePrefix = "arXiv",
    primaryClass = "hep-th",
    doi = "10.1016/j.physrep.2022.09.002",
    journal = "Phys. Rept.",
    volume = "989",
    pages = "1--50",
    year = "2022"
}

@article{Grana:2021zvf,
    author = "Gra\~na, Mariana and Herr\'aez, Alvaro",
    title = "{The Swampland Conjectures: A Bridge from Quantum Gravity to Particle Physics}",
    eprint = "2107.00087",
    archivePrefix = "arXiv",
    primaryClass = "hep-th",
    doi = "10.3390/universe7080273",
    journal = "Universe",
    volume = "7",
    number = "8",
    pages = "273",
    year = "2021"
}

@article{Agmon:2022thq,
    author = "Agmon, Nathan Benjamin and Bedroya, Alek and Kang, Monica Jinwoo and Vafa, Cumrun",
    title = "{Lectures on the string landscape and the Swampland}",
    eprint = "2212.06187",
    archivePrefix = "arXiv",
    primaryClass = "hep-th",
    month = "12",
    year = "2022"
}

@article{Kim:2019ths,
    author = "Kim, Hee-Cheol and Tarazi, Houri-Christina and Vafa, Cumrun",
    title = "{Four-dimensional $\mathbf{\mathcal{N}=4}$ SYM theory and the swampland}",
    eprint = "1912.06144",
    archivePrefix = "arXiv",
    primaryClass = "hep-th",
    doi = "10.1103/PhysRevD.102.026003",
    journal = "Phys. Rev. D",
    volume = "102",
    number = "2",
    pages = "026003",
    year = "2020"
}

@article{Katz:2020ewz,
    author = "Katz, Sheldon and Kim, Hee-Cheol and Tarazi, Houri-Christina and Vafa, Cumrun",
    title = "{Swampland Constraints on 5d $\mathcal{N}=1$ Supergravity}",
    eprint = "2004.14401",
    archivePrefix = "arXiv",
    primaryClass = "hep-th",
    doi = "10.1007/JHEP07(2020)080",
    journal = "JHEP",
    volume = "07",
    pages = "080",
    year = "2020"
}

@article{Montero:2020icj,
    author = "Montero, Miguel and Vafa, Cumrun",
    title = "{Cobordism Conjecture, Anomalies, and the String Lamppost Principle}",
    eprint = "2008.11729",
    archivePrefix = "arXiv",
    primaryClass = "hep-th",
    doi = "10.1007/JHEP01(2021)063",
    journal = "JHEP",
    volume = "01",
    pages = "063",
    year = "2021"
}

@article{Hamada:2021bbz,
    author = "Hamada, Yuta and Vafa, Cumrun",
    title = "{8d supergravity, reconstruction of internal geometry and the Swampland}",
    eprint = "2104.05724",
    archivePrefix = "arXiv",
    primaryClass = "hep-th",
    doi = "10.1007/JHEP06(2021)178",
    journal = "JHEP",
    volume = "06",
    pages = "178",
    year = "2021"
}

@article{Tarazi:2021duw,
    author = "Tarazi, Houri-Christina and Vafa, Cumrun",
    title = "{On The Finiteness of 6d Supergravity Landscape}",
    eprint = "2106.10839",
    archivePrefix = "arXiv",
    primaryClass = "hep-th",
    month = "6",
    year = "2021"
}

@article{Bedroya:2021fbu,
    author = "Bedroya, Alek and Hamada, Yuta and Montero, Miguel and Vafa, Cumrun",
    title = "{Compactness of brane moduli and the String Lamppost Principle in d \ensuremath{>} 6}",
    eprint = "2110.10157",
    archivePrefix = "arXiv",
    primaryClass = "hep-th",
    doi = "10.1007/JHEP02(2022)082",
    journal = "JHEP",
    volume = "02",
    pages = "082",
    year = "2022"
}

@article{Martucci:2022krl,
    author = "Martucci, Luca and Risso, Nicolo and Weigand, Timo",
    title = "{Quantum gravity bounds on $ \mathcal{N} $ = 1 effective theories in four dimensions}",
    eprint = "2210.10797",
    archivePrefix = "arXiv",
    primaryClass = "hep-th",
    doi = "10.1007/JHEP03(2023)197",
    journal = "JHEP",
    volume = "03",
    pages = "197",
    year = "2023"
}

@article{Baykara:2023plc,
    author = "Baykara, Zihni Kaan and Hamada, Yuta and Tarazi, Houri-Christina and Vafa, Cumrun",
    title = "{On the String Landscape Without Hypermultiplets}",
    eprint = "2309.15152",
    archivePrefix = "arXiv",
    primaryClass = "hep-th",
    reportNumber = "KEK-TH-2552",
    month = "9",
    year = "2023"
}

@article{Hayashi:2023hqa,
    author = "Hayashi, Hirotaka and Kim, Hee-Cheol and Kim, Minsung",
    title = "{Spectra of BPS Strings in 6d Supergravity and the Swampland}",
    eprint = "2310.12219",
    archivePrefix = "arXiv",
    primaryClass = "hep-th",
    month = "10",
    year = "2023"
}

@article{Witten:1982fp,
    author = "Witten, Edward",
    editor = "Shifman, Mikhail A.",
    title = "{An SU(2) Anomaly}",
    doi = "10.1016/0370-2693(82)90728-6",
    journal = "Phys. Lett. B",
    volume = "117",
    pages = "324--328",
    year = "1982"
}

@article{Witten:1985xe,
    author = "Witten, Edward",
    editor = "Salam, A. and Sezgin, E.",
    title = "{GLOBAL GRAVITATIONAL ANOMALIES}",
    reportNumber = "PRINT-85-0246 (PRINCETON)",
    doi = "10.1007/BF01212448",
    journal = "Commun. Math. Phys.",
    volume = "100",
    pages = "197",
    year = "1985"
}

@article{Basile:2023knk,
    author = "Basile, Ivano and Debray, Arun and Delgado, Matilda and Montero, Miguel",
    title = "{Global anomalies \& bordism of non-supersymmetric strings}",
    eprint = "2310.06895",
    archivePrefix = "arXiv",
    primaryClass = "hep-th",
    reportNumber = "IFT-23-129",
    month = "10",
    year = "2023"
}

@article{Guerrieri:2021ivu,
    author = "Guerrieri, Andrea and Penedones, Joao and Vieira, Pedro",
    title = "{Where Is String Theory in the Space of Scattering Amplitudes?}",
    eprint = "2102.02847",
    archivePrefix = "arXiv",
    primaryClass = "hep-th",
    doi = "10.1103/PhysRevLett.127.081601",
    journal = "Phys. Rev. Lett.",
    volume = "127",
    number = "8",
    pages = "081601",
    year = "2021"
}

@article{Guerrieri:2022sod,
    author = "Guerrieri, Andrea and Murali, Harish and Penedones, Joao and Vieira, Pedro",
    title = "{Where is M-theory in the space of scattering amplitudes?}",
    eprint = "2212.00151",
    archivePrefix = "arXiv",
    primaryClass = "hep-th",
    doi = "10.1007/JHEP06(2023)064",
    journal = "JHEP",
    volume = "06",
    pages = "064",
    year = "2023"
}

@article{Polchinski:2003bq,
	title        = {{Monopoles, duality, and string theory}},
	author       = {Polchinski, Joseph},
	year         = 2004,
	journal      = {Int. J. Mod. Phys. A},
	volume       = {19S1},
	pages        = {145--156},
	doi          = {10.1142/S0217751X0401866X},
	editor       = {Baer, H. and Belyaev, A.},
	eprint       = {hep-th/0304042},
	archiveprefix = {arXiv}
}

@article{Banks:2010zn,
	title        = {{Symmetries and Strings in Field Theory and Gravity}},
	author       = {Banks, Tom and Seiberg, Nathan},
	year         = 2011,
	journal      = {Phys. Rev. D},
	volume       = 83,
	pages        = {084019},
	doi          = {10.1103/PhysRevD.83.084019},
	eprint       = {1011.5120},
	archiveprefix = {arXiv},
	primaryclass = {hep-th},
	slaccitation = {%%CITATION = ARXIV:1011.5120;%%}
}

@article{Harlow:2018tng,
	title        = {{Symmetries in quantum field theory and quantum gravity}},
	author       = {Harlow, Daniel and Ooguri, Hirosi},
	year         = 2021,
	journal      = {Commun. Math. Phys.},
	volume       = 383,
	number       = 3,
	pages        = {1669--1804},
	doi          = {10.1007/s00220-021-04040-y},
	eprint       = {1810.05338},
	archiveprefix = {arXiv},
	primaryclass = {hep-th}
}

@article{McNamara:2019rup,
	title        = {{Cobordism Classes and the Swampland}},
	author       = {McNamara, Jacob and Vafa, Cumrun},
	year         = 2019,
	month        = 9,
	eprint       = {1909.10355},
	archiveprefix = {arXiv},
	primaryclass = {hep-th}
}

@article{Heidenreich:2020pkc,
    author = "Heidenreich, Ben and McNamara, Jacob and Montero, Miguel and Reece, Matthew and Rudelius, Tom and Valenzuela, Irene",
    title = "{Chern-Weil global symmetries and how quantum gravity avoids them}",
    eprint = "2012.00009",
    archivePrefix = "arXiv",
    primaryClass = "hep-th",
    reportNumber = "ACFI-T20-16",
    doi = "10.1007/JHEP11(2021)053",
    journal = "JHEP",
    volume = "11",
    pages = "053",
    year = "2021"
}

@article{Heidenreich:2021xpr,
    author = "Heidenreich, Ben and McNamara, Jacob and Montero, Miguel and Reece, Matthew and Rudelius, Tom and Valenzuela, Irene",
    title = "{Non-invertible global symmetries and completeness of the spectrum}",
    eprint = "2104.07036",
    archivePrefix = "arXiv",
    primaryClass = "hep-th",
    reportNumber = "ACFI-T21-03",
    doi = "10.1007/JHEP09(2021)203",
    journal = "JHEP",
    volume = "09",
    pages = "203",
    year = "2021"
}

@article{McNamara:2021cuo,
    author = "McNamara, Jacob",
    title = "{Gravitational Solitons and Completeness}",
    eprint = "2108.02228",
    archivePrefix = "arXiv",
    primaryClass = "hep-th",
    month = "8",
    year = "2021"
}

@phdthesis{McNamaraThesis,
    author = "McNamara, Jacob",
    title = "{The Kinematics of Quantum Gravity}",
    eprint = {https://nrs.harvard.edu/URN-3:HUL.INSTREPOS:37372201},
    school = "Harvard University",
    year = "2022"
}

@article{Stolz:2011zj,
    author = "Stolz, Stephan and Teichner, Peter",
    editor = "Sati, Hisham and Schreiber, Urs",
    title = "{Supersymmetric field theories and generalized cohomology}",
    eprint = "1108.0189",
    archivePrefix = "arXiv",
    primaryClass = "math.AT",
    pages = "279--340",
    month = "8",
    year = "2011"
}

@book{douglas2014topological,
  title={Topological Modular Forms},
  author={Douglas, C.L. and Francis, J. and Henriques, A.G. and Hill, M.A.},
  isbn={9781470418847},
  lccn={14029076},
  series={Mathematical Surveys and Monographs},
  url={https://books.google.de/books?id=n0W9BQAAQBAJ},
  year={2014},
  publisher={American Mathematical Society}
}

@InProceedings{hopkins,
author="Hopkins, Michael J.",
editor="Chatterji, S. D.",
title="Topological Modular Forms, the {W}itten Genus, and the Theorem of the Cube",
booktitle="Proceedings of the International Congress of Mathematicians",
year="1995",
publisher="Birkh{\"a}user Basel",
address="Basel",
pages="554--565",
isbn="978-3-0348-9078-6"
}

@article{Gukov:2018iiq,
    author = "Gukov, Sergei and Pei, Du and Putrov, Pavel and Vafa, Cumrun",
    title = "{4-manifolds and topological modular forms}",
    eprint = "1811.07884",
    archivePrefix = "arXiv",
    primaryClass = "hep-th",
    reportNumber = "CALT-TH-2018-034",
    doi = "10.1007/JHEP05(2021)084",
    journal = "JHEP",
    volume = "05",
    pages = "084",
    year = "2021"
}

@article{Gaiotto:2019asa,
    author = "Gaiotto, Davide and Johnson-Freyd, Theo and Witten, Edward",
    title = "{A Note On Some Minimally Supersymmetric Models In Two Dimensions}",
    eprint = "1902.10249",
    archivePrefix = "arXiv",
    primaryClass = "hep-th",
    month = "2",
    year = "2019"
}

@misc{gepner2023equivariant,
      title={On equivariant topological modular forms}, 
      author={David Gepner and Lennart Meier},
      year={2023},
      eprint={2004.10254},
      archivePrefix={arXiv},
      primaryClass={math.AT}
}

@article{Chua_2022,
    journal = "Journal of Homotopy and Related Structures",
	doi = {10.1007/s40062-021-00297-1},
    year = 2022,
	month = {1},
	publisher = {Springer Science and Business Media {LLC}},
	volume = {17},
	number = {1},
	pages = {23-75},
	author = {Dexter Chua},
	title = {\texorpdfstring{$C_2$}{C2}-equivariant topological modular forms}
}

@article{Bershadsky:1997sb,
    author = "Bershadsky, Michael and Vafa, Cumrun",
    title = "{Global anomalies and geometric engineering of critical theories in six-dimensions}",
    eprint = "hep-th/9703167",
    archivePrefix = "arXiv",
    reportNumber = "HUTP-97-A013",
    month = "3",
    year = "1997"
}

@article{Monnier:2018nfs,
    author = "Monnier, Samuel and Moore, Gregory W.",
    title = "{Remarks on the Green\textendash{}Schwarz Terms of Six-Dimensional Supergravity Theories}",
    eprint = "1808.01334",
    archivePrefix = "arXiv",
    primaryClass = "hep-th",
    doi = "10.1007/s00220-019-03341-7",
    journal = "Commun. Math. Phys.",
    volume = "372",
    number = "3",
    pages = "963--1025",
    year = "2019"
}

@article{Lee:2022spd,
    author = "Lee, Yasunori and Yonekura, Kazuya",
    title = "{Global anomalies in 8d supergravity}",
    eprint = "2203.12631",
    archivePrefix = "arXiv",
    primaryClass = "hep-th",
    reportNumber = "IPMU-22-0011, TU-1148",
    doi = "10.1007/JHEP07(2022)125",
    journal = "JHEP",
    volume = "07",
    pages = "125",
    year = "2022"
}

@article{Lee:2020gvu,
    author = "Lee, Seung-Joo and Lerche, Wolfgang and Lockhart, Guglielmo and Weigand, Timo",
    title = "{Quasi-Jacobi forms, elliptic genera and strings in four dimensions}",
    eprint = "2005.10837",
    archivePrefix = "arXiv",
    primaryClass = "hep-th",
    doi = "10.1007/JHEP01(2021)162",
    journal = "JHEP",
    volume = "01",
    pages = "162",
    year = "2021"
}

@article{Arkani-Hamed:2006emk,
    author = "Arkani-Hamed, Nima and Motl, Lubos and Nicolis, Alberto and Vafa, Cumrun",
    title = "{The String landscape, black holes and gravity as the weakest force}",
    eprint = "hep-th/0601001",
    archivePrefix = "arXiv",
    reportNumber = "HUTP-05-A0057",
    doi = "10.1088/1126-6708/2007/06/060",
    journal = "JHEP",
    volume = "06",
    pages = "060",
    year = "2007"
}

@article{Harlow:2022ich,
    author = "Harlow, Daniel and Heidenreich, Ben and Reece, Matthew and Rudelius, Tom",
    title = "{Weak gravity conjecture}",
    eprint = "2201.08380",
    archivePrefix = "arXiv",
    primaryClass = "hep-th",
    reportNumber = "ACFI-T22-01",
    doi = "10.1103/RevModPhys.95.035003",
    journal = "Rev. Mod. Phys.",
    volume = "95",
    number = "3",
    pages = "035003",
    year = "2023"
}

\end{document}